\documentclass[apj]{emulateapj}

\usepackage{amsmath}
\usepackage{lscape}
\usepackage{apjfonts}
\usepackage{graphicx}
\usepackage{natbib}
\usepackage{color}
\usepackage{fancybox}
\usepackage{subfigure}
\slugcomment{Received 2013 December 23; accepted 2014 May 27; ApJ in press}

\newcommand{\angstrom}{{\rm \AA}}

\newcommand{\hbeta}{H{$\beta$}}
\newcommand{\halpha}{H{$\alpha$}}

\newcommand{\CIV}{C{\sevenrm\,IV}}
\newcommand{\FeII}{Fe{\sevenrm\,II}}

\newcommand{\MgII}{Mg{\sevenrm\,II}}

\newcommand{\OIII}{[O{\sevenrm\,III}]}

\newcommand{\OIIIb}{[O{\sevenrm\,III}]\,$\lambda$5007}
\newcommand{\OIIIc}{[O{\sevenrm\,III}]\,$\lambda\lambda$4959,5007}

\font\sevenrm=cmr7 scaled 1000

\begin{document}

\title{Constraining Sub-Parsec Binary Supermassive Black Holes
in Quasars with Multi-Epoch Spectroscopy. II. The Population
with Kinematically Offset Broad Balmer Emission
Lines\altaffilmark{1}}

\shorttitle{Sub-parsec BBHs in Quasars. II}

\shortauthors{LIU ET AL.}
\author{Xin Liu\altaffilmark{2,9}, Yue Shen\altaffilmark{3,9},
Fuyan Bian\altaffilmark{4,5,10}, Abraham
Loeb\altaffilmark{6,7}, and Scott Tremaine\altaffilmark{8}}
%\author{the binary team}

\altaffiltext{1}{Based, in part, on data obtained at the MMT,
ARC 3.5\,m, and FLWO 1.5\,m Telescopes.}

\altaffiltext{2}{Department of Physics and Astronomy,
University of California, Los Angeles, CA 90095, USA;
xinliu@astro.ucla.edu}

\altaffiltext{3}{Carnegie Observatories, 813 Santa Barbara
Street, Pasadena, CA 91101, USA}

\altaffiltext{4}{Research School of Astronomy and Astrophysics,
Australian National University, Canberra, ACT 2611, Australia}

\altaffiltext{5}{Steward Observatory, University of Arizona,
933 N. Cherry Avenue, Tucson, AZ 85721, USA}

\altaffiltext{6}{Harvard-Smithsonian Center for Astrophysics,
60 Garden Street, Cambridge, MA 02138, USA}

\altaffiltext{7}{Institute for Theory and Computation, Harvard
University, 60 Garden Street, Cambridge, MA 02138, USA}

\altaffiltext{8}{Institute for Advanced Study, Princeton, NJ
08540, USA}

\altaffiltext{9}{Hubble Fellow}

\altaffiltext{10}{Stromlo Fellow}

\begin{abstract}
A small fraction of quasars have long been known to show bulk
velocity offsets (of a few hundred to thousands of km s$^{-1}$)
in the broad Balmer lines with respect to the systemic redshift
of the host galaxy. Models to explain these offsets usually
invoke broad-line region gas kinematics/asymmetry around single
black holes (BHs), orbital motion of massive ($\sim$sub-parsec
(sub-pc)) binary black holes (BBHs), or recoil BHs, but
single-epoch spectra are unable to distinguish between these
scenarios. The line-of-sight (LOS) radial velocity (RV) shifts
from long-term spectroscopic monitoring can be used to test the
BBH hypothesis. We have selected a sample of 399 quasars with
kinematically offset broad \hbeta\ lines from the Sloan Digital
Sky Survey (SDSS) Seventh Data Release quasar catalog, and have
conducted second-epoch optical spectroscopy for 50 of them.
Combined with the existing SDSS spectra, the new observations
enable us to constrain the LOS RV shifts of broad \hbeta\ lines
with a rest-frame baseline of a few years to nearly a decade.
While previous work focused on objects with extreme velocity
offset ($> 10^3$ km s$^{-1}$), we explore the parameter space
with smaller (a few hundred km s$^{-1}$) yet significant
offsets (99.7\% confidence). Using cross-correlation analysis,
we detect significant (99\% confidence) radial accelerations in
the broad \hbeta\ lines in 24 of the 50 objects, of
$\sim$10--$200$ km s$^{-1}$ yr$^{-1}$ with a median measurement
uncertainty of $\sim$10 km s$^{-1}$ yr$^{-1}$, implying a high
fraction of variability of the broad-line velocity on
multi-year timescales. We suggest that 9 of the 24 detections
are sub-pc BBH candidates, which show consistent velocity
shifts independently measured from a second broad line (either
\halpha\ or \MgII ) without significant changes in the
broad-line profiles. Combining the results on the general
quasar population studied in Paper I, we find a tentative
anti-correlation between the velocity offset in the first-epoch
spectrum and the average acceleration between two epochs, which
could be explained by orbital phase modulation when the time
separation between two epochs is a non-negligible fraction of
the orbital period of the motion causing the line displacement.
We discuss the implications of our results for the
identification of sub-pc BBH candidates in offset-line quasars
and for the constraints on their frequency and orbital
parameters.
\end{abstract}


%%%%%%%%%%%%%%%%%%%%%%%%%%%%%%
%%%%%%%%%%%%%%%%%%%%%%%%%%%%%%
\section{Introduction}\label{sec:intro}

This paper describes a search for temporal radial velocity (RV)
shifts (i.e., accelerations) of quasar broad emission lines as
evidence for massive, sub-parsec (sub-pc) binary black holes
(BBHs). Our working hypothesis is that only one of the two BHs
in the binary is active and powering its own broad-line region
(BLR), and that the binary separation is sufficiently large
compared to the BLR size that the broad line velocity traces
the binary motion, yet small enough that the acceleration can
be detected over the temporal baseline of our observations. In
\citet[][hereafter Paper I]{shen13}, we have reported results
for the general quasar population. Here, in the second paper of
this series, we target a sample of quasars with offset broad
Balmer emission lines, to probe a different and complementary
parameter space. To date, no convincing case of a sub-pc BBH
has been found\footnote{The BL Lac object OJ~287 (at $z =
0.306$) has quasi-periodic optical outbursts at 12-yr intervals
\citep{sillanpaa96} and has been suggested to host a BBH with a
binary separation of 0.06 pc
\citep[e.g.,][]{lehto96,valtaoja00,valtonen08}, but alternative
scenarios remain viable
\citep[e.g.,][]{sillanpaa88,katz97,hughes98,villata98,igumenshchev99,villforth10}.
The radio galaxy 3C 66B (at $z=0.02$) has been suggested to
host a BBH with a binary separation of $<0.02$ pc based on the
elliptical motion of the unresolved radio core with a period of
1.05 yr. The hypothetical binary would have coalesced in
$\sim5$ yr by gravitational radiation \citep{sudou03}, and this
rapid decay rate has subsequently been ruled out using pulsar
timing observations \citep{jenet04}.}.

\subsection{Motivation to Search for Sub-pc
BBHs}\label{subsec:motivation}

Theoretical models for the observed evolution of galaxies and
BHs suggest that BBHs should be common
\citep{begelman80,roos81,milosavljevic01,yu02,colpi09,DEGN}.
 are expected to be the most abundant at binary separations
between $\sim1$ pc and $\sim10^{-3}$ pc, where the orbital
decay has slowed due to loss cone depletion
\citep[e.g.,][]{begelman80,gould00}, and the emission of
gravitational waves (GWs) is not yet efficient
\citep{thorne76,Centrella2010}. Whether or not the BBH orbit
enters the GW regime, and the relative importance of gas and
stellar dynamical processes in facilitating such decays, are
still subject to active debate
\citep[e.g.,][]{gould00,Escala2005,merritt05,Mayer2007,Dotti2012,Vasiliev2013}.
A determination of the frequency of sub-pc BBHs would help
constrain the orbital decay rate and inform the expected
abundance of low-frequency GW sources
\citep[e.g.,][]{sesana07,trias08,amaro12}. It would also test
models of BH assembly and growth in hierarchical cosmologies
\citep[e.g.,][]{volonteri09,Kulkarni2012}.

\subsection{Evidence for Massive BH Pairs on Various
Scales}\label{subsec:evidence}

Merger models have been successful in reproducing the observed
demographics and spatial distribution of galaxies and quasars
across cosmic time
\citep[e.g.,][]{kauffmann00,volonteri03,wyithe03,hopkins08,shen09}.
At wide separations, from tens of kpc to $\sim$kpc scales,
there is strong direct evidence for active BH pairs in merging
or merged galaxies \citep[e.g.,][]
{komossa03,ballo04,bianchi08,comerford09,comerford11a,comerford11b,wang09,smith09,green10,liu10,liu10b,liu11a,liu11c,liu12a,shen10b,shen10c,piconcelli10,fabbiano11,fu11,fu11b,greene11,koss11,koss12,mazzarella11,mcgurk11};
their observed frequency can be reasonably well explained by
merger models \citep{yu11,vanwassenhove12}.

At $\sim$pc to sub-pc separations, however, CSO 0402+379
remains the only secure case known: this is a compact
flat-spectrum double radio source \citep{rodriguez06}
serendipitously discovered by the Very Long Baseline Array,
with a projected separation of $\sim 7$ pc. While sub-pc BBHs
are largely elusive, there is {\it indirect} evidence, at least
in local massive elliptical galaxies, that they may have had a
strong dynamical impact on the stellar structures at the
centers of galaxies by scouring out flat cores
\citep[e.g.,][]{ebisuzaki91,faber97,graham04,kormendy09,DEGN}.

Sub-pc BBHs may be abundant but either mostly inactive or just
difficult to resolve. Direct imaging searches in the radio have
proven extremely challenging \citep[e.g.,][]{burke11}, largely
because of both insufficient resolving power (even the
unparalleled resolution of the very long baseline
interferometry can only resolve very nearby sources, e.g.,
$z<0.01$ for $\sim$pc or $z<0.001$ for $\sim0.1$pc separations)
and the small probability for both BHs to be simultaneously
bright in the radio. On the other hand, the dynamical signature
of binary orbital motion, in principle, can be used to identify
candidate sub-pc BBHs
\citep[e.g.,][]{komberg68,gaskell83,loeb10,shen10,eracleous11,popovic11},
in analogy to spectroscopic binary stars
\citep[e.g.,][]{Abt1976}. In particular, it has long been
proposed that BBH candidates may be selected from quasars whose
broad emission lines are offset from the systemic velocities of
their host galaxies
\citep[e.g.,][]{gaskell83,peterson87,halpern88,bogdanovic09}.
Large spectroscopic surveys have enabled the selection of
statistical samples of quasars with offset broad lines
\citep{bonning07,boroson10,shen11,tsalmantza11,eracleous11}, as
well as individual interesting cases
\citep[e.g.,][]{boroson09,shields09}.

However, there are alternative, and perhaps more natural,
explanations for broad-line velocity offsets, such as gas
motion in the accretion disk around a single BH
\citep[so-called ``disk emitters'';
e.g.,][]{Chen1989,eracleous95,Shapovalova11,eracleous03,strateva03}.
Single-peaked but offset broad emission lines may be disks with
high emissivity asymmetry or the other peak may be too weak to
identify \citep[e.g.,][]{chornock10}. Yet another scenario is a
recoiled BH
\citep[e.g.,][]{campanelli07,bonning07,loeb07,shields09} from
the anisotropic GW emission following BBH coalescence
\citep[e.g.,][]{Fitchett1983,Baker2006}, which carries the
inner part of its accretion disk and a BLR with it, and could
fuel a continuing quasar phase for millions of years
\citep[e.g.,][]{loeb07,Blecha2011}. The recoil scenario is
perhaps less likely for offsets $\gtrsim$ a few hundred km
s$^{-1}$ \citep[e.g.,][]{bogdanovic07,Dotti2010}, since recoil
velocities tend to be smaller than this when the BH spin axes
are aligned, which tends to happen in gas-rich mergers. Another
mechanism for even larger kicks is slingshot ejection of a BH
from a triple system, formed when a new galaxy merger occurs
before a pre-existing BBH has coalesced
\citep{hoffman07,Kulkarni2012}.

\subsection{Summary of Previous Radial Velocity
Monitoring}\label{subsec:sum_previous}

The temporal RV shift of broad emission lines offers a
promising test for the BBH hypothesis
\citep[e.g.,][]{eracleous97,shen10}. Dedicated long-term (i.e.,
more than a few years) spectroscopic monitoring programs are
needed, as the binary orbital periods are typically several
decades to several centuries \citep[e.g.,][]{yu02,loeb10}.
Depending on properties of the broad emission lines in
single-epoch spectra, there are the following strategies for
looking for BBHs using temporal RV shifts:
\begin{enumerate}

\item[1] monitor quasars with double-peaked broad lines
    (which, by definition, usually show extreme (i.e., $>$
    a few thousand km s$^{-1}$) offset velocities in both
    peaks);

\item[2] monitor quasars having single-peaked broad lines
    with significant velocity offsets;

\item[3] monitor the much larger population of quasars
    having single-peaked broad lines without offsets.

\end{enumerate}

Most earlier spectroscopic monitoring work has focused on
quasars with double-peaked broad lines, most notably 3C 390.3
\citep{gaskell96a,eracleous97,Shapovalova11}, Arp 102B
\citep{halpern88}, NGC 5548
\citep{peterson87,Shapovalova2004,Sergeev2007}, and NGC 4151
\citep{Shapovalova2010,Bon2012}. Long-term spectroscopic
monitoring studies of quasars with double-peaked broad lines
suggest that most of them are likely explained by emission from
gas in the outer accretion disk rather than BBHs
\citep[e.g.,][]{halpern88,eracleous99,eracleous03}. The
long-term line profile/velocity changes in such objects are
likely caused by transient dynamical processes (e.g., shocks
from tidal perturbations) or physical changes (e.g., emissivity
variation and/or precession of fragmented spiral arms) in the
line emitting region in the outer accretion disk around single
BHs \citep{gezari07,lewis10}. Furthermore, there are reasons to
expect that most BBHs do not exhibit double-peaked broad lines,
because there is limited parameter space, if any, for two
well-separated broad-line peaks to be associated with two
physically distinct BLRs \citep[e.g.,][]{shen10}.

While double-peaked broad lines are likely not a promising
diagnostic to look for BBHs, the case remains open for
single-peaked broad-line offsets. In principle, the probability
of having one BH active is much higher than having both BHs
simultaneously active, and the allowed binary parameter space
is also larger than in the case with double-peaked broad lines.
Recently \citet{eracleous11} carried out the first systematic
spectroscopic followup study of quasars with offset broad
\hbeta\ lines. The authors identified 88 quasars with
broad-line offset velocities of $\gtrsim 10^3$ km s$^{-1}$ and
conducted second-epoch spectroscopy of 68 objects. They found
significant (at 99\% confidence) velocity shifts in 14 objects,
with accelerations in the range of [$-120$, 120] km s$^{-1}$
yr$^{-1}$. \citet{Decarli2013} also obtained second-epoch
spectra for 32 Sloan Digital Sky Survey \citep[SDSS;][]{york00}
quasars selected to have peculiar broad-line profiles
\citep{tsalmantza11}, such as large velocity offsets
($\gtrsim1000$ km s$^{-1}$) and/or double-peaked or asymmetric
line profiles. However, the conclusions from
\citet{Decarli2013} are relatively weak, as they measured
velocity shifts using model fits to the emission-line profiles
rather than the more powerful cross-correlation approach
adopted by \citet{eracleous11}, Paper I, and this work.
Nevertheless, at least for BBH candidates with symmetric line
profiles, the line fitting method should still be a sensible
(albeit less sensitive) tool to look for velocity shifts.

Finally, for the general quasar population (i.e., without
broad-line offsets), Paper I presents the first statistical
spectroscopic monitoring study based on the broad \hbeta\ lines
\citep[see also][for a similar study but based on \MgII\
$\lambda$2800]{Ju2013}.

\begin{figure*}
  \centering
    \includegraphics[width=59mm]{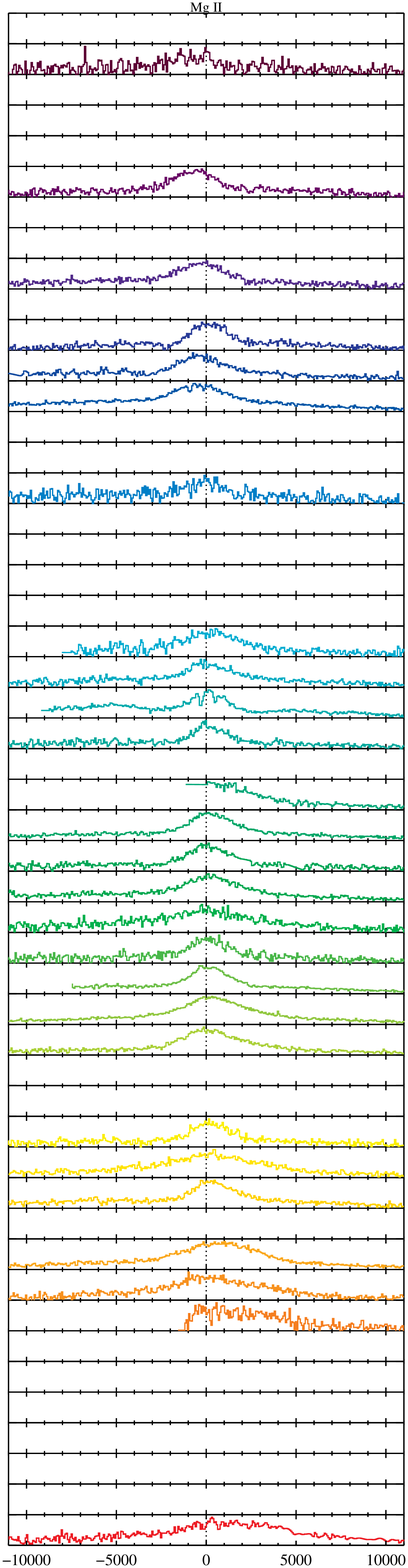}
    \includegraphics[width=59mm]{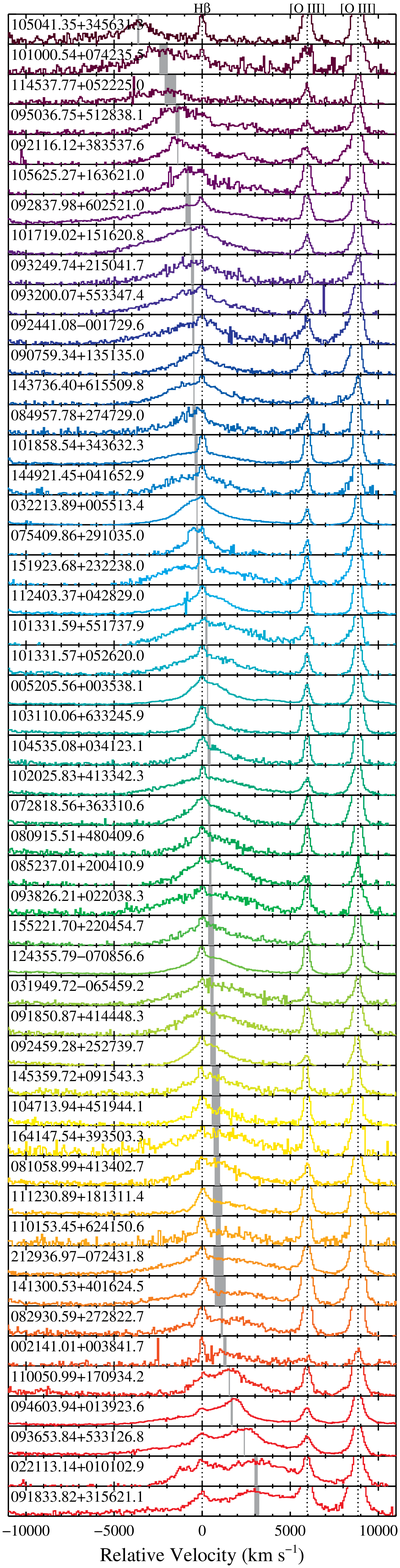}
    \includegraphics[width=59mm]{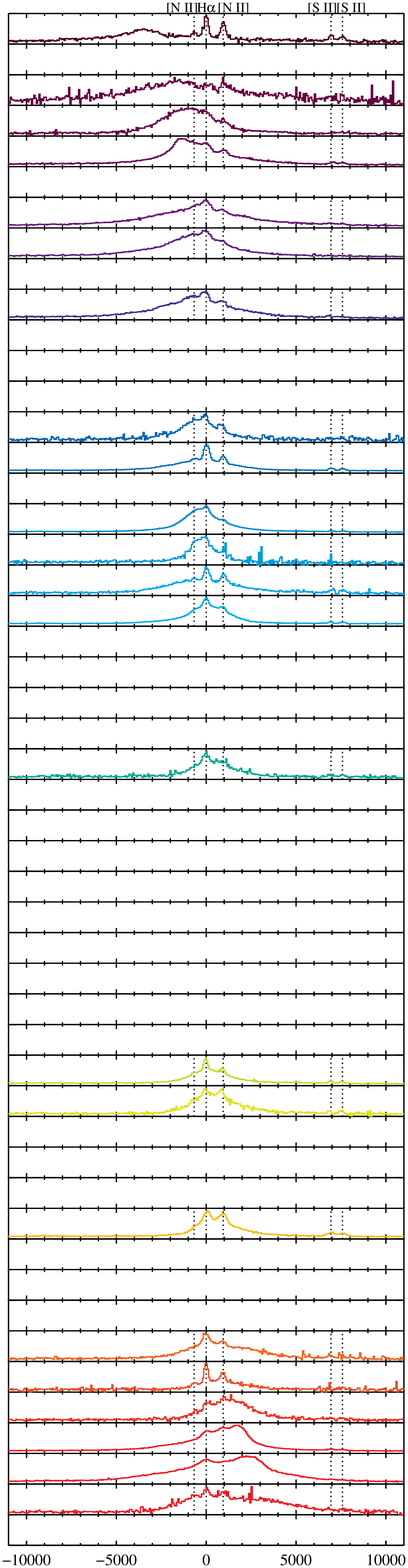}
    \caption{Examples of quasars with offset broad Balmer emission lines.
    Spectra are shown in arbitrarily normalized flux density,
    in the \MgII\ $\lambda$2800 (left column),
    \hbeta\ (middle column), and \halpha\ (right column) region.
    Objects are ordered according to broad \hbeta\ peak velocity offsets,
    as indicated by the gray shaded areas (with 1 $\sigma$ uncertainties).
    Wavelength scales are in relative velocity centered on the systemic velocity.
    Labels in the middle column are SDSS designations with R.A. and decl. for each object.
    }
    \label{fig:egspec}
\end{figure*}
\subsection{This Work}

We have compiled a large sample (399 objects) of significantly
(99.7\% confidence) offset broad-line quasars \citep[see
also][]{shen11}. They were selected from the quasar catalog
\citep{schneider10} in the Seventh Data Release
\citep[DR7;][]{SDSSDR7} of the SDSS. Their broad \hbeta\ lines
(as well as \halpha\ or \MgII , when available) show
significant velocity offsets of $\gtrsim$ a few hundred km
s$^{-1}$ relative to the systemic redshift determined from
narrow emission lines (Section \ref{sec:sample}). To mitigate
contamination by disk emitters, we focus on objects with
well-defined single-peaked offset broad lines. As a pilot
program, we have obtained a second-epoch optical spectrum,
separated by $\sim$5--10 yr (rest frame) from the original SDSS
spectrum, for a subset of 50 objects in the sample (Section
\ref{sec:ss}).  We measure the temporal velocity shifts of the
broad emission lines (Section \ref{sec:result}) using a
cross-correlation method (Section \ref{sec:analysis}) to
constrain the BBH hypothesis and model parameters (Section
\ref{sec:discuss}). We have detected significant (99\%
confidence) shifts in 24 objects (Section
\ref{subsec:detection}), of which we suggest that 9 are strong
BBH candidates (Section \ref{subsubsec:candidate}), and future
observations can definitively test this suggestion. Our results
have implications for the general approach of identifying BBH
candidates in offset-broad-line quasars, for the orbital
evolution of sub-pc BBHs, and for the forecast of low-frequency
GW sources (Section \ref{sec:discuss}).

Our program has two major differences from previous work.
First, we probe a parameter space in broad-line velocity offset
that is complementary to both the general quasar population
studied in Paper I and the population with more extreme offset
velocities examined in other statistical studies
\citep{eracleous11,Decarli2013}. Therefore by selection our
sample would probe a different parameter space in the BBH
scenario, since the velocity offset is determined by BH masses,
binary separations, orbital phases, and inclinations (e.g., the
Appendix). Second, in contrast to previous studies on
offset-line quasars which often adopt some minimum threshold on
the measured offset velocity, we apply the threshold on the
statistical significance of the measured offset velocity (where
the measurement uncertainty depends on signal-to-noise ratio
(S/N) of the spectrum and on line widths; Paper I). As a
result, our sample includes small but significant velocity
offsets. Combined with the results of Paper I, the improved
statistics and the enlarged dynamic range enable us to study
the statistical relation between velocity offset and
acceleration (Section \ref{subsec:vacorr}), to examine whether
the observations are consistent with the BBH hypothesis.
Understanding the origin of the broad-line velocity offsets in
single-epoch spectra is important for addressing the selection
biases of offset-line quasars to put their monitoring results
in the context of the general population.

Throughout this paper, we assume a concordance cosmology with
$\Omega_m = 0.3$, $\Omega_{\Lambda} = 0.7$, and $H_{0}=70$ km
s$^{-1}$ Mpc$^{-1}$, and use the AB magnitude system
\citep{oke74}. Following Paper I, we adopt ``offset'' to refer
to the velocity difference between two lines in single-epoch
spectra, and ``shift'' to denote changes in the line velocity
between two epochs. We quote velocity offsets relative to the
observer, i.e., negative values mean blueshifts. We define
accelerations relative to the original SDSS observations, where
positive means moving toward longer wavelengths. All time
intervals and relative velocities are in the quasar rest frames
by default, unless noted otherwise.

%%%%%%%%%%%%%%%%%%%%%%%%%%%%%%%%%%%%%%%%%%%%%%%%%%%%%%%%%%%%
\section{SDSS Quasars with Offset Broad Balmer Emission Lines}\label{sec:sample}

A fraction of quasars have long been noticed to have bulk
velocity offsets (both blueshifts and redshifts with absolute
velocities $\sim$a few hundred to thousand km s$^{-1}$)
in the broad permitted emission lines (e.g., \hbeta\ and
\halpha ) with respect to the systemic velocity (determined
from narrow emission lines and/or stellar absorption features)
of the host galaxies \citep[e.g.,][]{osterbrock82}. Quasars
with offset broad lines are rare \citep[$\sim$a few percent,
depending on offset velocity; e.g.,][]{bonning07}, but large
spectroscopic redshift surveys such as the SDSS have increased
the inventory of such objects by orders of magnitude
\citep[e.g.,][]{boroson10,shen11,tsalmantza11}. In this
section, we first describe the selection and general properties
of a sample of SDSS quasars with offset broad Balmer emission
lines (Section \ref{subsec:selection}). We then discuss
selection biases and uncertainties (Section
\ref{subsec:incompleteness}).

%%%%%%%%%%%%%%%%%%%%%%%%%%%%%%%%%%
\subsection{Sample Selection}\label{subsec:selection}

We start with the SDSS DR7 quasar catalog \citep{schneider10},
adopting the spectral measurements of \citet{shen11}. The
catalog contains 105,783 spectroscopically confirmed quasars at
redshifts $0.065 <z<5.46$. These quasars have luminosities $M_i
< -22.0$ and at least one broad emission line with FWHM larger
than 1000 km s$^{-1}$. The SDSS DR7 provides optical spectra
covering $\lambda = 3800$--9180 \angstrom\ with moderate
spectral resolution ($R\sim1850$--2200) and S/N ($\sim$ 15
pixel$^{-1}$, with the pixel size being $69$ km ${\rm
s}^{-1}$). Among the SDSS DR7 quasars, 20,774 are at $z<0.83$,
where SDSS spectra cover \hbeta\ and \OIIIc\ (hereafter \OIII\
for short). As we discuss in detail below, from this parent
sample of 20,774 objects we select a subset of 399 with offset
broad Balmer emission lines, based on the spectral region
around \hbeta\ and \OIII . Our selection was a combination of
automated spectral fitting \citep{shen08,shen11} and visual
examination.  Here and throughout, we refer to the 399 objects
as the ``offset'' sample.  Below, we first briefly describe the
spectral model fitting (Section \ref{subsubsec:model}). We
refer to \citet{shen11} and Paper I for details. We then
discuss the selection procedure in Sections
\ref{subsubsec:voff}--\ref{subsubsec:visual} and discuss the
properties of the samples in Section \ref{subsubsec:property}.

%%%%%%%%%%%%%%%%%%%%%%%
\subsubsection{Spectral Model Fitting}\label{subsubsec:model}

To measure velocity offsets between the broad and narrow
emission lines, we first construct models of the spectra. Each
SDSS spectrum was decomposed using a combination of models for
the power-law continuum, \FeII\ emission complex, broad
emission lines, and narrow emission lines. First, the line-less
regions are fit by a pseudo-continuum model using an \FeII\
template plus a power-law continuum. After subtracting the
pseudo-continuum, the narrow and broad emission lines are then
fit using multiple Gaussians, where the widths and mean
velocities of different narrow lines are constrained to be the
same. The fitting was performed separately around each broad
emission line (\halpha , \hbeta , and \MgII ). For the \hbeta\
region, \OIII\ is occasionally fit by two Gaussians for the
core and wing (possibly associated with narrow-line region
outflows) components. In these cases, the width and mean
velocity of the narrow \hbeta\ emission line are constrained to
be the same as those of the core \OIII\ component. All fits
have been checked by visual inspection to ensure that models
reproduce data well.

%%%%%%%%%%%%%%%%%%%%%%%
\subsubsection{Measuring Velocity Offsets of Broad Emission Lines}\label{subsubsec:voff}

Using the spectral models, we measure the offset of the broad
emission lines relative to the systemic velocity. The systemic
redshift is estimated from the core component of \OIII , which
may be different (by a median offset of $32$ km s$^{-1}$ with a
standard deviation of $125$ km s$^{-1}$) from the nominal
redshift listed by the DR7 catalog based on the SDSS
spectroscopic pipeline \citep{SDSSDR0}. Our adopted systemic
redshift agrees with the improved redshift for SDSS quasars
from \citet{hewett10} within uncertainties. As discussed in
Paper I, we focus on the \hbeta--\OIII\ region. \OIII\ allows a
good estimate of the systemic redshift, and also provides
empirical constraints on the profile of the narrow \hbeta\
component. While \halpha\ is stronger than \hbeta\ and
therefore offers better S/N for each object, it would restrict
us to a $\sim$five times smaller parent sample (i.e., 3873
quasars in total at $z<0.36$). Compared to \hbeta , \CIV\ and
\MgII\ are less well understood in terms of their BLR
structure; furthermore, \CIV\ is more asymmetric and more
likely to be associated with a non-virial outflowing component.
Despite these caveats, we also compare \hbeta\ with independent
measurements from \MgII\ and \halpha\ when available to control
systematics and check for consistency.

%%%%%%%%%%%%%%%%%%%%%%%
\subsubsection{Automatic Pipeline Selection}\label{subsubsec:pipeline}

We first examine whether there is a significant (99.7\%
confidence) velocity offset in broad \hbeta\ relative to the
systemic redshift using the spectral model from automatic
pipeline fitting. Our selection criteria are the following:
\begin{enumerate}

\item[1] Median S/N pixel$^{-1}$ $>5$ in rest-frame
    4750--4950 \angstrom .

\item[1] Broad \hbeta\ rest-frame equivalent width
    $>3\sigma_{{\rm EW}}$.

\item[3] $|V_{{\rm off}}| > 3 \sigma_{V_{{\rm off}}}$,
    where $\sigma_{V_{{\rm off}}}$ is the total velocity
    offset error propagated from the velocity uncertainties
    of both broad and narrow emission lines.

\end{enumerate}
The uncertainty of the velocity offset, $\sigma_{V_{{\rm
off}}}$, taking into account both photon noise in the spectrum
and the ambiguity in subtracting a narrow-line component, was
estimated from Monte Carlo simulations \citep{shen11}.

We have experimented with two measures of the velocity offset
$V_{{\rm off}}$ of the broad emission line with respect to the
systemic redshift: the line centroid and line peak. These two
measures generally give consistent results for objects whose
broad-line profile is well-fit by a single Gaussian. But for
objects with more complex broad-line profiles, which are
usually well-fit by multiple Gaussians, the two measures can
give different results. We demand that $|V_{{\rm off}}| > 3
\sigma_{V_{{\rm off}}}$ for both the line centroid and the line
peak. The automatic selection yielded 1212 objects with
measurable velocity offsets in the broad \hbeta\ out of the
20,574 objects.

\begin{figure}
  \centering
    \includegraphics[width=88mm]{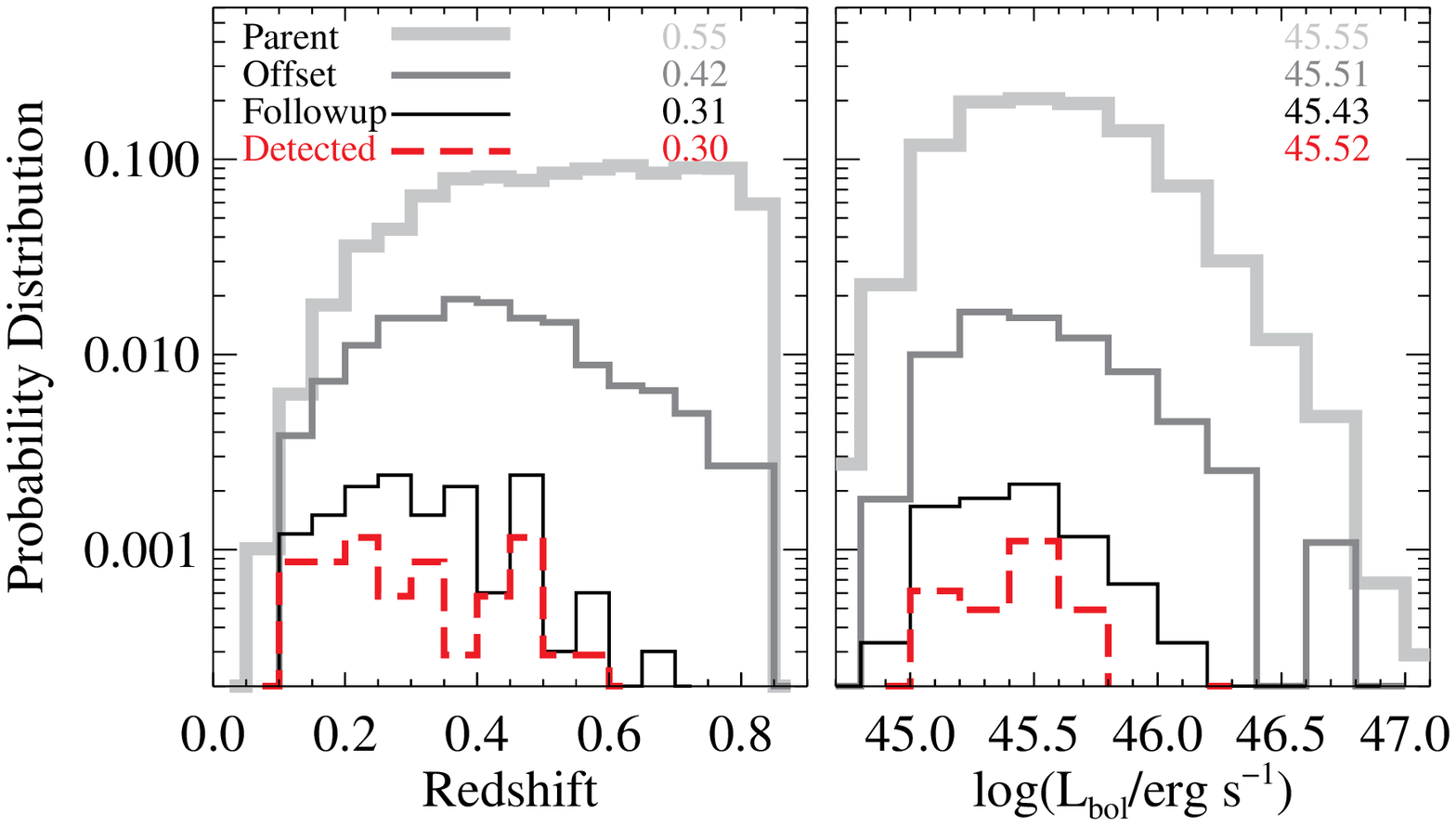}
    \includegraphics[width=88mm]{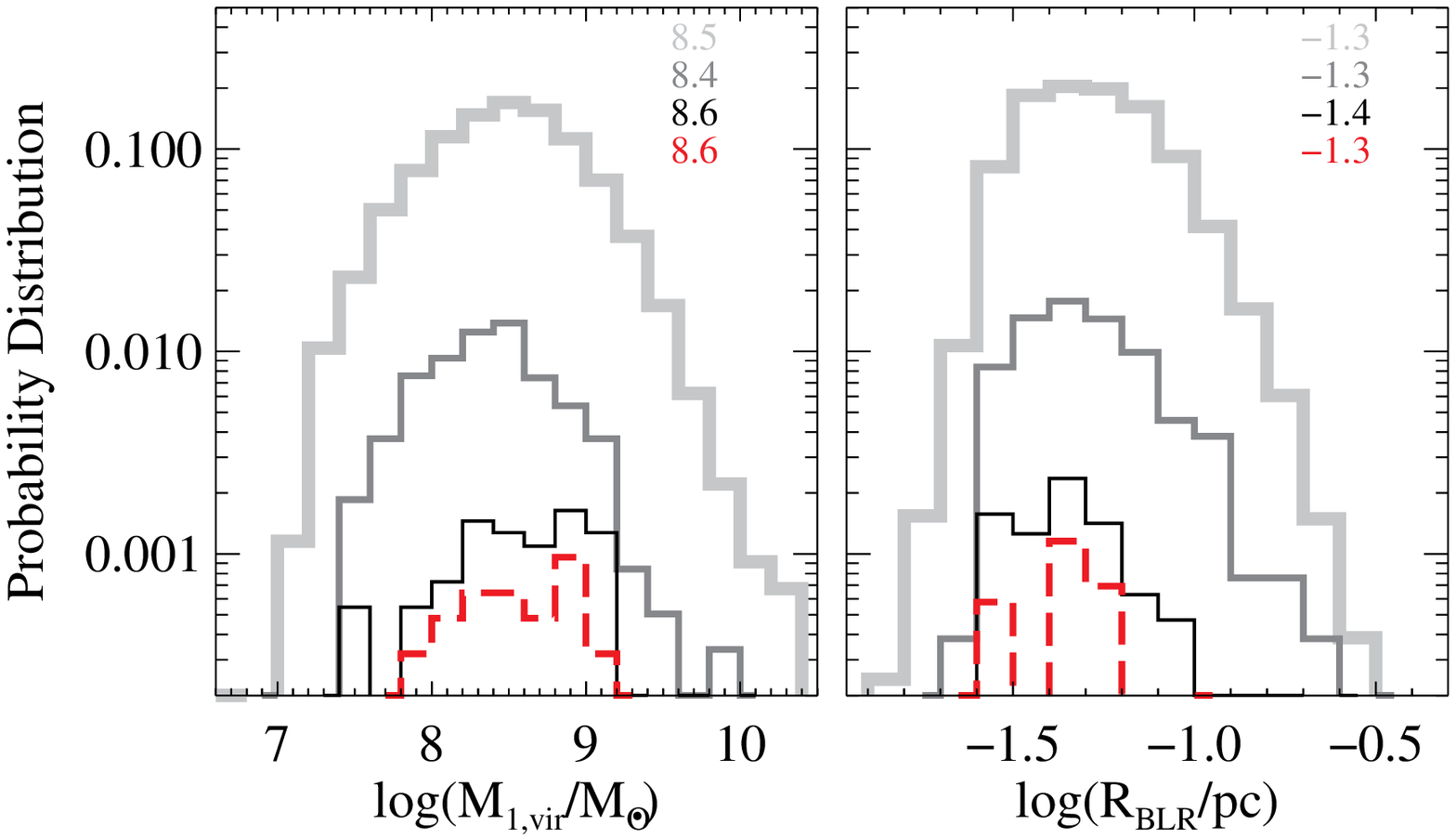}
    \caption{Distribution of quasar properties (redshift, bolometric luminosity,
    virial BH mass estimate, and BLR size) for the parent,
    offset, followup, and velocity-shift detected sample.
    Labeled on each panel are median values for different samples.
    }
    \label{fig:zmag}
\end{figure}
%%

%%%%%%%%%%%%%%%%%%%%%%%
\subsubsection{Visual Inspection}\label{subsubsec:visual}

We then visually inspected all the 1212 objects to verify the
pipeline measurements. We removed unreliable fits caused by
noise, weak broad \hbeta\ components, inconsistent results in
\halpha\ or \MgII , and/or problematic systemic redshift due to
weak, broad, winged and/or double-peaked \OIII\ without a
reliable narrow \hbeta ; we rejected narrow-line Seyfert 1
galaxies \citep[with broad \hbeta\ FWHM $< 2000$ km s$^{-1}$,
the ratio of \OIIIb\ to \hbeta\ smaller than three, and strong
optical \FeII\ emission; e.g.,][]{Osterbrock1985,Goodrich1989}
whose broad \hbeta\ often have blue wings
\citep[e.g.,][]{Boroson1992}, which may be attributed to
outflowing gas in the \hbeta\ emitting part of the BLR around
small BHs with high Eddington ratios
\citep[e.g.,][]{Boroson2002}; we also rejected objects with
prominent double-peaked broad emission lines, which are most
likely due to accretion disk emission around single BHs as
discussed above. This visual inspection yielded 399 objects,
which constitute our final ``offset'' sample. We present the
full offset sample in Table \ref{tab:sample} with basic
measurements.  Figure \ref{fig:egspec} shows SDSS spectra of 50
examples from the offset sample.

\begin{figure}
  \centering
    \includegraphics[width=85mm]{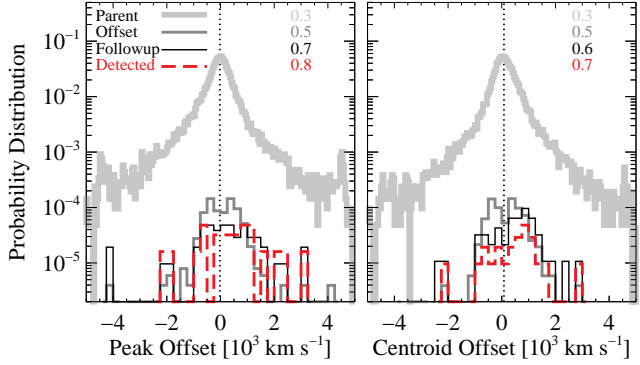}
    \caption{Distribution of broad \hbeta\ velocity offset
    for the parent, offset, followup, and velocity-shift detected sample.
    The left panel shows the peak velocity and the right panel shows the centroid.
    Labeled on the plot are median values (in units of $10^{3}$ km s$^{-1}$) of
    $|V_{{\rm off}} - V^0_{{\rm off}}|$ for different samples,
    where $V^0_{{\rm off}}$ denotes median values of $V_{{\rm off}}$.
    Vertical dotted lines indicate $V^0_{{\rm off}}$ for the parent sample
    (which is consistent with zero for the peak velocity,
    or $80\pm6$ km s$^{-1}$ for the centroid velocity).}
    \label{fig:voff}
\end{figure}
%%

%%%%%%%%%%%%%%%%%%%%%%%%%%%%%%%%%%%%%%%%%%%%%%
%%%%%%%%%%%%%%%%%%%%%%%
\subsubsection{Sample Properties}\label{subsubsec:property}

Figure \ref{fig:zmag} shows the basic quasar properties
(redshift, luminosity, virial BH mass, and BLR size) for the
parent sample and the offset sample. The bolometric luminosity
and virial BH mass estimates \citep[e.g.,][]{Shen2013} were
taken from \citet{shen11}. The BLR sizes {\bf $R$} were
estimated from the 5100 \angstrom\ continuum luminosity
assuming the empirical $R$--$L_{5100}$ relation in
\citet{Bentz2009}. The quasar luminosity distribution of the
offset sample is similar to the parent sample, whereas the
redshift is lower than that of the parent sample (median
redshift of 0.42 and 0.55, respectively). The estimated virial
BH masses and BLR sizes of the offset sample are both similar
to those of the parent quasar sample.

Figure \ref{fig:voff} shows broad \hbeta\ peak and centroid
velocity offsets for the parent and offset samples. By
construction, the offset sample has larger $|V_{{\rm off}}|$
than the parent sample (median value of $\sim500$ km s$^{-1}$
compared to $\sim300$ km s$^{-1}$). %%
%%%%%%%%%%%%%%%%%%%%%%%
The offset sample also has larger broad \hbeta\ emission line
offsets relative to the general quasar population studied in
Paper I: most of our offset objects have $300 < |V_{{\rm off}}|
< 1000$ km s$^{-1}$, whereas most ordinary quasars have
$|V_{{\rm off}}| < 300$ km s$^{-1}$. By comparison, the
majority of the \citet{eracleous11} sample has peak $|V_{{\rm
off}}| > 1000$ km s$^{-1}$.

For the parent sample the velocity centroid offset distribution
is centered at $80\pm6$ km s$^{-1}$, likely due to
gravitational redshift and transverse Doppler shift in BLR
clouds close to the BH
\citep[e.g.,][]{Zheng1990,Corbin1996,Tremaine2014a}. For the
offset sample the observed velocity peak and centroid offset
distributions are centered around $\sim200\pm40$ km s$^{-1}$,
i.e., redshifted relative to the systemic velocity. Subtracting
the $80\pm6$ km s$^{-1}$ measured from the parent sample
(assuming it is due to gravitational redshift and transverse
Doppler shift), the corrected distribution center is
$\sim120\pm40$ km s$^{-1}$. This residual redshift is likely a
selection bias, caused by our rejection of broad \hbeta\ blue
wings (Section \ref{subsubsec:visual}). The rejection was meant
to eliminate narrow-line Seyfert 1s, but objects with small
blue velocity offsets (i.e., $\sim100$ km s$^{-1}$) may be
mistaken for blue wings and got rejected more often than those
with small red offsets.

\begin{figure}
  \centering
    \includegraphics[width=85mm]{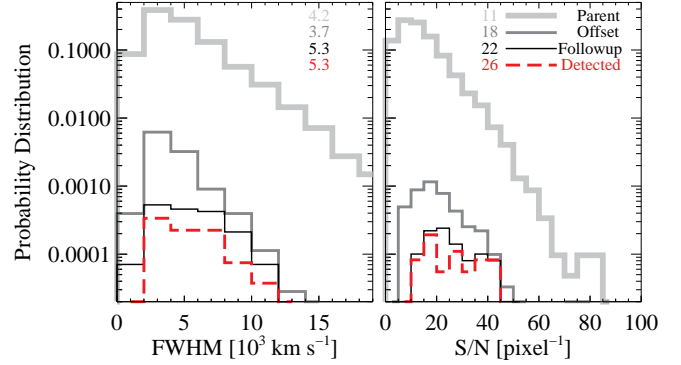}
    \caption{Distribution of broad \hbeta\ FWHMs and median S/N pixel$^{-1}$
    of the SDSS spectra for the parent, offset, followup, and
    velocity-shift detected sample.
    The median values for different samples are indicated.
    }
    \label{fig:fwhm}
\end{figure}
%%

%% offset sample - example
%% full table electronic version only
%%

%----------------------------------------------------------------------------------------------
%\begin{landscape}
\begin{deluxetable*}{lccccccc}
%\tabletypesize{\footnotesize} %
\tabletypesize{\scriptsize} %
%\tablewidth{0pc} %
\tablewidth{\textwidth} %
\tablecaption{SDSS Quasars with Kinematically Offset Broad Balmer Emission Lines
\label{tab:sample} %
} %
\tablehead{
\colhead{~~~~~~~~~~~~SDSS Designation~~~~~~~~~~~~} &
\colhead{$z_{{\rm sys}}$} & \colhead{Plate} & \colhead{Fiber} &
\colhead{MJD} & \colhead{$V^{{\rm peak}}_{{\rm off}}$} &
\colhead{$\sigma_{V_{{\rm off}}}$} & \colhead{$V^{{\rm cen}}_{{\rm off}}$} \\
\colhead{(1)} & \colhead{(2)} & \colhead{(3)} & \colhead{(4)} &
\colhead{(5)} & \colhead{(6)} & \colhead{(7)} & \colhead{(8)}%
} %
\startdata
001224.01$-$102226.5\dotfill & 0.2288 & 0651 & 072 & 52141 &$-$1952  & 119 &$-$638 \\
001247.93$-$084700.5\dotfill & 0.2203 & 0652 & 326 & 52138 &~~$-$176 &~~39 & ~~283 \\
001257.25+011527.3\dotfill   & 0.5047 & 0389 & 379 & 51795 &~~~~717  & 162 & ~~753 \\
002043.58+141249.4\dotfill   & 0.5880 & 0753 & 253 & 52233 &~~~~379  &~~65 & ~~375 \\
\enddata
\tablecomments{Column 1: SDSS names with J2000 coordinates given in the form of
``hhmmss.ss+ddmmss.s''; Column 2: systemic redshift from \citet{hewett10};
Columns 3--5: Plate ID, fiber ID, and MJD of the SDSS spectrum;
Column 6: broad \hbeta\ peak velocity offset in km s$^{-1}$.
Positive (negative) value means redshift (blueshift);
Column 7: 1$\sigma$ uncertainty of broad \hbeta\ peak velocity
offset in km s$^{-1}$, taking into account both statistical
and systematic errors estimated from Monte Carlo simulations \citep{shen11};
Column 8: broad \hbeta\ centroid velocity offset in km s$^{-1}$. \\
(This table is available in its entirety in a machine-readable
form in the online journal. A portion is shown here for guidance
regarding its form and content.)}
\end{deluxetable*}
%\clearpage
%\end{landscape}
%-------------

%%%%%%%%%%%%%%%%%%%%%%%
\subsection{Selection Biases and Incompleteness}\label{subsec:incompleteness}

We discuss possible selection biases to put our offset quasar
sample into the context of the general quasar population.
First, the visual inspection is subjective and by no means
complete, and the resulting bias cannot be easily quantified.
Second, only about half of the objects in the parent quasar
catalog were selected uniformly using the final quasar target
selection algorithm \citep{richards02}; the remaining objects
were selected via earlier algorithms or serendipitously
\citep{schneider10}, and in this case the selection function
cannot be easily quantified. Nevertheless, we do not believe
that any of our results depend significantly on these
complications in the selection function.

Third, the S/N threshold of our automatic selection is somewhat
arbitrary and may bias against less luminous quasars.
Nevertheless, the effect is mainly on redshift and not on
quasar luminosity (Figure \ref{fig:zmag}). However, the parent
quasar catalog was defined to have $M_i < -22.0$, so that the
offset sample is inherently more incomplete at lower
luminosities and by extension lower masses. This may be an
important bias considering that the figure of merit of
detecting BBHs may depend on quasar luminosity (Paper I).

Figure \ref{fig:fwhm} shows the broad \hbeta\ line widths and
S/N measured from the SDSS spectra. The median broad \hbeta\
FWHM of the offset sample is $\sim10$\% smaller than that of
the parent quasar sample. The median S/N of the offset sample
is $\sim60$\% higher than that of the parent quasar sample. As
shown in Paper I, the measurement errors in the velocity shift
of broad emission lines increase with increasing line width and
decreasing S/N. These uncertainties would cause different
sensitivities in the measured velocity shifts. We will return
to uncertainties in the velocity shift measurement in Section
\ref{subsec:simu}.

Finally, for BBHs in quasars, by selection objects with offset
broad lines may be different from the general population in
terms of binary separation, orbital phase, inclination, and BH
mass. The resulting effects must be properly accounted for to
translate the detection rate into constraints on the binary
fraction in the general quasar population. In Section
\ref{subsec:vacorr} we discuss these possible effects in the
context of detecting BBH candidates.

%% note for observations:
%% 100412: mostly clear; seeing ~ 1.1 - 1.6
%% 100416: partly cloudy; seeing ~ 1.2 - 3.3
%% 100601: partly cloudy; seeing ~1.4 - 4.0
%%

%----------------------------------------------------------------------------------------------
%\begin{landscape}
\begin{deluxetable*}{lcccccccccc}
%\tabletypesize{\footnotesize} %
\tabletypesize{\scriptsize} %
%\tablewidth{0pc} %
\tablewidth{\textwidth} %
\tablecaption{Followup Spectroscopy of SDSS Quasars with
Offset Broad Balmer Emission Lines
\label{tab:obs} %
} %
\tablehead{\colhead{} & \colhead{} & \colhead{$r$} & \colhead{} &
\colhead{} & \colhead{$t_{{\rm exp}}$} & \colhead{S/N} &
\colhead{$\Delta t$} &
\colhead{$V_{{\rm ccf}}$} & \colhead{$a_{{\rm ccf}}$} & \colhead{} \\
\colhead{No.} & \colhead{~~~SDSS Designation~~~} &
\colhead{(mag)} & \colhead{Spec} & \colhead{UT}
& \colhead{(s)} & \colhead{(pixel$^{-1}$)} & \colhead{(yr)} &
\colhead{(km
s$^{-1}$)} & \colhead{(km s$^{-1}$ yr$^{-1}$)} & \colhead{Category} \\
\colhead{(1)} & \colhead{(2)} & \colhead{(3)} & \colhead{(4)} &
\colhead{(5)} & \colhead{(6)} & \colhead{(7)} & \colhead{(8)} &
\colhead{(9)} & \colhead{(10)} & \colhead{(11)} %
} %
\startdata
%01 & 000710.01$+$005329.0 & $0.3158$ & 0388 & 445 & 51793 & 16.93 & BCS  & 111226 & 3600 & 6   \\
01 & 001224.02$-$102226.5 & 16.97 & BCS  & 110731 & 1200 & 16 & 8.31 & $-193^{+55}_{-51}$  & $-23^{+7}_{-6}$   & 2 \\
02 & 001257.25$+$011527.3 & 18.78 & BCS  & 111227 & 2400 & 10 & 7.70 & $-213^{+90}_{-89}$  & $-27^{+11}_{-12}$ & 3 \\
03 & 002141.02$+$003841.7 & 18.65 & FAST & 120126 & 9000 & 5  & 8.70 &$-103^{+205}_{-221}$ & $-11^{+22}_{-26}$ & \nodata \\
04 & 004712.58$-$084330.8 & 18.53 & BCS  & 111227 & 1200 & 13 & 7.33 & $131^{+53}_{-50}$   & $17^{+8}_{-6}$    & 3 \\
05 & 011110.04$-$101631.8 & 16.74 & BCS  & 111227 & 1200 & 35 & 8.87 & $27^{+63}_{-55}$    & $3^{+7}_{-6}$     & \nodata \\
06 & 014219.00$+$132746.6 & 18.12 & FAST & 111129 & 16200& 14 & 9.03 &$-379^{+146}_{-139}$ & $-42^{+17}_{-15}$ & 3 \\
07 & 020011.53$-$093126.2 & 18.02 & BCS  & 111226 & 2400 & 5  & 7.79 &$365^{+442}_{-563}$  & $46^{+57}_{-71}$  & \nodata \\
08 & 022113.14$+$010102.9 & 18.15 & BCS  & 111227 & 3600 & 3  & 8.41 &$179^{+277}_{-548}$  & $21^{+33}_{-64}$  & \nodata \\
09 & 032838.28$-$000341.7 & 19.21 & BCS  & 111227 & 1200 & 16 & 6.44 & $-627^{+67}_{-69}$  & $-97^{+10}_{-11}$ & 3 \\
%11 & 034931.03$-$062621.0 & $0.2874$ & 0463 & 177 & 51908 & 18.31 & BCS  & 111226 & 2400 & 2 & & &    \\
% garbage -- too noisy; barely see broad emission
10 & 073915.36$+$401445.7 & 18.41 & FAST & 120126 & 18000 & 7 & 8.84 &$206^{+283}_{-313}$  & $23^{+32}_{-35}$  & \nodata \\
11 & 074541.67$+$314256.7 & 15.81 & FAST & 120103 & 1800 & 15 & 7.10 &$117^{+174}_{-180}$  & $16^{+25}_{-24}$  & \nodata \\
12 & 080202.79$+$101943.1 & 18.18 & FAST & 120127 & 18000 & 17& 4.23 & $6^{+56}_{-57}$     & $1\pm13$          & \nodata \\
13 & 081032.73$+$565105.3 & 18.92 & FAST & 120103 & 18000 & 16& 5.60 & $41^{+103}_{-108}$  & $7^{+18}_{-19}$   & \nodata \\
%16 & 082708.53$+$425017.8 & $0.3330$ & 0761 & 436 & 52266 & 18.57 & BCS  & 111226 & 2400 & 2 & & &    \\
% garbage -- too noisy
14 & 082930.60$+$272822.7 & 18.10 & FAST & 111104 & 25200 & 23& 6.24 &$-213^{+129}_{-127}$ & $-34^{+21}_{-20}$ & 1 \\
15*& 084716.04$+$373218.1 & 18.45 & DIS  & 100416 & 2700 & 17 & 5.76 & $55^{+46}_{-39}$    & $10^{+8}_{-7}$    & 1 \\
%% seeing 1.6--2.2, noted "IQ got much worse" in obslog
16 & 085237.02$+$200411.0 & 18.10 & DIS  & 100416 & 2700 & 17 & 3.12 &$-296^{+162}_{-120}$ & $-95^{+52}_{-38}$ & 1 \\
17 & 091833.82$+$315621.2 & 17.94 & DIS  & 100416 & 3600 & 28 & 4.47 &$738^{+110}_{-102}$  & $165^{+24}_{-23}$ & 3 \\
%%17 & 091833.82$+$315621.2 & $0.4517$ & 1592 & 139 & 52990 & 17.94 & FAST & 111229 & 21600& 12 & 5.62 &$165^{+322}_{-167}$  & $29^{+57}_{-29}$  & \nodata \\
18$\dag$& 091858.15$+$232555.4 & 17.60 & FAST & 120130 & 16200 & 2 & 3.70 & $-41^{+564}_{-571}$ & $-11^{+130}_{-154}$ & \nodata \\
% maybe ok -- hb noisy, but MgII much better
19 & 091930.32$+$110854.0 & 17.25 & FAST & 111228 & 18000& 22 & 5.89 & $82^{+185}_{-154}$  & $14^{+31}_{-26}$  & \nodata \\
20$\dag$*& 091941.13$+$534551.4 & 18.82 & DIS  & 100416 & 3600 & 11 & 5.94 & $204^{+132}_{-99}$ & $59^{+22}_{-17}$ & 3 \\
%% seeing 2.4--3.3
21 & 092837.98$+$602521.0 & 17.01 & FAST & 120313 & 5400 & 25 & 8.87 & $220^{+154}_{-207}$ & $24^{+18}_{-23}$  & 1 \\
22 & 093653.84$+$533126.8 & 16.88 & FAST & 111130 & 19800& 40 & 8.26 & $-303^{+75}_{-81}$  & $-36^{+9}_{-10}$  & 2 \\
%26 & 095257.64$+$411954.0 & $0.3512$ & 1217 & 314 & 52672 & 18.82 & FAST & 120421 & 1800 & 2   \\
% garbage -- too noisy; weird continuum shape at 5200; ha emission undetected
23 & 101000.54$+$074235.5 & 18.99 & FAST & 111229 & 18000& 15 & 6.01 & $165^{+74}_{-85}$   & $27^{+12}_{-14}$  & 3 \\
24 & 102106.05$+$452331.9 & 18.34 & DIS  & 110601 & 3600 & 22 & 6.38 & $0^{+95}_{-89}$     & $0\pm14$          & \nodata \\
%%24$\ddag$&J102106.05$+$452331.9 & $0.3644$ & 0944 & 146 & 52614 & 18.34& FAST & 111230 & 10800& 16 & 6.82 & $6^{+90}_{-92}$     & $1\pm13$          & \nodata \\
%% note that the problem is not noisy, but the chi square curve of hb is not reasonable.
25 & 103059.09$+$310255.8 & 16.77 & FAST & 120119 & 1800 & 13 & 5.97 & $289^{+130}_{-133}$ & $48\pm22$         & 1 \\
26 & 104448.81$+$073928.6 & 18.31 & FAST & 120101 & 14400& 22 & 7.50 & $689^{+156}_{-124}$ & $92^{+19}_{-17}$  & 2 \\
27 & 110051.02$+$170934.3 & 18.48 & DIS  & 110601 & 3600 & 21 & 3.20 & $-269^{+92}_{-94}$  & $-84\pm29$        & 1 \\
28 & 111230.90$+$181311.4 & 18.13 & FAST & 120101 & 9000 & 17 & 4.10 & $-124^{+88}_{-86}$  & $-30^{+22}_{-21}$ & 1 \\
29 & 112007.43$+$423551.4 & 17.31 & FAST & 120119 & 18000& 31 & 6.60 & $-89^{+71}_{-74}$   & $-13\pm11$        & 2 \\
30 & 113640.91$+$573840.0 & 17.33 & FAST & 120129 & 9000 & 10 & 7.81 & $-138^{+208}_{-249}$& $-17^{+26}_{-32}$ & \nodata \\
31 & 115449.42$+$013443.5 & 17.89 & DIS  & 100416 & 2700 & 13 & 6.22 & $-96^{+188}_{-317}$ & $-15^{+29}_{-51}$ & \nodata \\
32 & 122018.44$+$064119.6 & 17.51 & FAST & 120313 & 5400 & 16 & 5.52 & $731^{+66}_{-921}$  & $132^{+12}_{-166}$& \nodata \\
33$\ddag$& 122811.89$+$514622.8 & 18.43 & DIS  & 110601 & 4500 & 16 & 7.10 & $-13^{+62}_{-63}$  & $-1^{+7}_{-9}$ & \nodata \\
34*& 130534.49$+$181932.9 & 16.68 & DIS  & 100412 & 2700 & 24 & 1.98 & $236^{+79}_{-76}$   & $119\pm39$        & 1 \\
%% seeing ~ 1.5;
%% also noted in obs log that "drive oscillation on azimuth, had to pause and then offset back"
35 & 130712.34$+$340622.5 & 17.51 & FAST & 120419 & 5400 & 15 & 6.27 & $172^{+143}_{-144}$ & $27\pm23$         & 3 \\
%37 & 134042.85$+$092214.3 & $0.0389$ & 1804 & 253 & 53886 & 17.43 & DIS  & 100412 & 2700 & 13   \\
% S/N ok but not in sdss quasar catalog
36 & 134548.50$+$114443.5 & 17.03 & FAST & 120419 & 5400 & 16 & 7.22 & $427^{+104}_{-109}$ & $59^{+14}_{-15}$  & 1 \\
37 & 141300.53$+$401624.5 & 18.19 & FAST & 120421 & 9000 & 10 & 6.62 &$-296^{+253}_{-323}$ & $-44^{+38}_{-49}$ & 3 \\
38 & 142314.18$+$505537.3 & 17.67 & BCS  & 110730 & 1200 & 3  & 6.71 & $-152^{+483}_{-414}$& $-23^{+72}_{-62}$ & \nodata \\
% noisy -- barely ok
% note: eyeball result
39 & 163020.78$+$375656.4 & 17.98 & BCS  & 110731 & 1800 & 17 & 6.01 & $0^{+56}_{-52}$     & $0^{+9}_{-8}$     & \nodata \\
%45 & 164841.24$+$411650.2 & $0.4240$ & 0631 & 320 & 52079 & 18.47 & BCS  & 110731 & 1800 & 3   \\
% garbage -- flat subtraction seems problematic! -- still pattern noise
40 & 165219.48$+$290319.2 & 18.09 & BCS  & 110731 & 1800 & 6  & 5.34 & $158^{+208}_{-804}$ & $29^{+39}_{-150}$ & \nodata \\
%47 & 170352.89$+$335545.7 & $0.3353$ & 0972 & 080 & 52435 & 18.78 & BCS  & 110731 & 1800 & 2   \\
% garbage -- too noisy
%48 & 171617.23$+$273410.1 & $0.2888$ & 0980 & 261 & 52431 & 19.06 & BCS  & 110731 & 1800 & 3   \\
% garbage -- broad line too weak, barely detected
41 & 211234.89$-$005926.9 & 18.36 & BCS  & 110731 & 1800 & 9  & 7.57 &$-117^{+128}_{-153}$ & $-15^{+16}_{-20}$ & \nodata \\
42 & 212936.97$-$072431.9 & 17.80 & FAST & 111128 & 7200 & 11 & 7.17 &$-296^{+165}_{-404}$ & $-41^{+23}_{-56}$ & 3 \\
43 & 213040.16$-$082160.0 & 18.25 & BCS  & 110731 & 1800 & 16 & 7.93 & $55^{+77}_{-91}$    & $6\pm10$          & \nodata \\
44 & 220537.72$-$071114.6 & 18.27 & BCS  & 110731 & 1800 & 4  & 7.17 &$-120^{+601}_{-580}$ & $-17^{+84}_{-81}$ & \nodata \\
45 & 224903.29$-$080841.8 & 18.78 & BCS  & 110731 & 1800 & 3  & 6.89 &$-172^{+250}_{-229}$ & $-25^{+36}_{-33}$ & \nodata \\
% garbage -- noisy, plus, pattern noise from bad flat field
46 & 230248.88$+$134553.5 & 18.88 & BCS  & 110730 & 1800 & 3  & 7.13 & $75^{+482}_{-169}$  & $10^{+68}_{-23}$  & \nodata \\
47 & 230323.47$-$100235.4 & 17.90 & BCS  & 110731 & 1800 & 15 & 8.36 & $-13^{+33}_{-39}$   & $-1^{+3}_{-5}$    & \nodata \\
48 & 230845.60$-$091124.0 & 17.20 & FAST & 110904 & 10800& 44 & 8.36 & $137^{+108}_{-97}$  & $16^{+13}_{-12}$  & 2 \\
49 & 232124.44$+$134930.1 & 18.59 & FAST & 111028 & 16200& 10 & 6.43 &$-345^{+621}_{-690}$ &$-54^{+97}_{-107}$ & \nodata \\
% note: eyeball result
50 & 234852.50$-$091400.8 & 18.78 & BCS  & 110731 & 1800 & 3  & 6.18 &$-627^{+972}_{-408}$ &$-101^{+157}_{-66}$& \nodata \\
% note: eyeball result
\enddata
\tablecomments{Column 1: *objects whose \OIII\ lines show nonzero
velocity shifts, which have been subtracted from the
broad-line velocity shift measurements (assuming that the shift in \OIII\
is due to wavelength calibration errors); $\dag$objects whose broad
\hbeta\ are too noisy and whose \MgII\ measurements are
adopted; $\ddag$objects whose broad \hbeta\ are too noisy
and whose \halpha\ measurements are adopted; Column 2: SDSS
names with J2000 coordinates given in the form of
``hhmmss.ss+ddmmss.s''; Column 3: SDSS $r$-band PSF magnitude;
Column 4: spectrograph used for the followup observations;
Column 5: UT date of the followup observations;
Column 6: total exposure time of the followup observations;
Column 7: median S/N pixel$^{-1}$ around the broad \hbeta\ region
of the followup spectra. The pixel size is 1.2 \angstrom\ for the BCS spectra,
0.62 (0.56) \angstrom\ in the blue (red) channel of DIS, and 1.5 \angstrom\ for FAST;
Column 8: rest-frame time baseline between the followup and
SDSS observations; Columns 9 and 10: velocity shift and acceleration,
based on the broad \hbeta\ line unless otherwise indicated in Column 1.
Positive (negative) values indicate that the followup spectrum is redshifted
(blueshifted) relative to the original SDSS spectrum. The
quoted uncertainties enclose the 2.5$\sigma$ confidence range.
Column 11: we classify the detections into three categories
(see Section \ref{subsec:detection})-- ``1'' for BBH candidates, ``2'' for
broad-line variability, and ``3'' for ambiguous cases.}
\end{deluxetable*}
%\clearpage
%\end{landscape}
%-------------

\begin{figure}
  \centering
    \includegraphics[width=88mm]{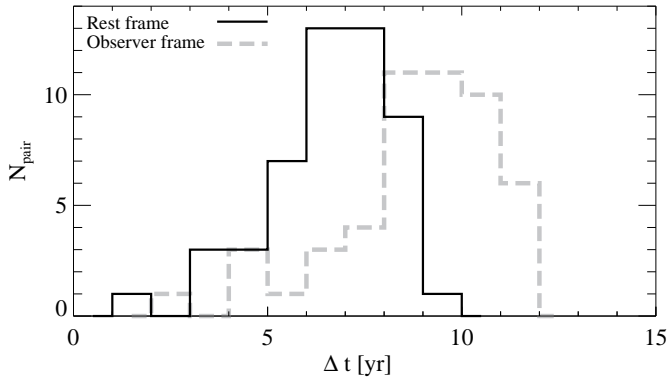}
    \caption{Time baseline between the followup and SDSS observations.
    For 88\% of the targets the rest frame $\Delta t>5$ yr.}
    \label{fig:dmjd}
\end{figure}

%%%%%%%%%%%%%%%%%%%%%%%%%%%%%%%%%%%%%%%%%%%%%%%%%%%%%%%%%%%%%%%%%
%%%%%%%%%%%%%%%
%%%%%%%%%%%%%%%%%%%%%%%%%%%%%%%%%%%%%%%%%%%%%%%%%%%%%%%%%%%%%%%%%
%%%%
%%%%%%%%%%%%%%%%%%%%%%%%%%%%%%%%%%%%%%%%%%%%%%
\section{Second-Epoch Optical Spectroscopy}\label{sec:ss}

We have conducted second-epoch optical spectroscopy for $50$
objects drawn from the offset sample presented in Section
\ref{sec:sample}. We list all targets with second-epoch
spectroscopy in Table \ref{tab:obs}. We preferentially selected
brighter targets whenever available for better S/N. Our targets
have median SDSS $r\sim 18.3$ mag. This ``followup'' sample is
representative of the offset sample in general except that it
is at lower redshifts (Figures \ref{fig:zmag}--\ref{fig:fwhm}).
Figure \ref{fig:dmjd} shows the distribution of rest-frame time
intervals $\Delta t$ between the two epochs. In the following
we describe details of the followup observations and data
reduction.

\begin{figure*}
  \centering
    \includegraphics[width=34mm]{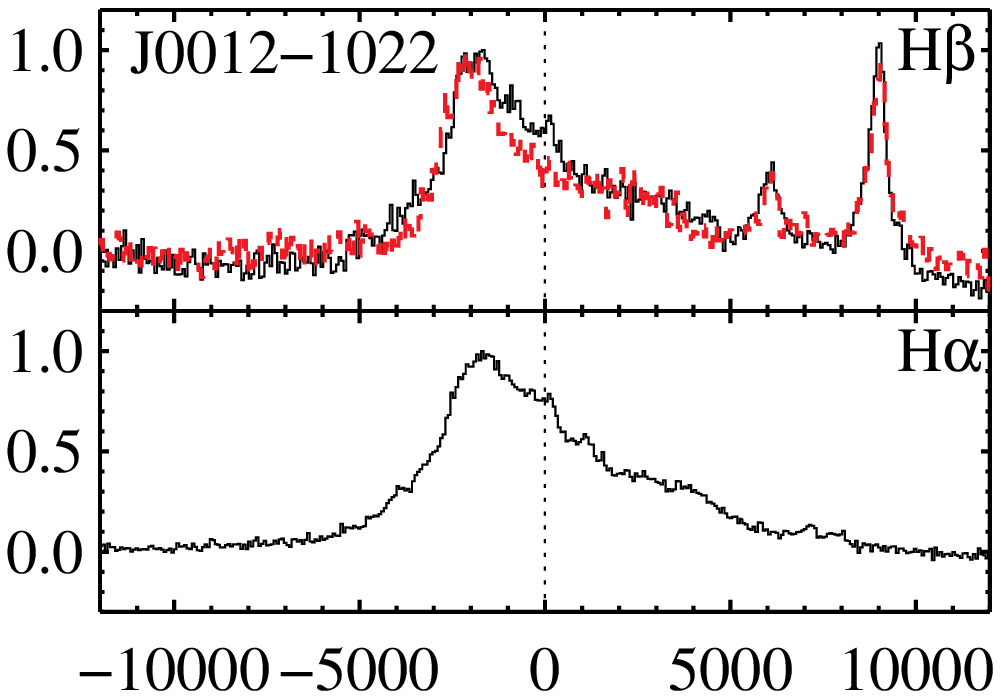}
    \includegraphics[width=34mm]{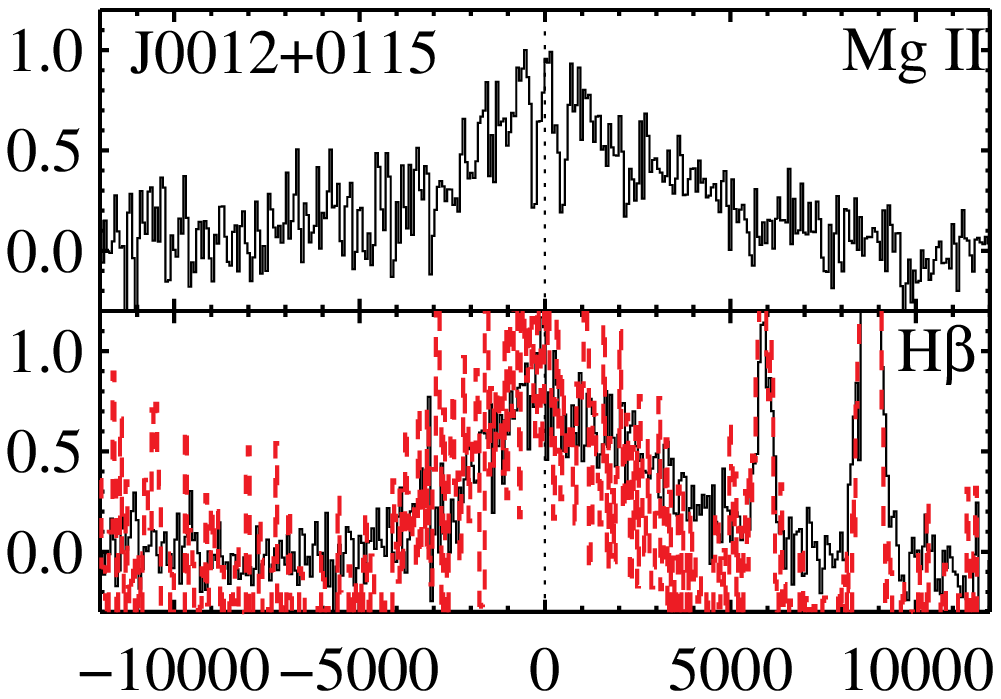}
    \includegraphics[width=34mm]{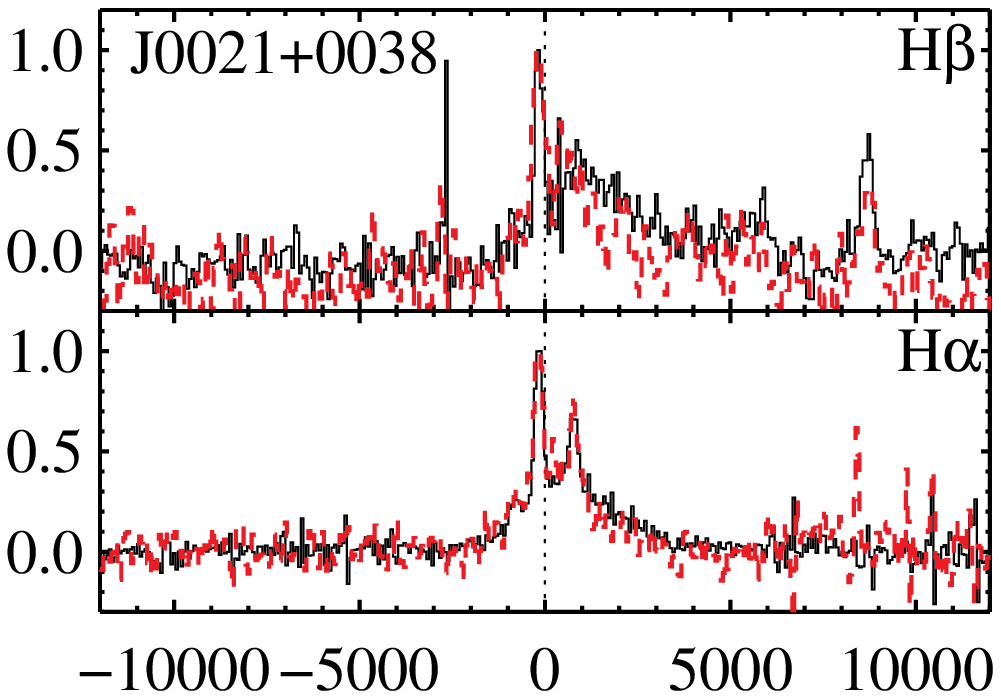}
    \includegraphics[width=34mm]{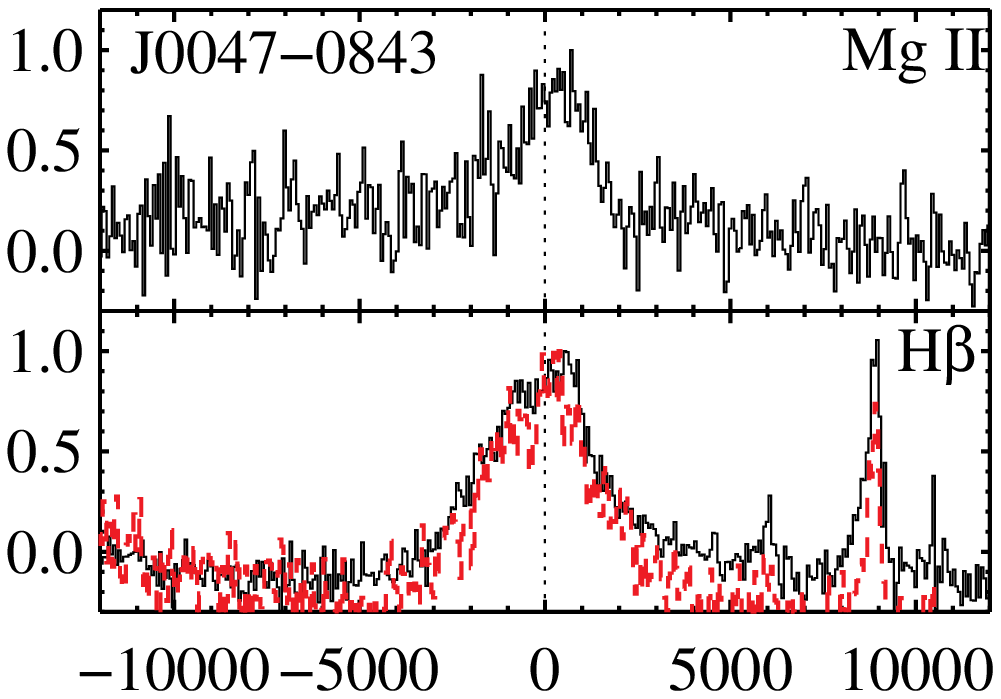}
    \includegraphics[width=34mm]{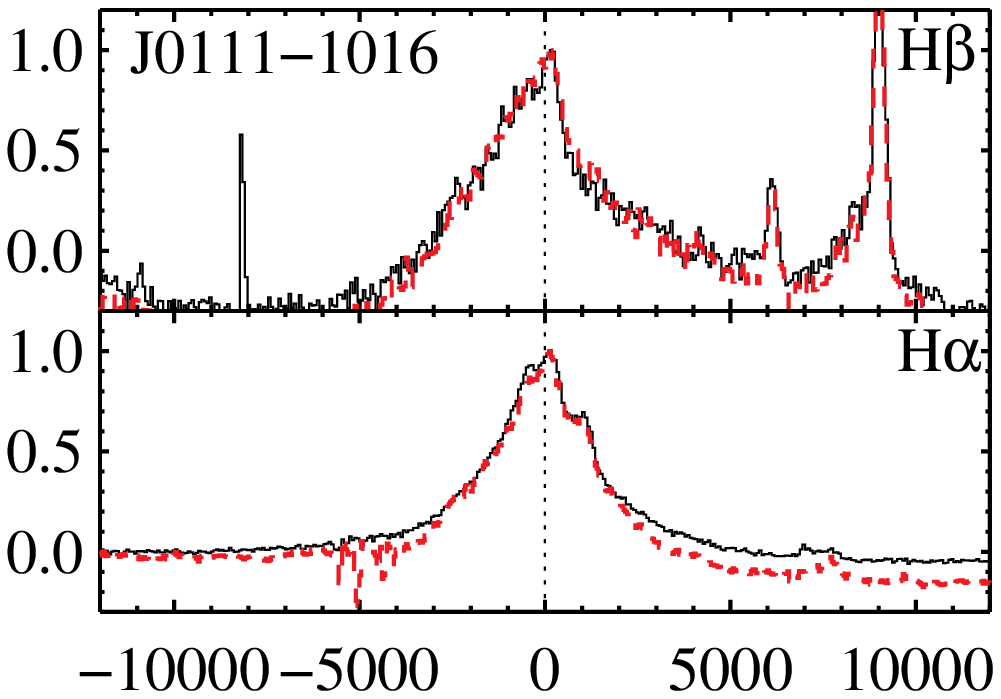}
    \includegraphics[width=34mm]{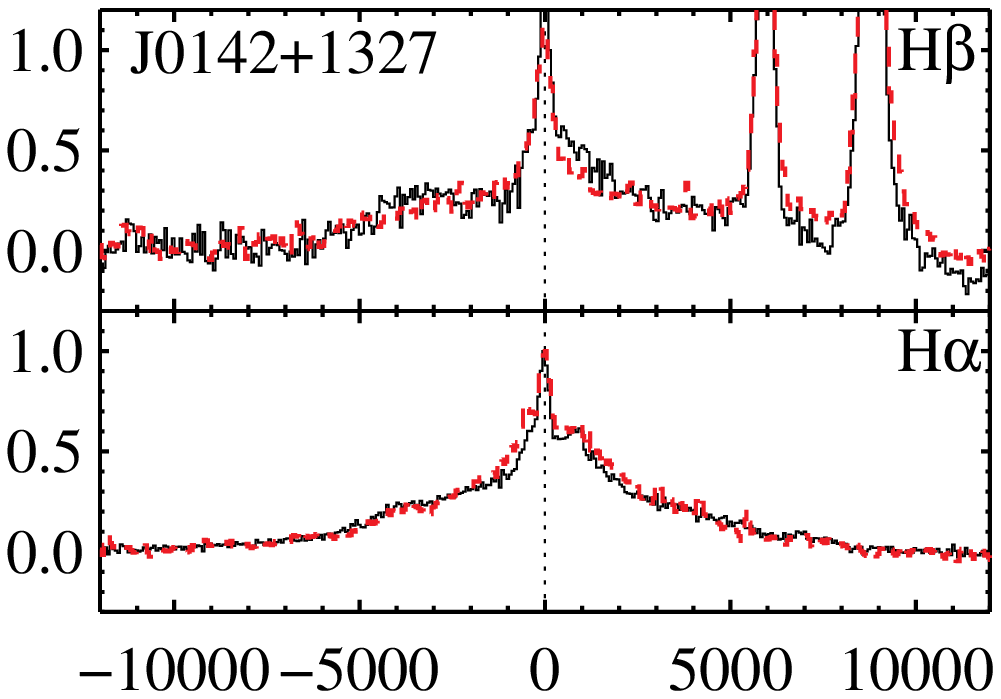}
    \includegraphics[width=34mm]{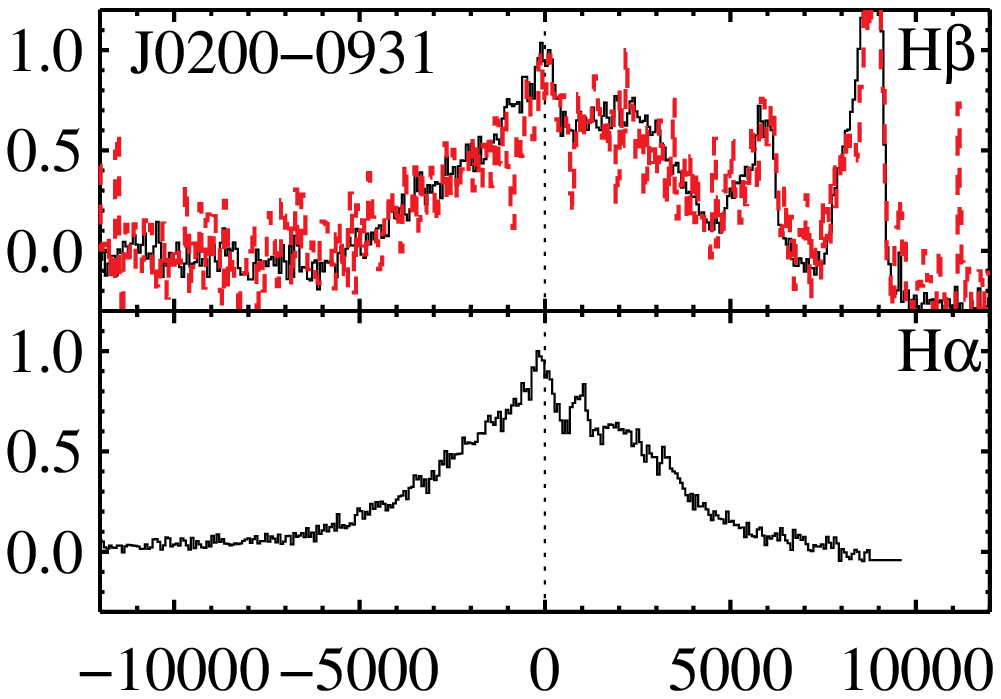}
    \includegraphics[width=34mm]{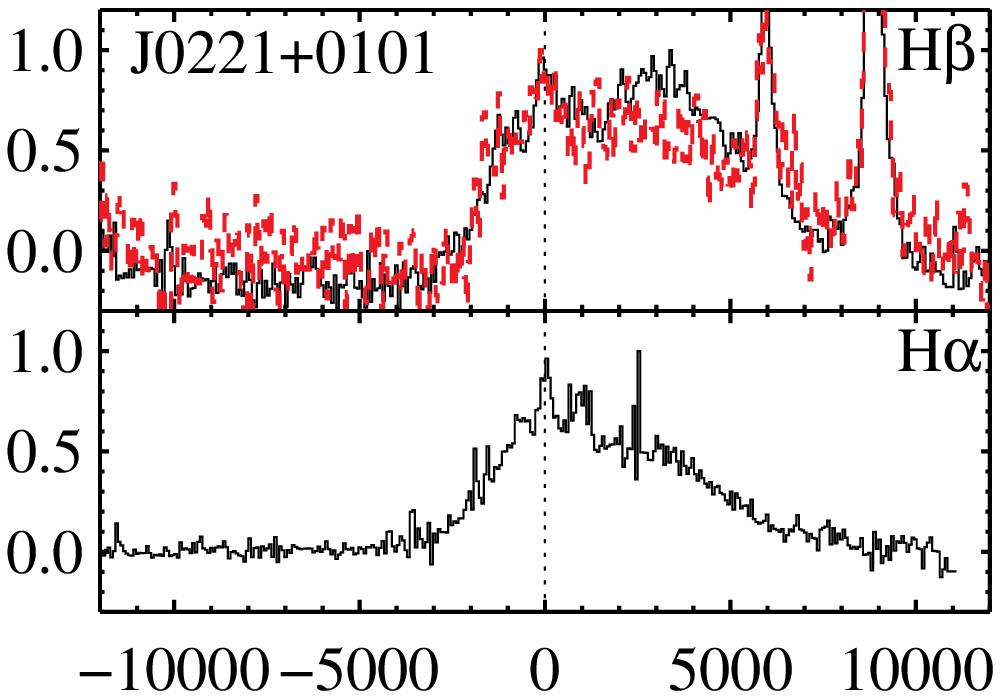}
    \includegraphics[width=34mm]{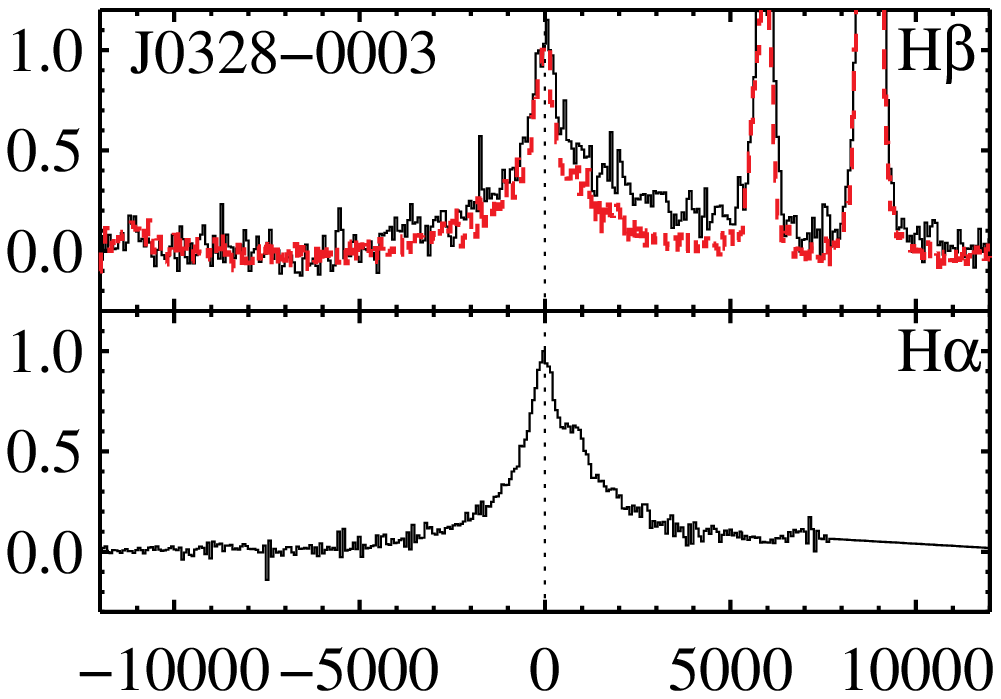}
    \includegraphics[width=34mm]{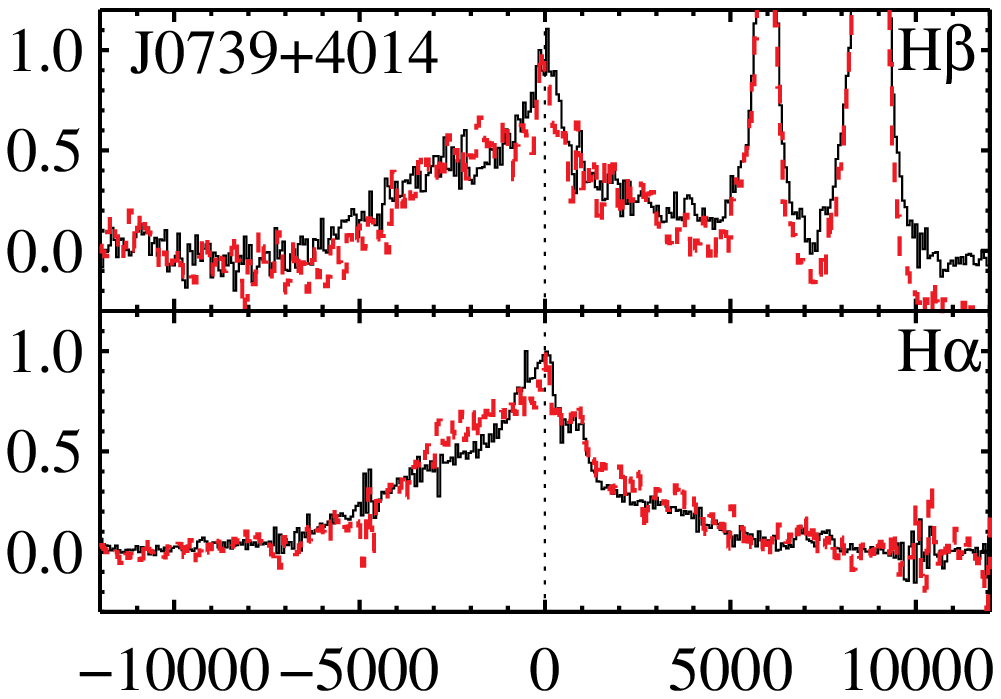}
    \includegraphics[width=34mm]{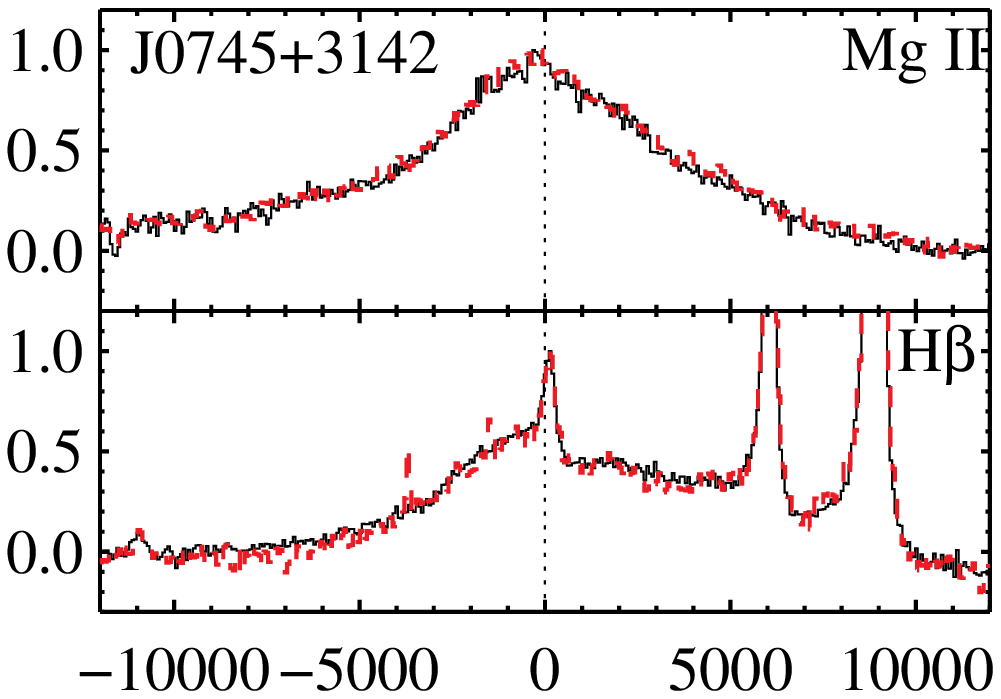}
    \includegraphics[width=34mm]{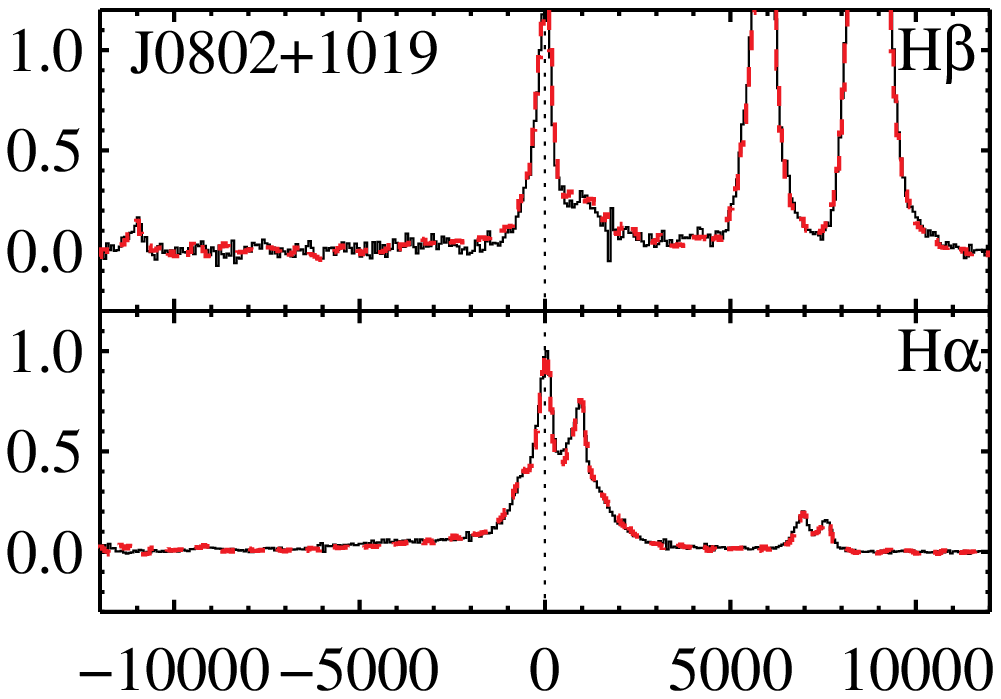}
    \includegraphics[width=34mm]{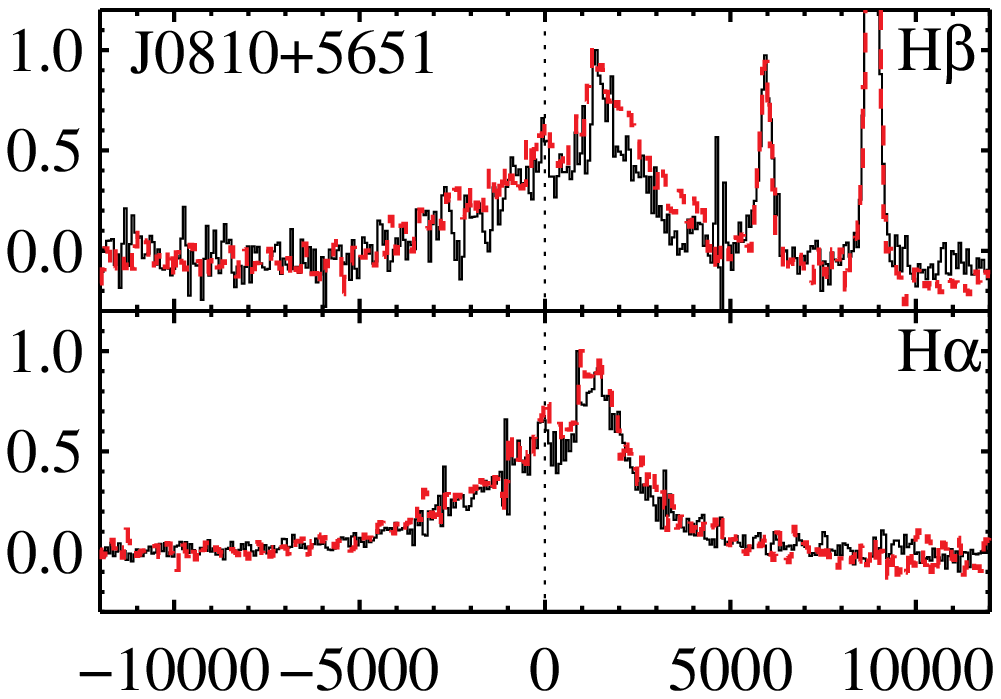}
    \includegraphics[width=34mm]{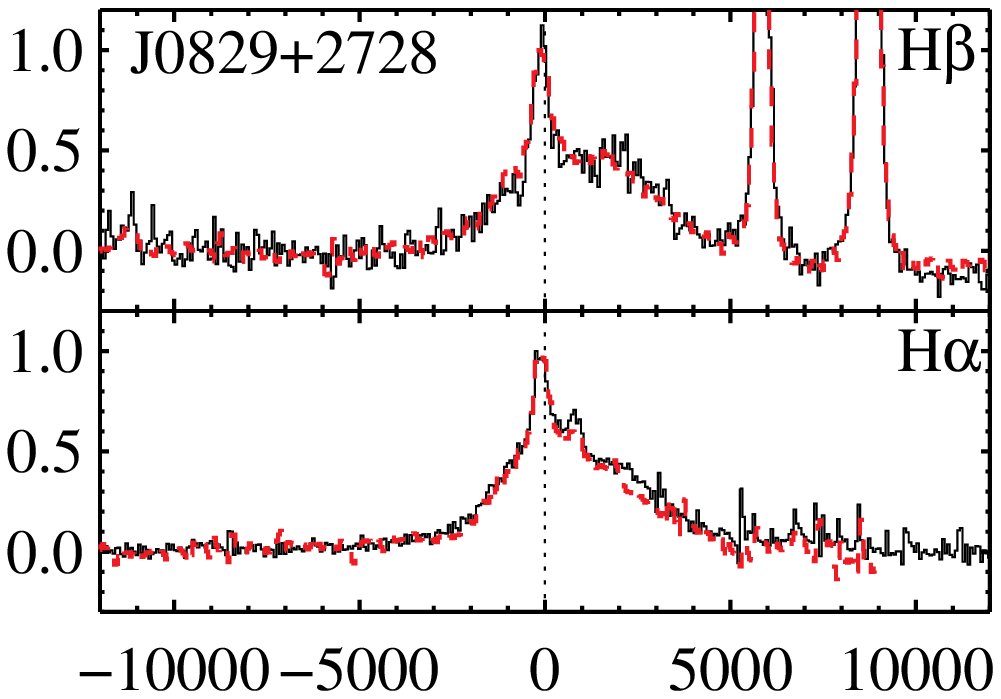}
    \includegraphics[width=34mm]{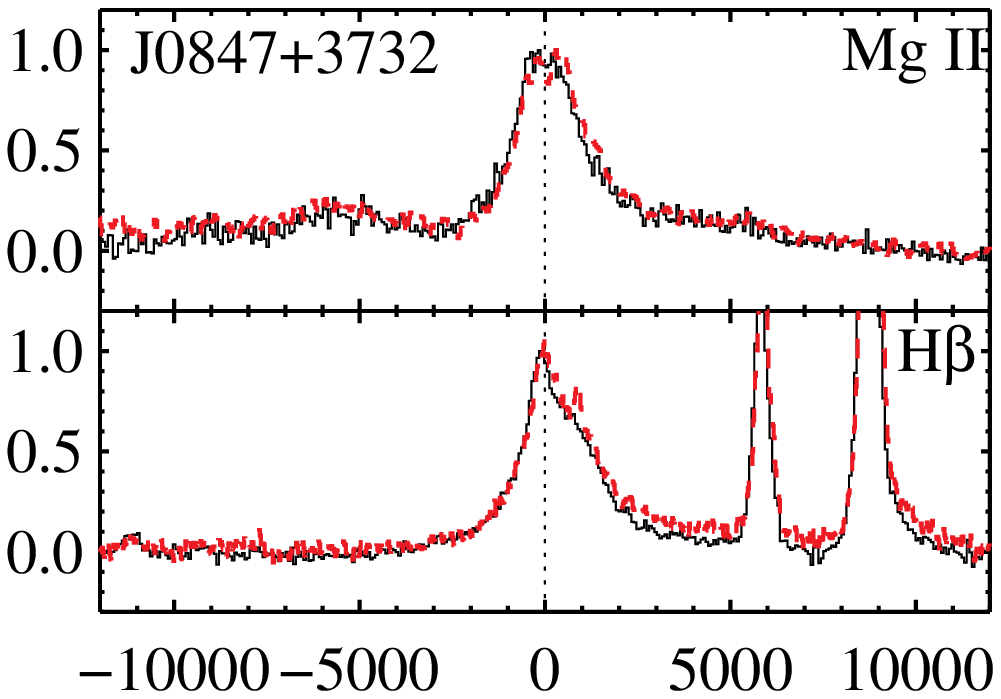}
    \includegraphics[width=34mm]{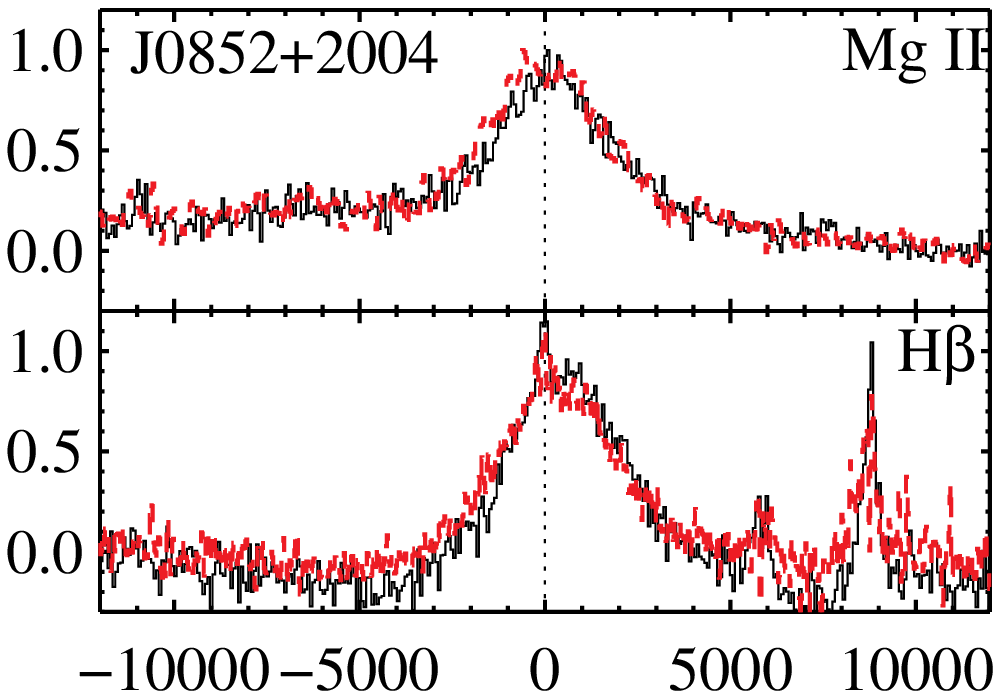}
    \includegraphics[width=34mm]{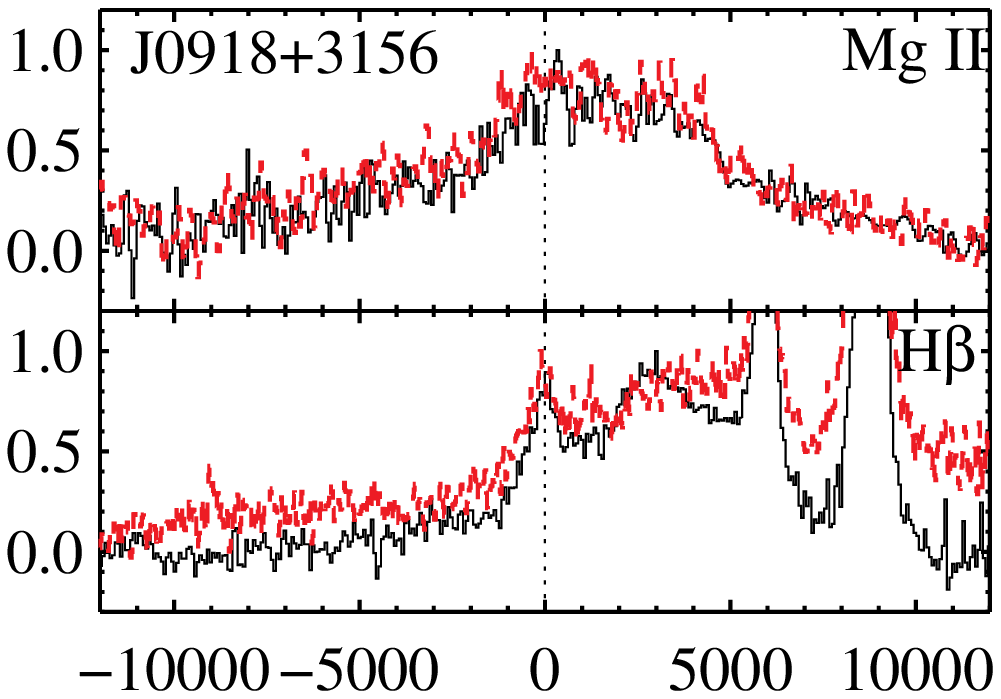}
    \includegraphics[width=34mm]{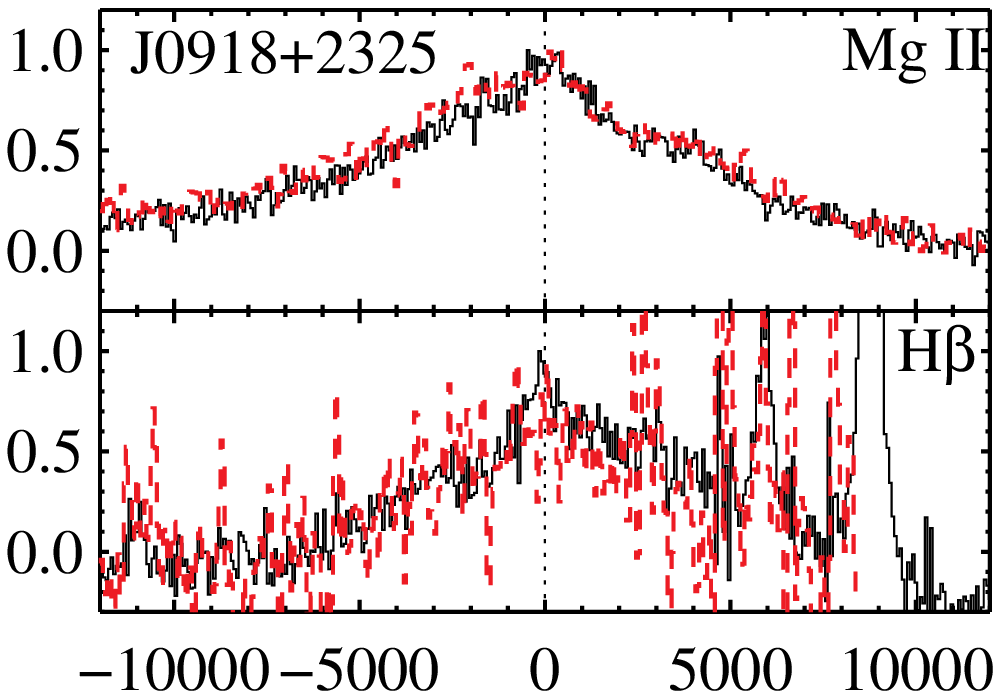}
    \includegraphics[width=34mm]{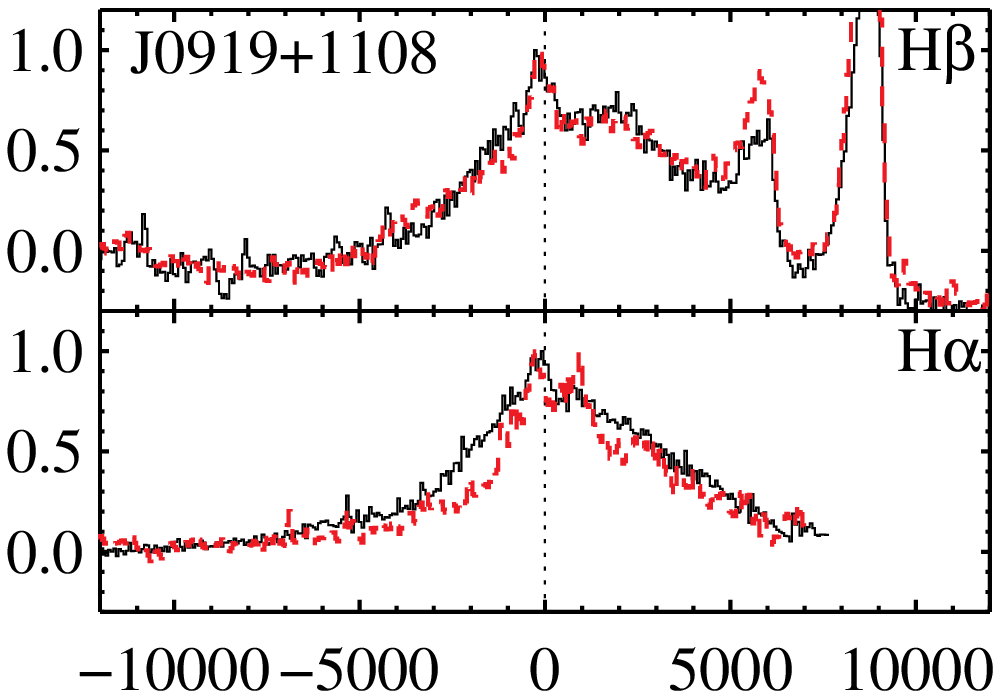}
    \includegraphics[width=34mm]{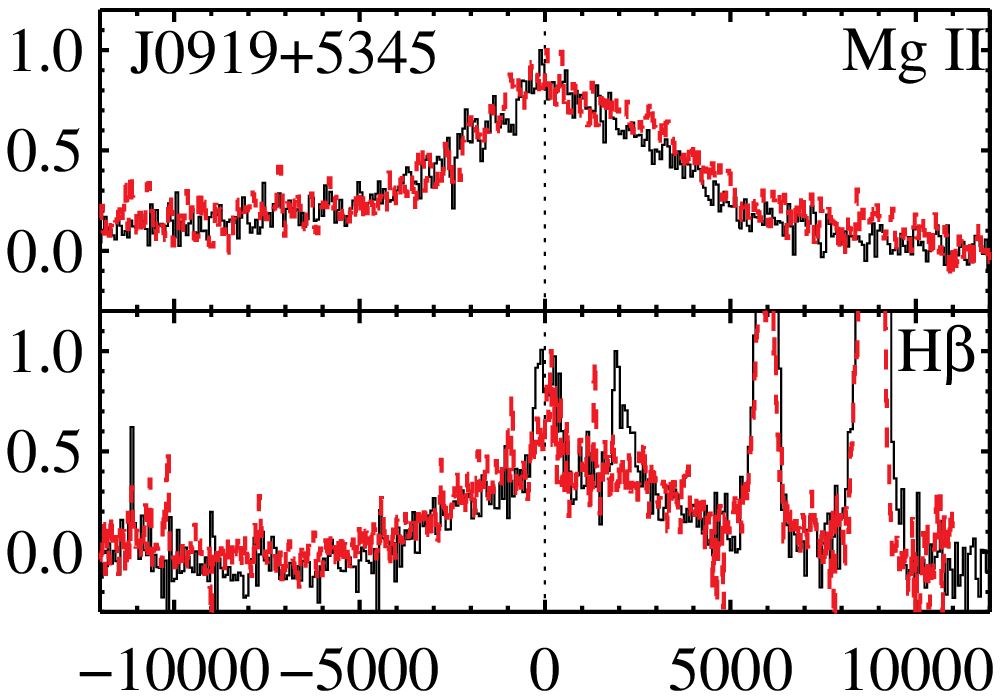}
    \includegraphics[width=34mm]{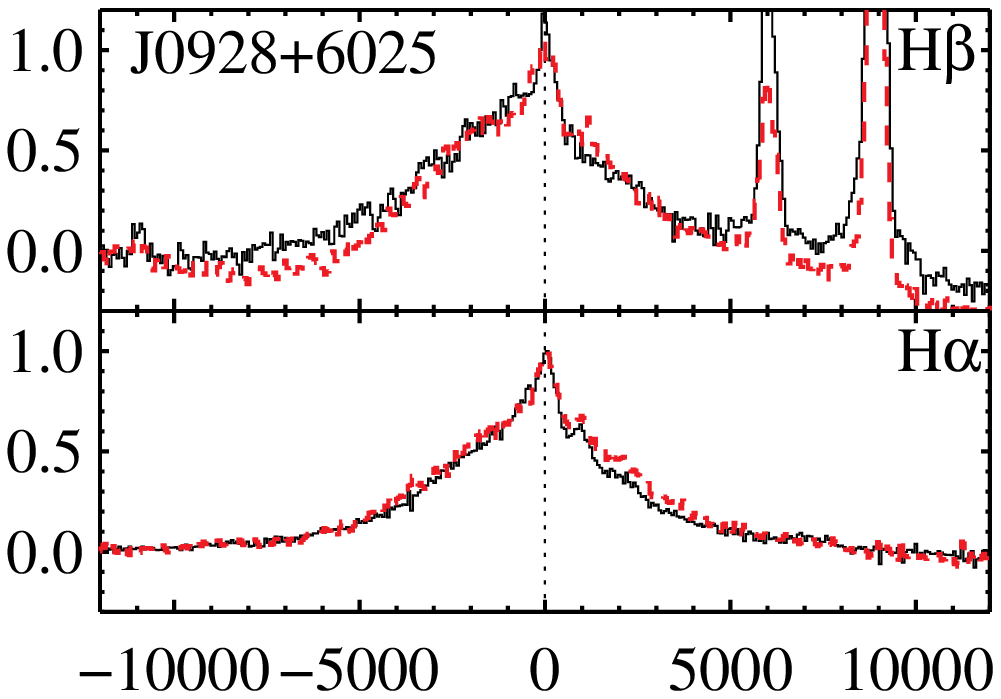}
    \includegraphics[width=34mm]{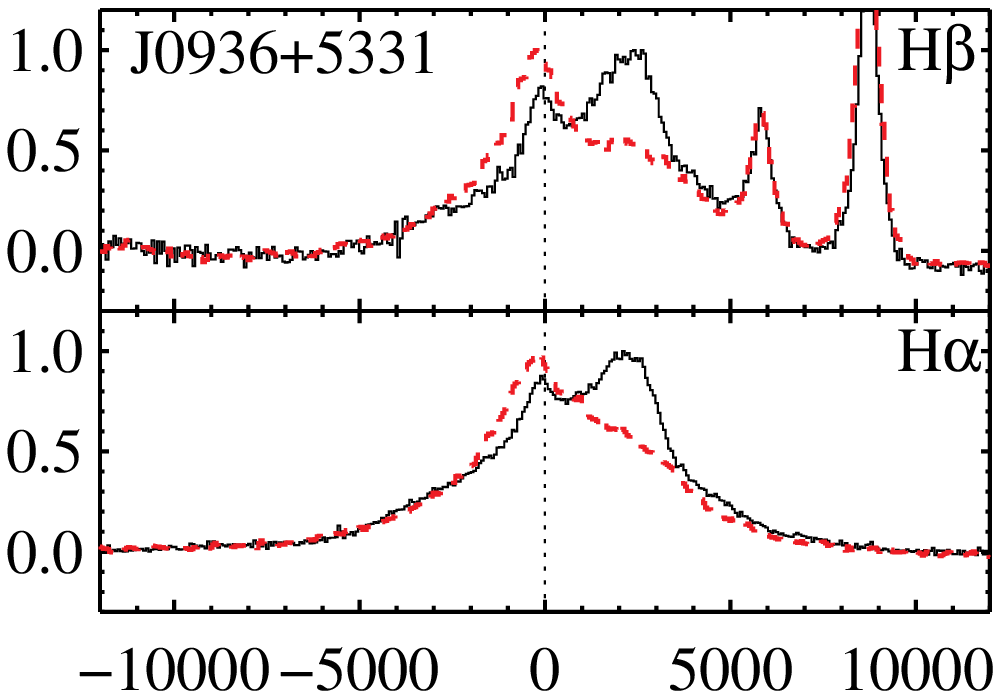}
    \includegraphics[width=34mm]{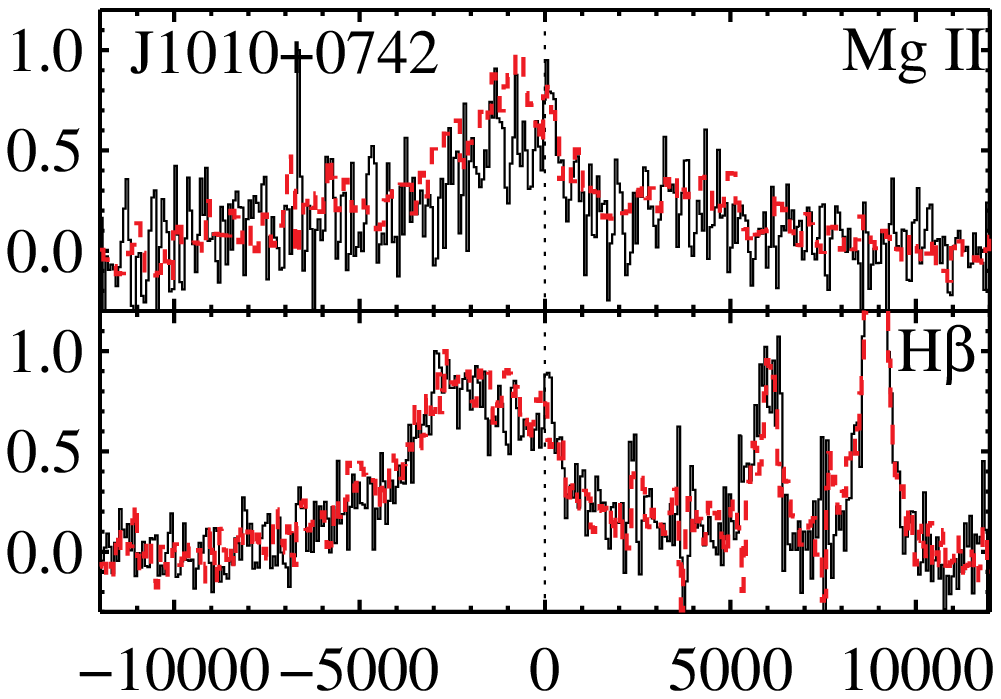}
    \includegraphics[width=34mm]{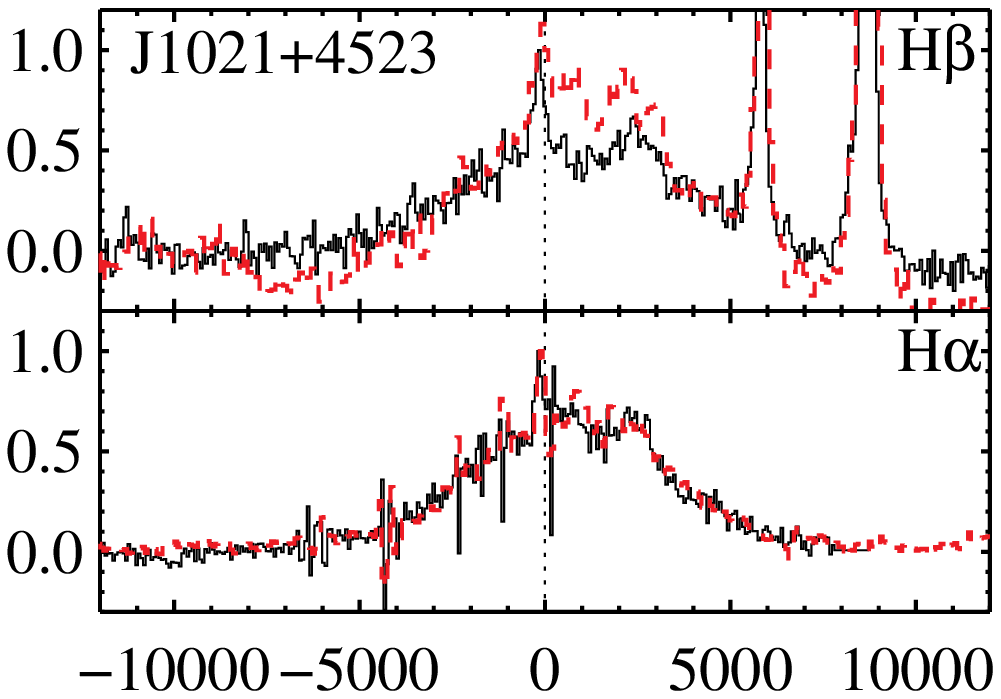}
    \includegraphics[width=34mm]{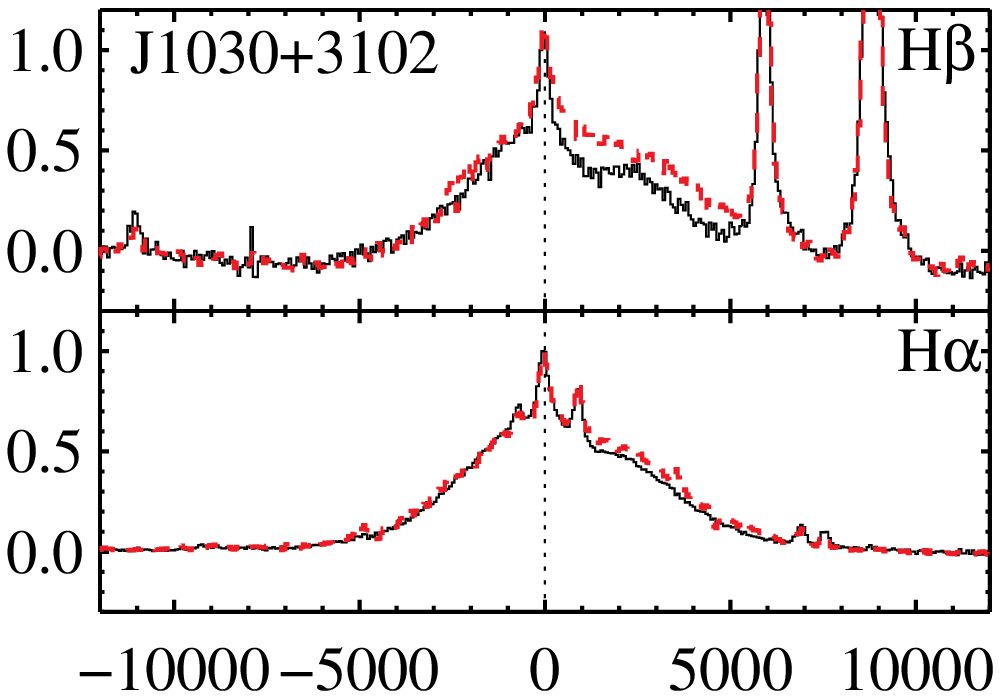}
    \includegraphics[width=34mm]{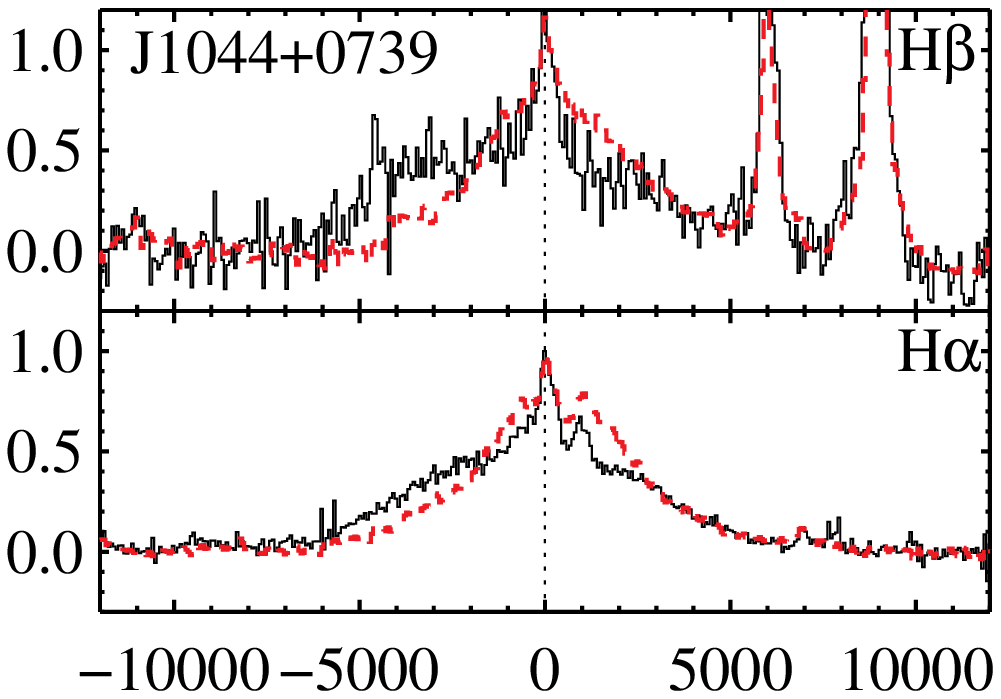}
    \includegraphics[width=34mm]{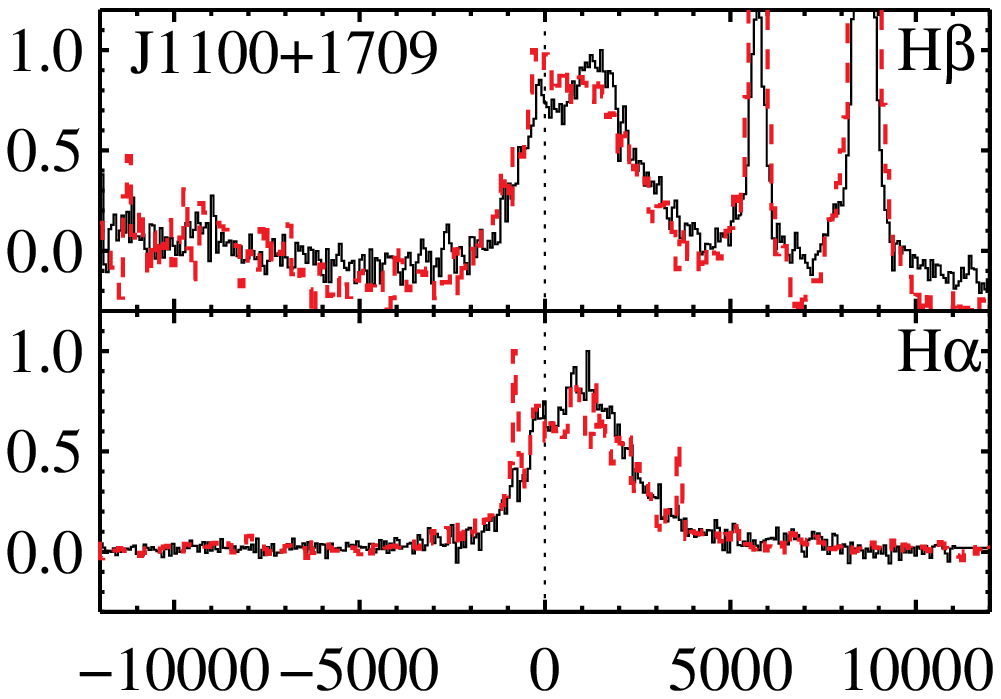}
    \includegraphics[width=34mm]{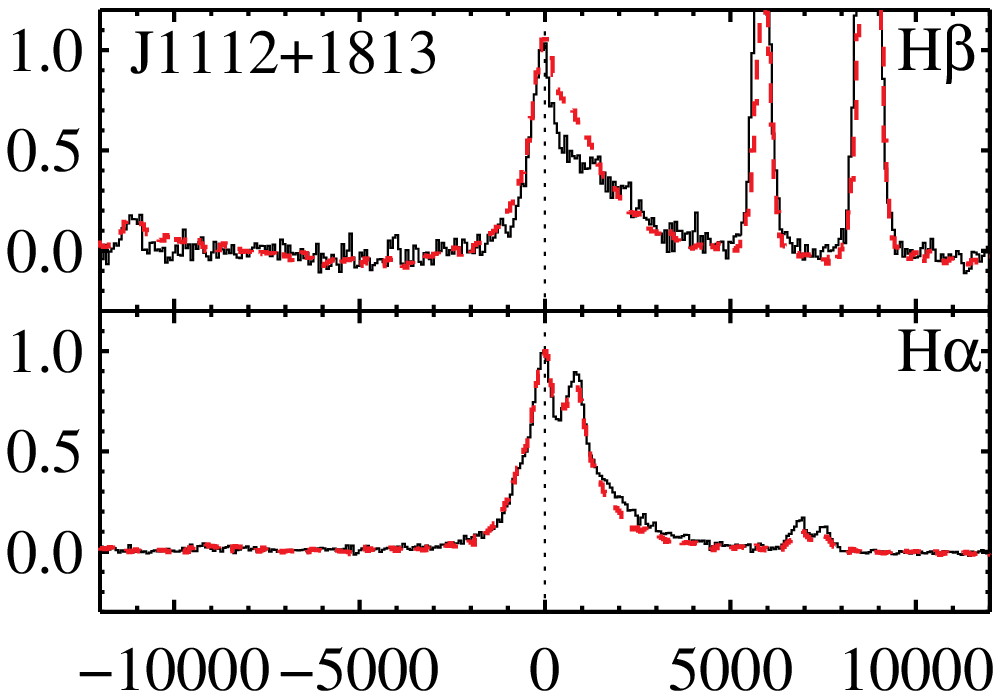}
    \includegraphics[width=34mm]{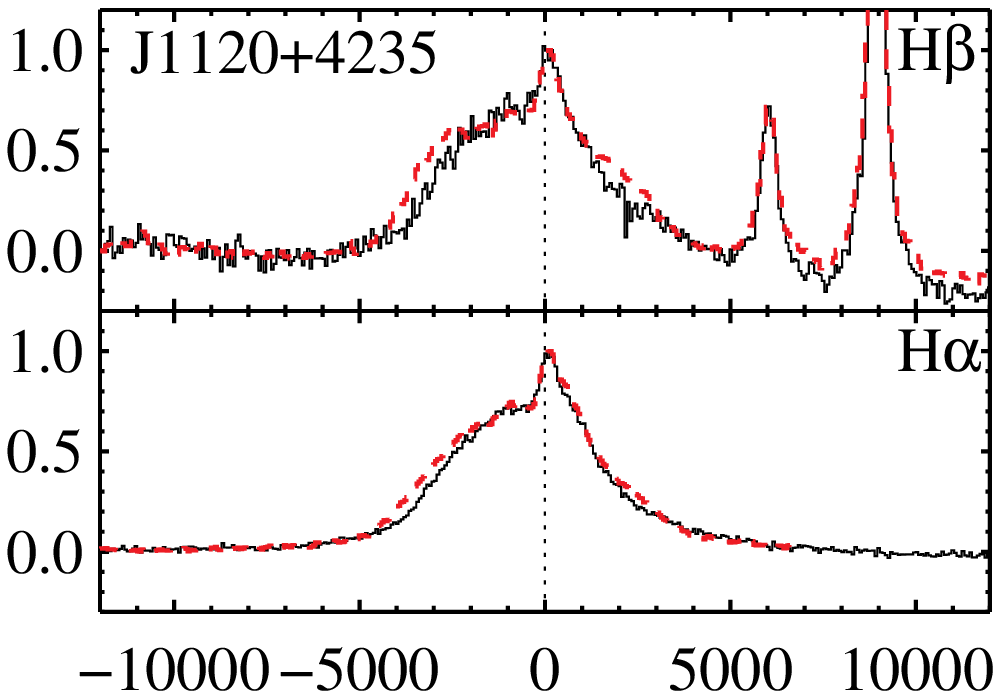}
    \includegraphics[width=34mm]{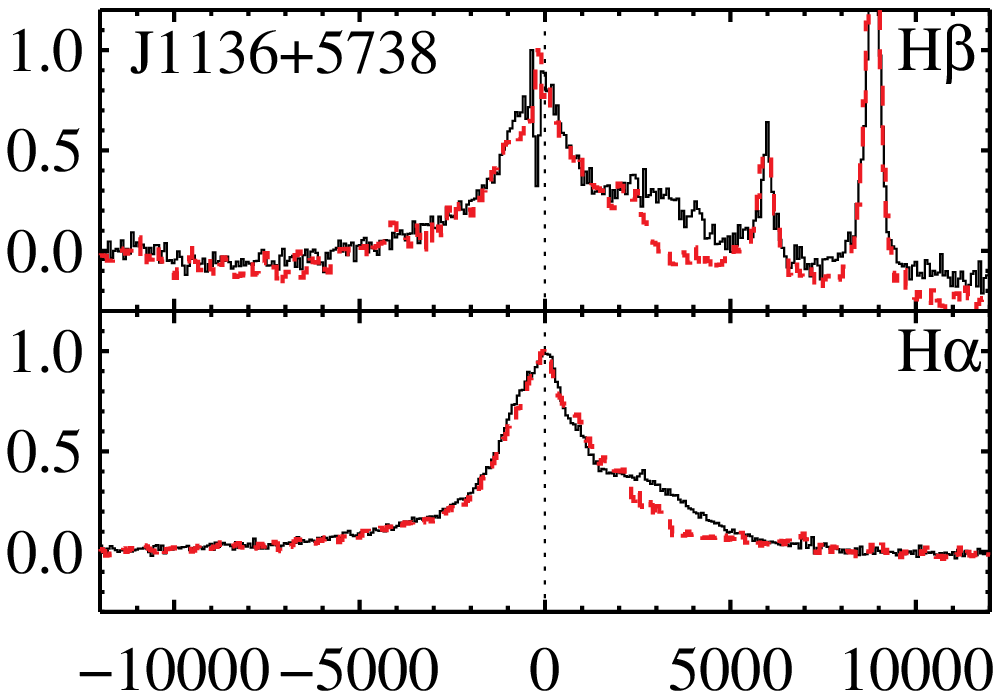}
    \includegraphics[width=34mm]{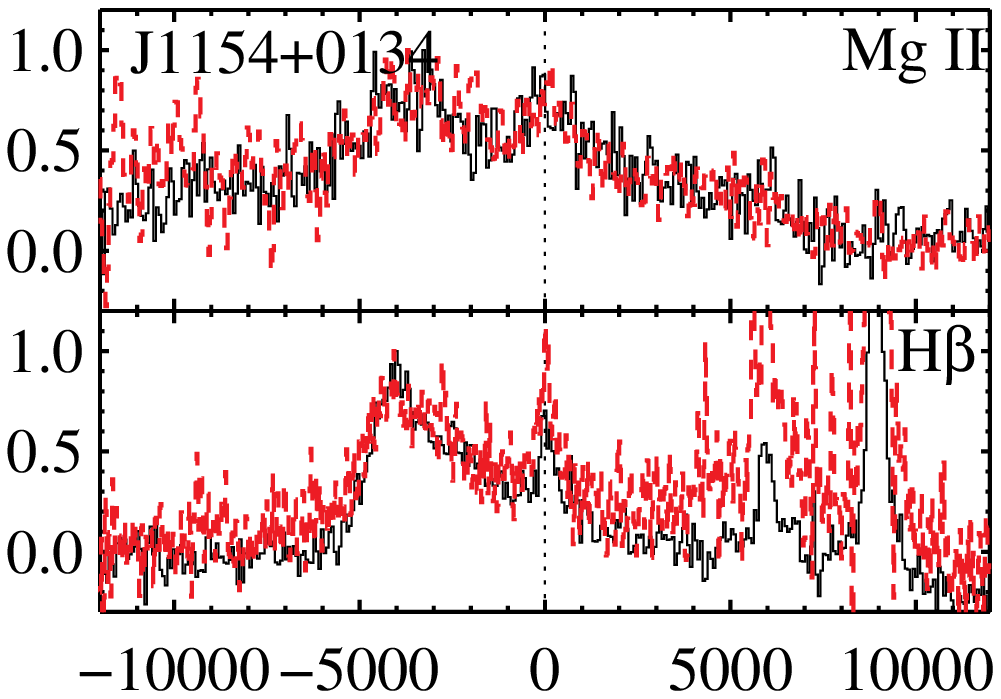}
    \includegraphics[width=34mm]{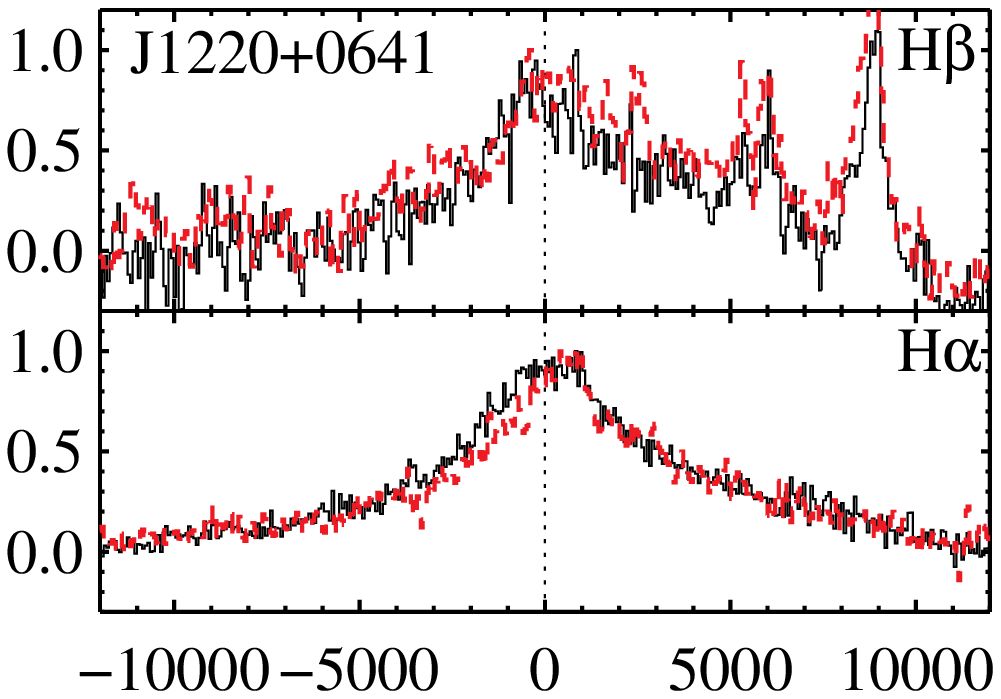}
    \includegraphics[width=34mm]{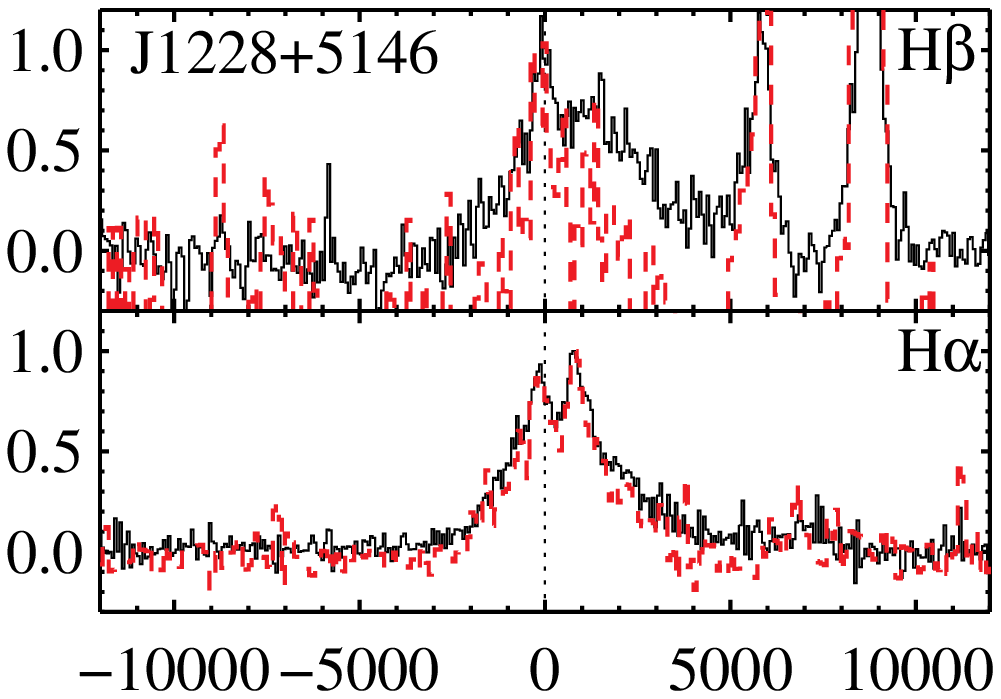}
    \includegraphics[width=34mm]{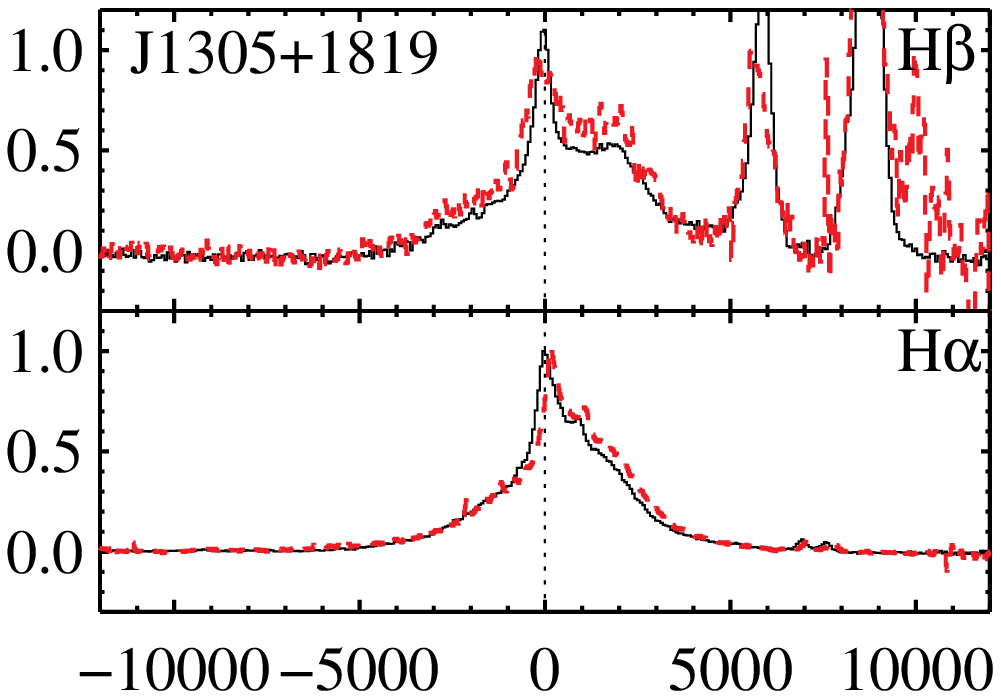}
    \includegraphics[width=34mm]{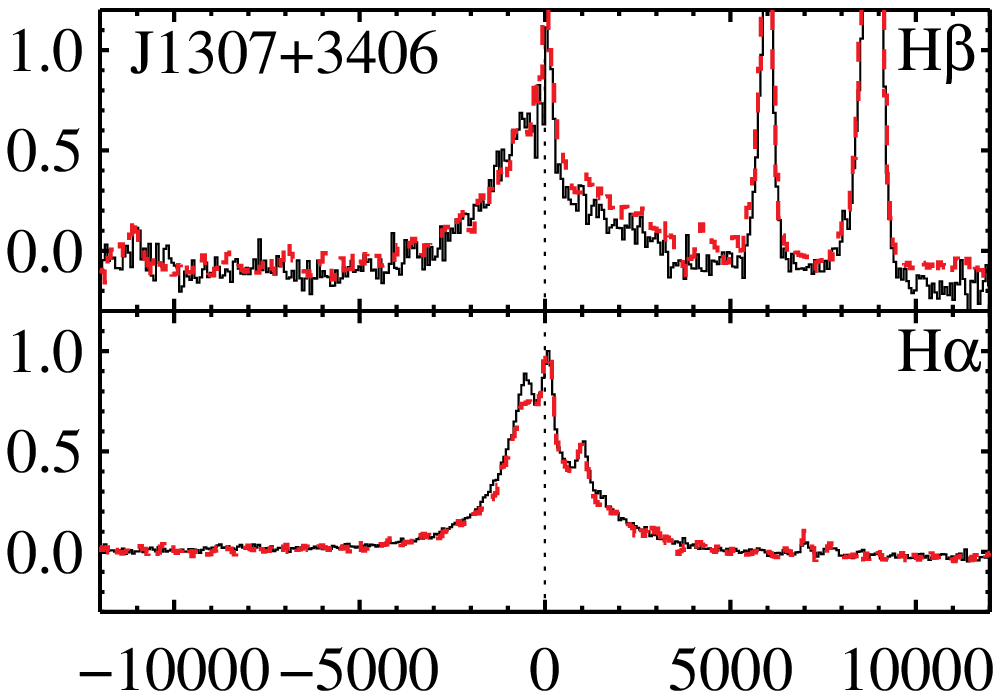}
    \includegraphics[width=34mm]{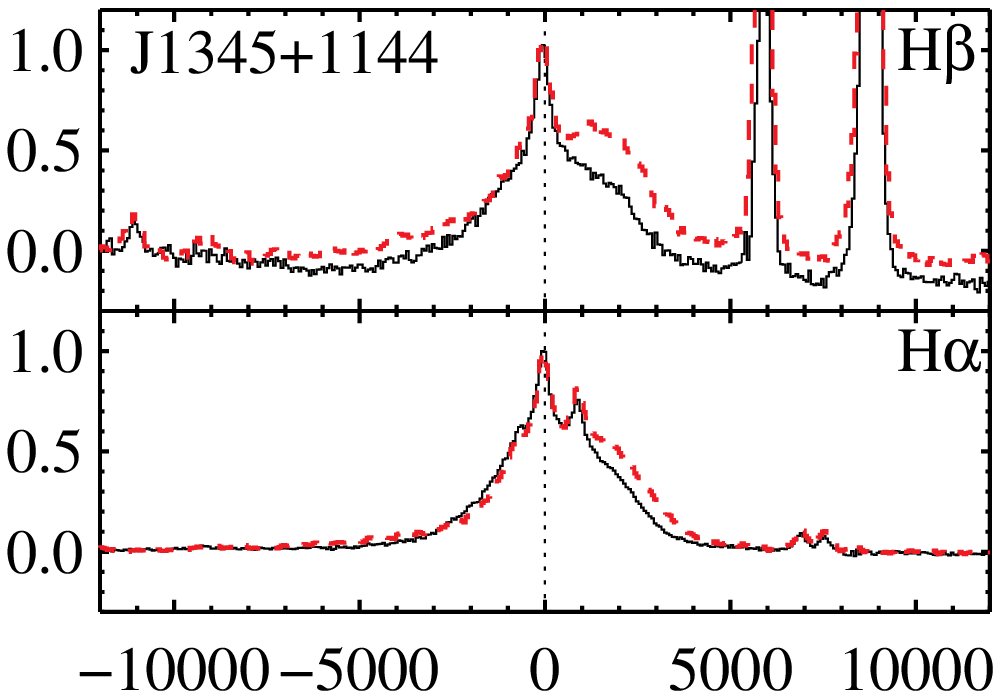}
    \includegraphics[width=34mm]{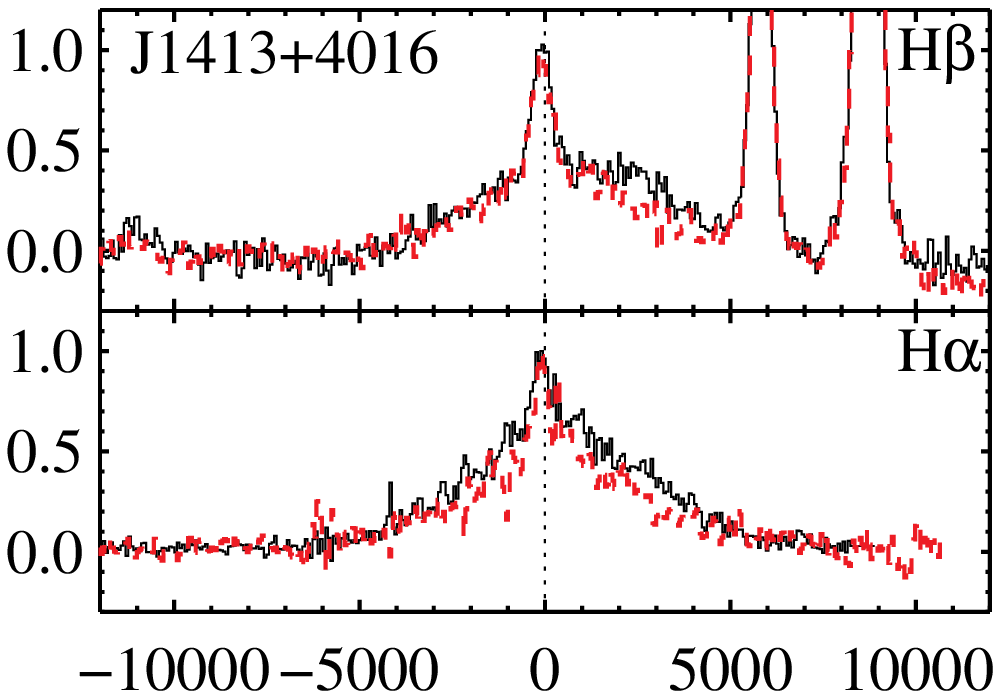}
    \includegraphics[width=34mm]{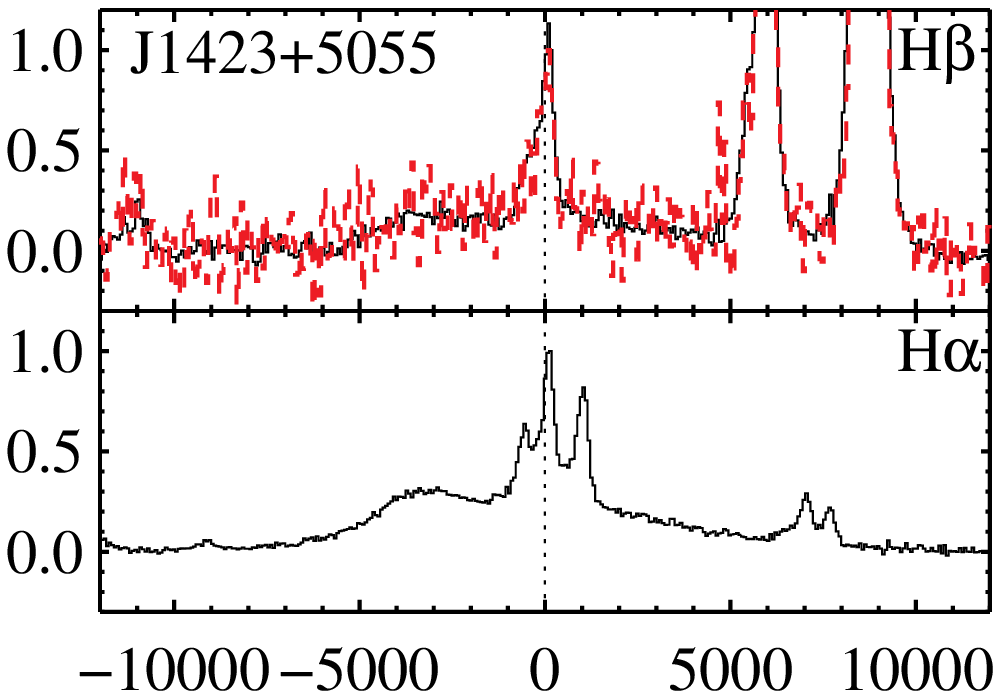}
    \includegraphics[width=34mm]{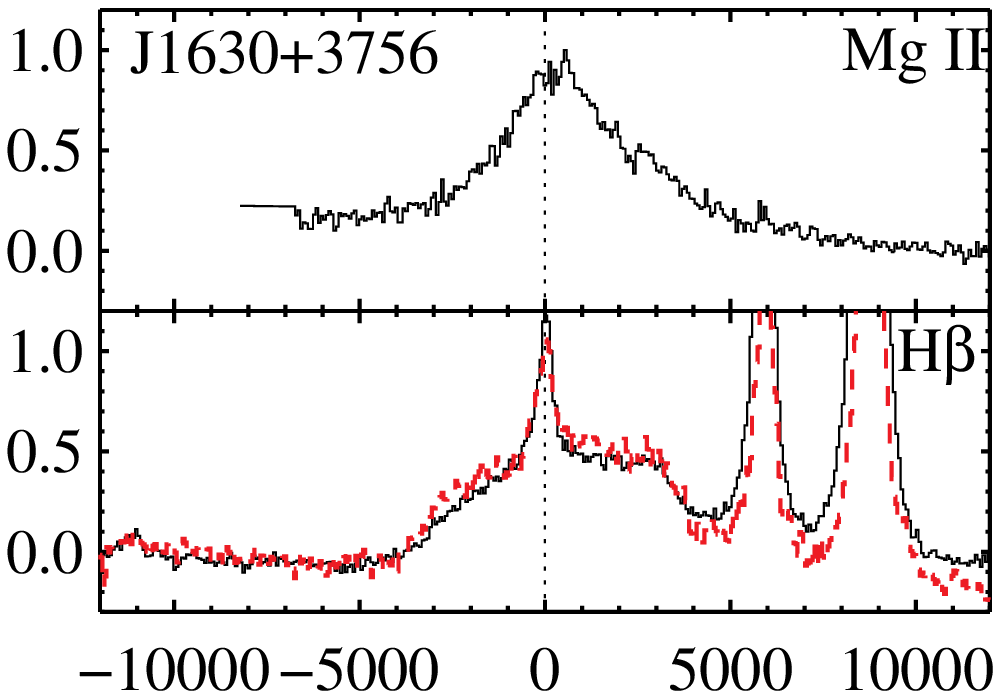}
    \includegraphics[width=34mm]{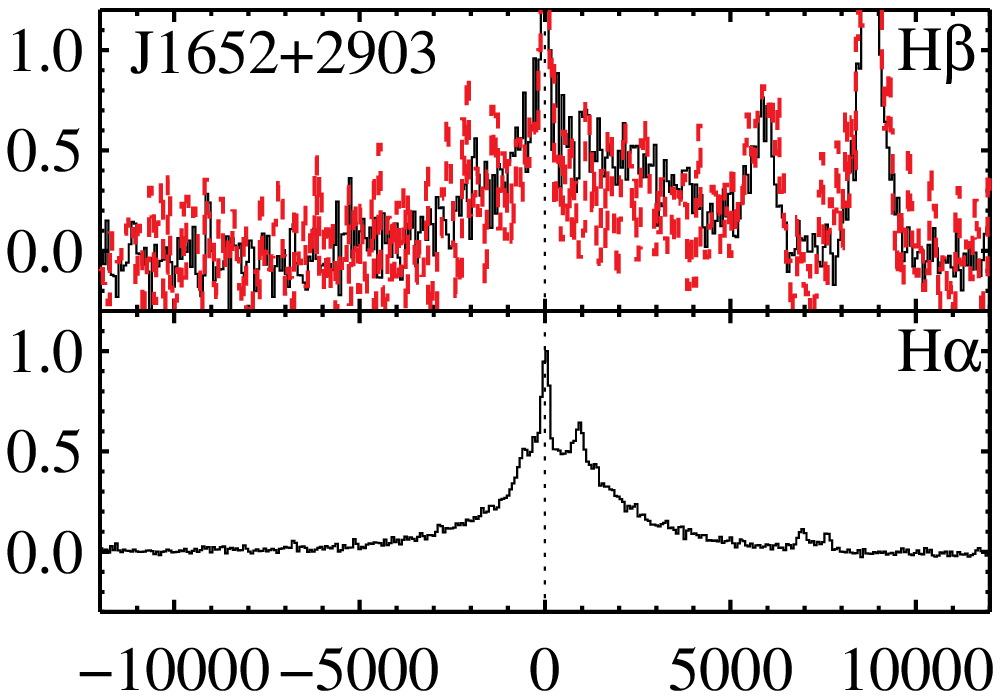}
    \includegraphics[width=34mm]{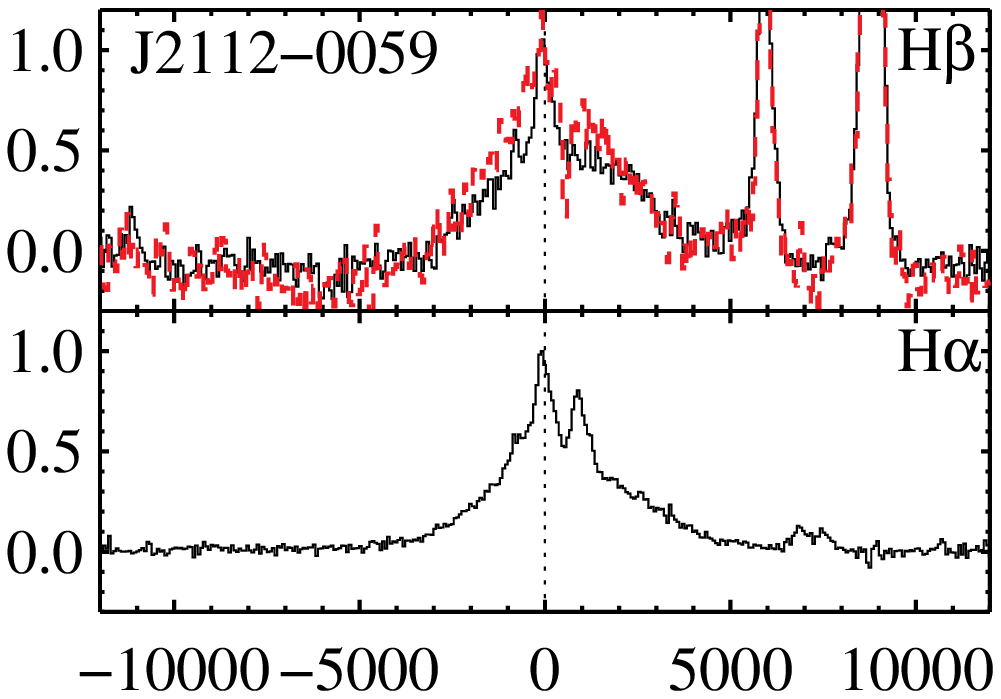}
    \includegraphics[width=34mm]{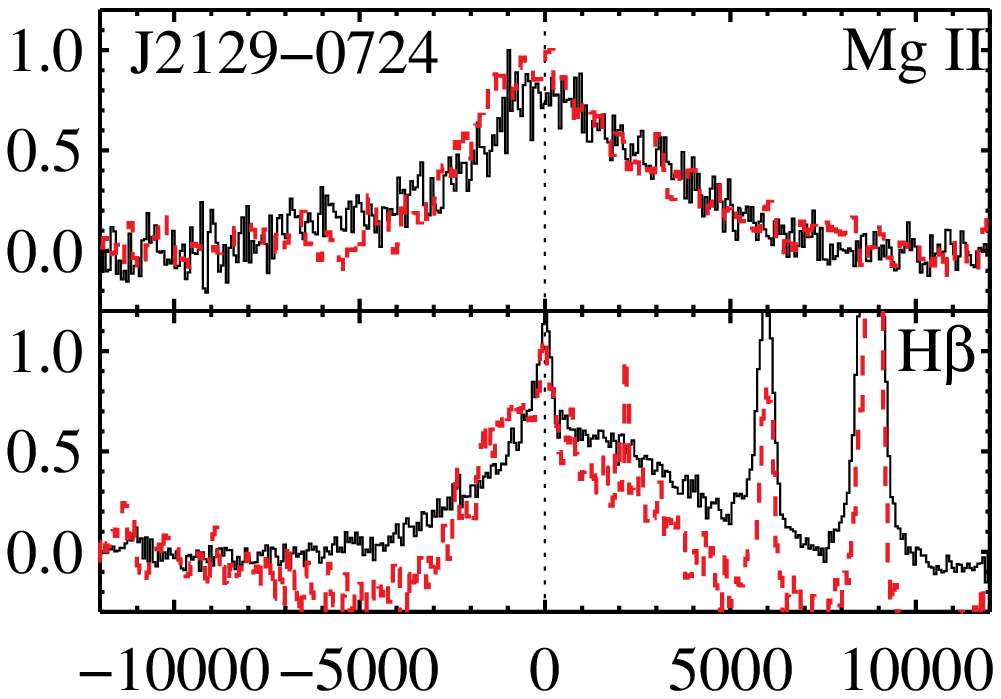}
    \includegraphics[width=34mm]{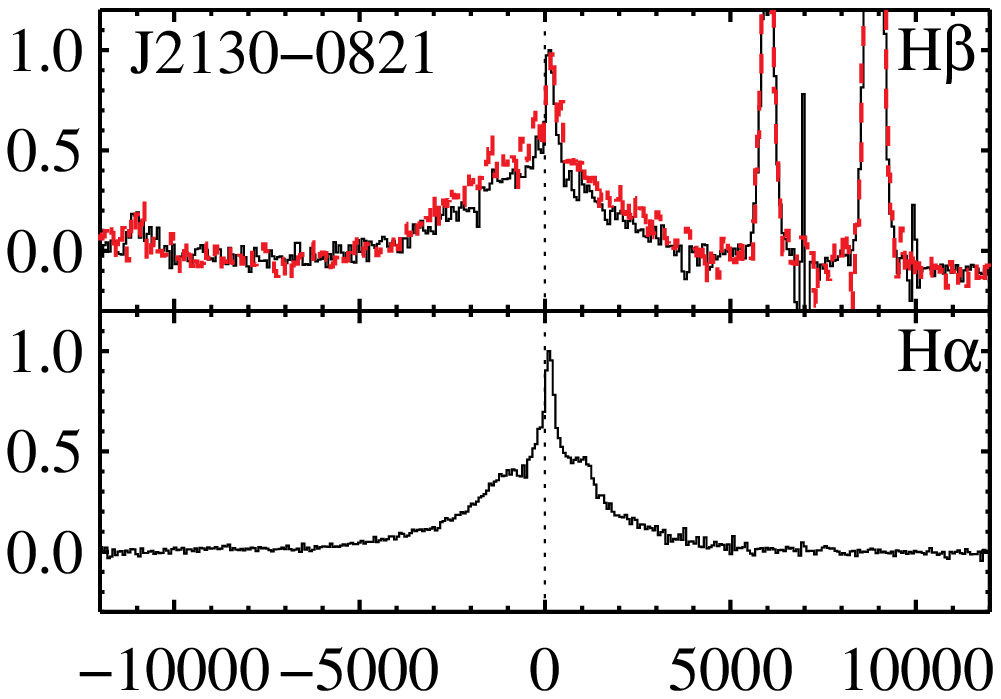}
    \includegraphics[width=34mm]{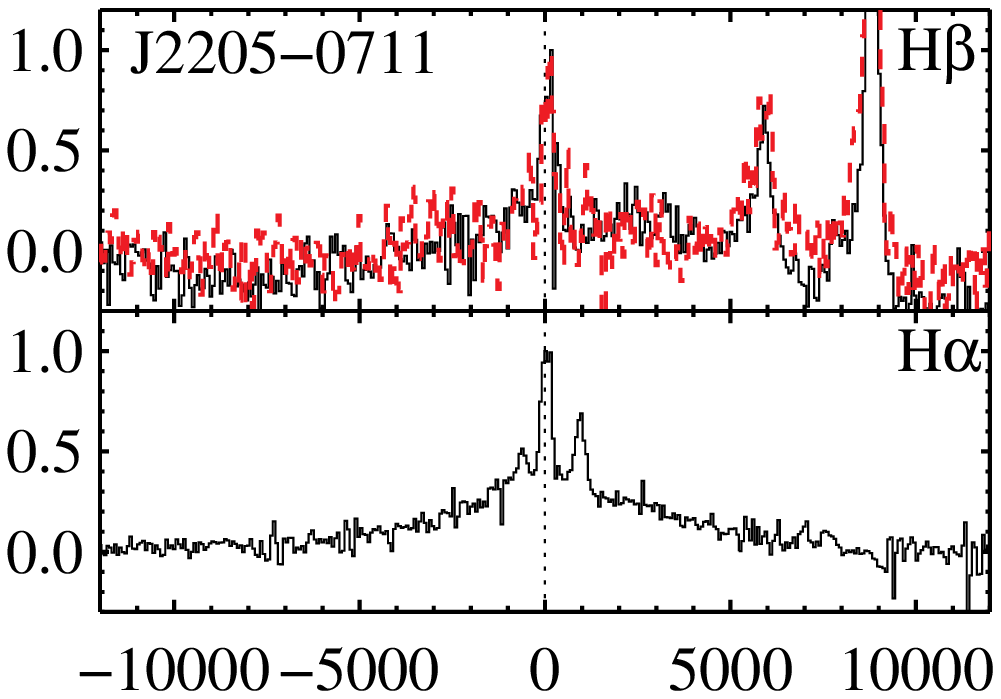}
    \includegraphics[width=34mm]{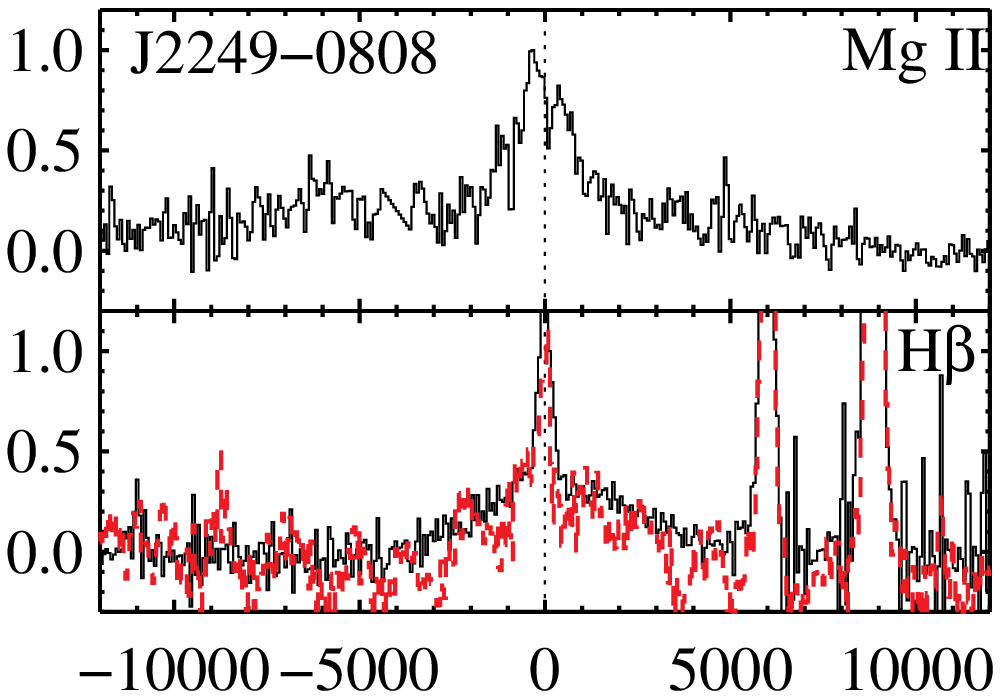}
    \subfigure[]{\includegraphics[width=34mm]{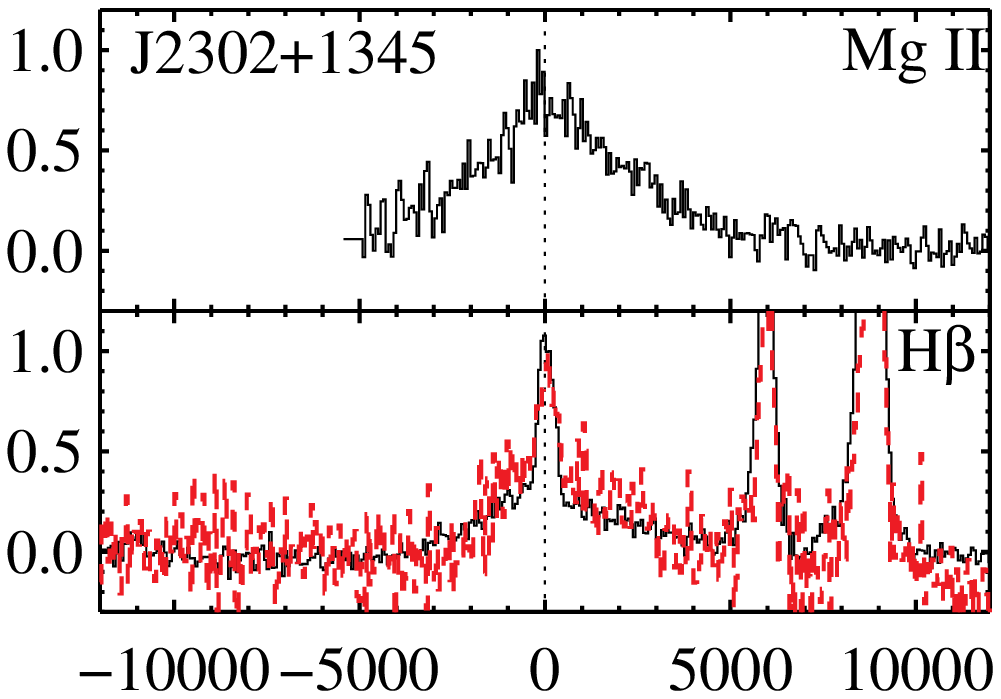}}
    \subfigure[]{\includegraphics[width=34mm]{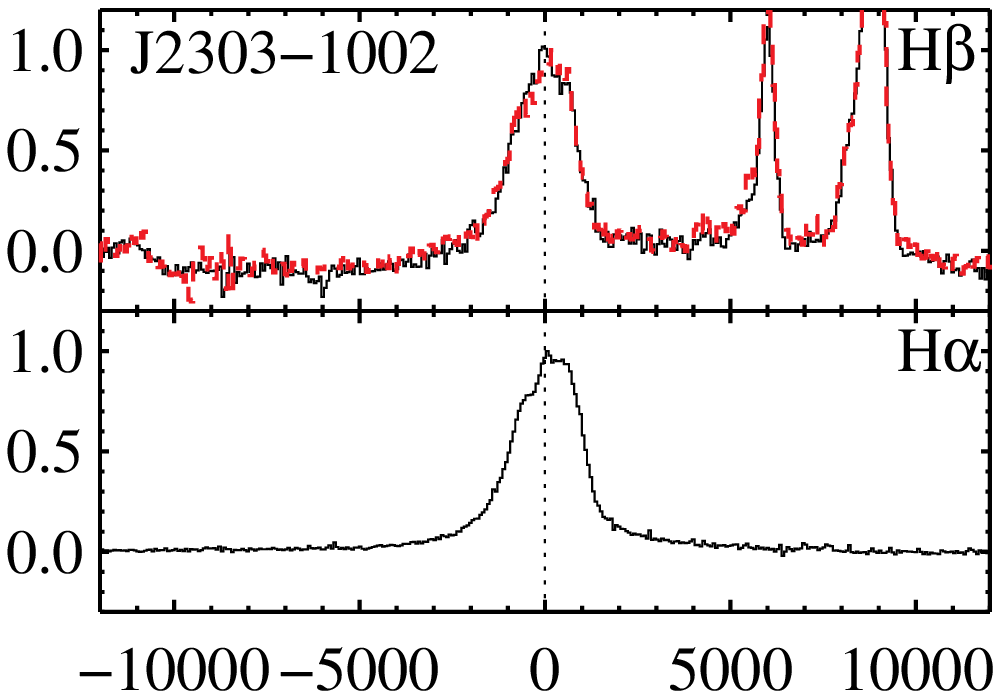}}
    \subfigure[Relative Velocity (km
    s$^{-1}$)]{\includegraphics[width=34mm]{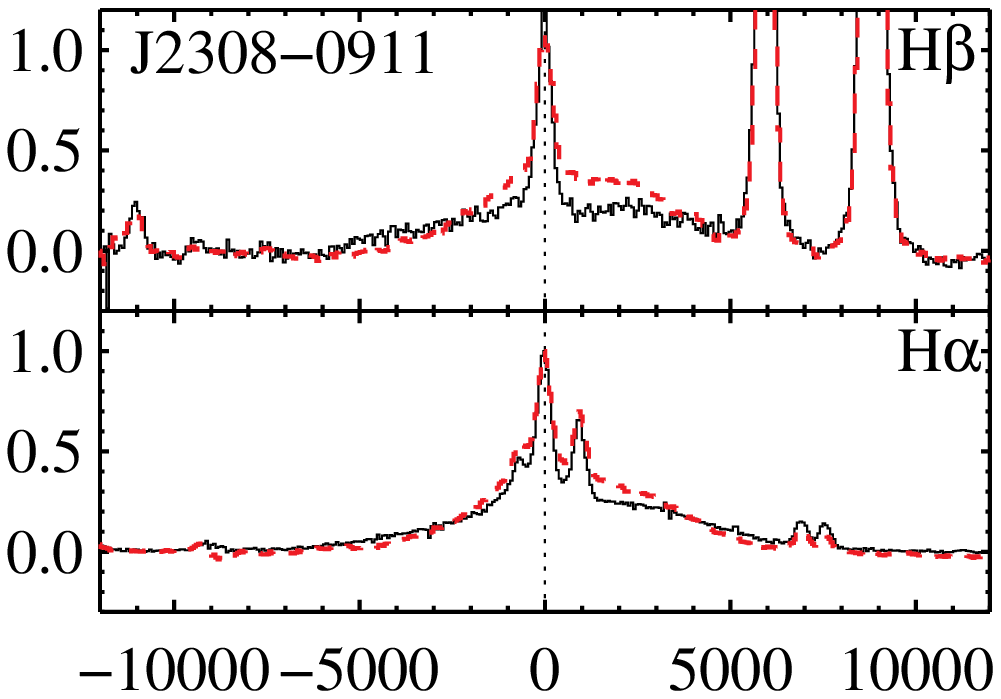}}
    \subfigure[]{\includegraphics[width=34mm]{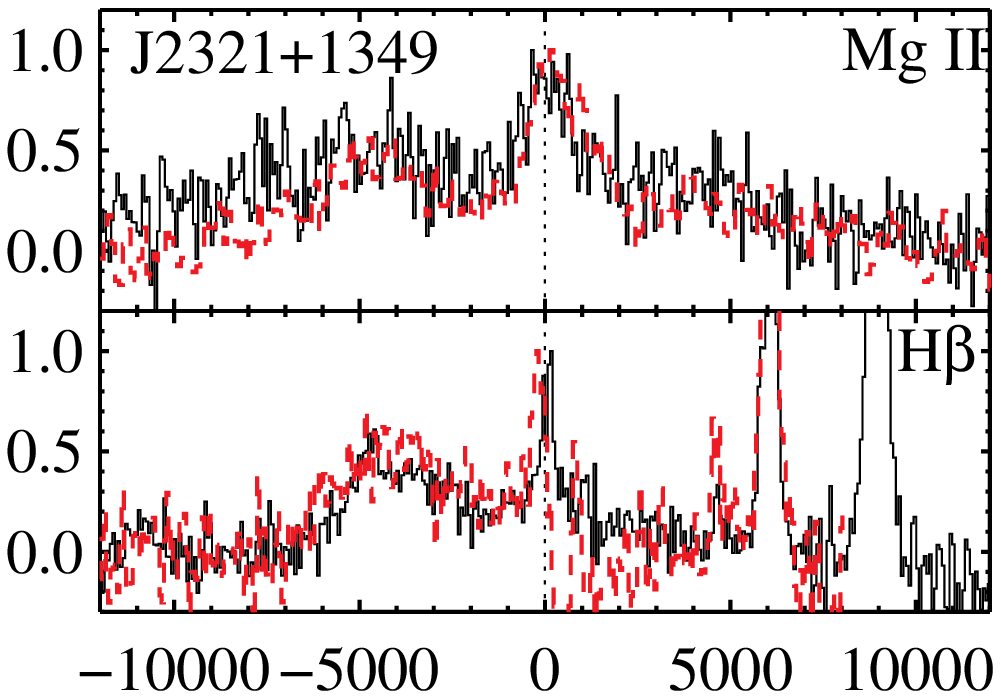}}
    \subfigure[]{\includegraphics[width=34mm]{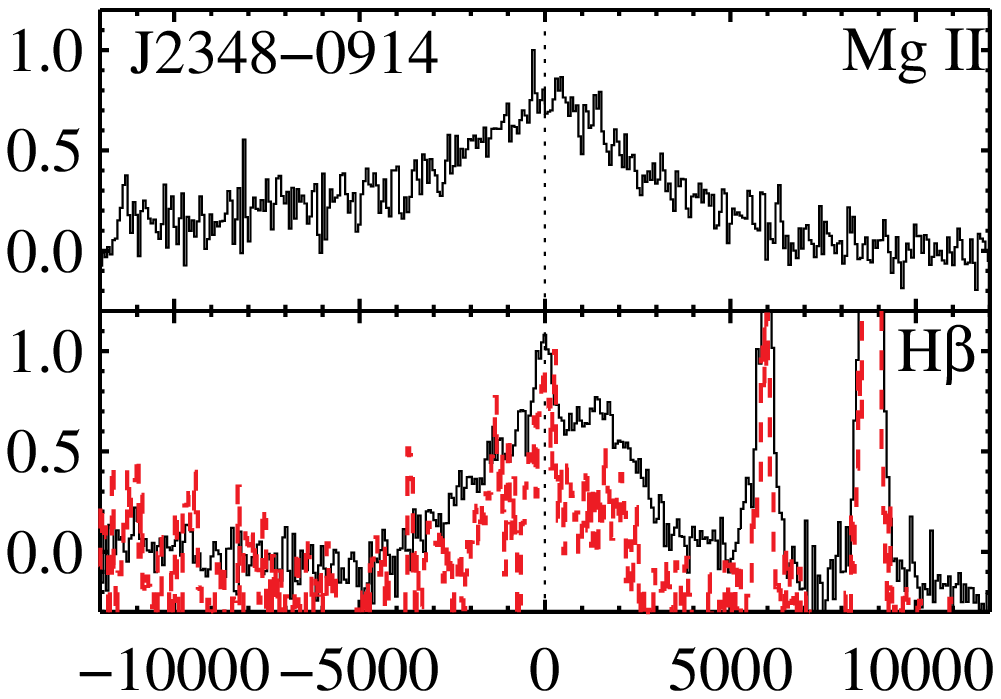}}
    \caption{Multi-epoch spectra (normalized for display purposes) of SDSS quasars
    with offset broad Balmer emission lines.
    The SDSS spectra are in solid black and followup spectra are in dashed red.
    For each object, the top (bottom) panel shows the \hbeta\ (\halpha )
    or \MgII\ (\hbeta ) region, centered on the systemic redshift.}
    \label{fig:1dspec}
\end{figure*}

%%%%%%%%%%%%%%%%%%%%%
%
\subsection{ARC 3.5\,m/DIS}

We observed nine targets using the Dual Imaging
Spectrograph\footnote{http://www.apo.nmsu.edu/arc35m/Instruments/DIS}
(DIS) on the Apache Point Observatory 3.5\,m telescope on the
nights of 2010 April 12, 16 and 2011 June 1 UT. The sky was
non-photometric with varied seeing conditions
($1\farcs1$--$1\farcs6$ for April 12, $1\farcs2$--$3\farcs3$
for April 16, and $1\farcs5$--$4\farcs0$ for June 1). We
adopted a $1\farcs5\times$6$'$ slit and the B1200+R1200
gratings centered at 4400 (or 5300) and 7200 (or 6700)
\angstrom . This wavelength set up covers the \hbeta --\OIIIb\
and \halpha\ (or \MgII ) regions for targets at various
redshifts. The slit was oriented at the parallactic angle at
the time of observation. The spectral resolution is 1.8 (1.3)
\angstrom\ in FWHM (corresponding to $\sigma_{{\rm
inst}}\sim$30--50 km s$^{-1}$) with a pixel scale of 0.62
(0.56) \angstrom\ pixel$^{-1}$ in the blue (red) channel. Total
exposure time varied between 2700 s and 4500 s (Table
\ref{tab:obs}).

%%%%%%%%%%%%%%%%%%%%%%%%%%%%%%%%%%%
%
\subsection{MMT/BCS}

We observed 17 targets with the Blue Channel
Spectrograph\footnote{http://www.mmto.org/node/222} (BCS) on
the 6.5\,m MMT Telescope on the nights of 2011 July 30, 31, and
December 26 and 27 UT. The sky was non-photometric with varied
seeing conditions ($\lesssim 1$\arcsec\ for July 30 and 31,
3\arcsec --5\arcsec\ for December 26, and $\sim1$\arcsec\ for
December 27). We adopted a $0\farcs75\times$180\arcsec\ slit
and the 500 lines mm$^{-1}$ grating centered at 6000 \angstrom\
with the UV36 filter. The slit was oriented at the parallactic
angle at the time of observation. The spectral coverage was
3134 \angstrom\ with a spectral resolution of $R=2800$
($\sigma_{{\rm inst}}\sim$50 km s$^{-1}$) and a pixel scale of
1.2 \angstrom\ pixel$^{-1}$. Total exposure time varied between
1200 s and 3600 s (Table \ref{tab:obs}).

%%%%%%%%%%%%%%%%%%%%%%%%%%%%%%%%%%%
%
\subsection{FLWO 1.5\,m/FAST}

We observed 24 targets using the FAST spectrograph
\citep{fabricant98} on the 1.5\,m Tillinghast Telescope at the
Fred Lawrence Whipple Observatory (FLWO). Observations were
carried out in queue mode over 19 nights from 2011 September to
2012 April. We employed the 300 lines mm$^{-1}$ grating with a
$2''\times180''$ slit oriented at the parallactic angle at the
time of observation. The spectral coverage was $\sim4000$
\angstrom\ with a spectral resolution of $\sim4$ \angstrom\
($\sigma_{{\rm inst}}\sim$90 km s$^{-1}$) and a pixel scale of
1.5 \angstrom\ pixel$^{-1}$. The spectrograph was centered at
different wavelengths depending on target redshift to cover
their \hbeta\ and \halpha\ (or \MgII ) regions. Table
\ref{tab:obs} lists the total exposure time for each object.

%%%%%%%%%%%%%%%%%%%%%%%%%%%%%%%%%%%
%
\subsection{Data Reduction}\label{subsec:redux}

We reduced the second-epoch optical spectra following standard
IRAF\footnote{IRAF is distributed by the National Optical
Astronomy Observatory, which is operated by the Association of
Universities for Research in Astronomy (AURA) under cooperative
agreement with the National Science Foundation.} procedures
\citep{tody86}, with special attention to accurate wavelength
calibration. Wavelength solutions were obtained using multiple
HeNeAr lamp lines with rms of a few percent in a single
exposure. Bright sky lines in the spectrum were used to correct
for small errors in the absolute wavelength scale, which was
corrected to heliocentric velocity. Flux calibration and
telluric correction were applied after extracting
one-dimensional spectra for each individual frame. We then
combined all the frames to get a co-added spectrum for each
target. Table \ref{tab:obs} lists the S/N achieved in the final
followup spectrum for each object. Figure \ref{fig:1dspec}
shows the second-epoch spectra compared against the original
SDSS observations.

To measure the temporal velocity shift of the broad lines using
cross-correlation analysis (ccf; Section \ref{subsec:cc}; see
also Paper I), we have re-sampled the second-epoch spectra so
that they share the exact same wavelength grids (in vacuum) as
the original SDSS spectra, which are linear on a logarithmical
scale (i.e., linear in velocity space) with a pixel scale of
$10^{-4}$ in log-wavelength, corresponding to 69 km s$^{-1}$.

\begin{figure}
  \centering
    \includegraphics[width=90mm]{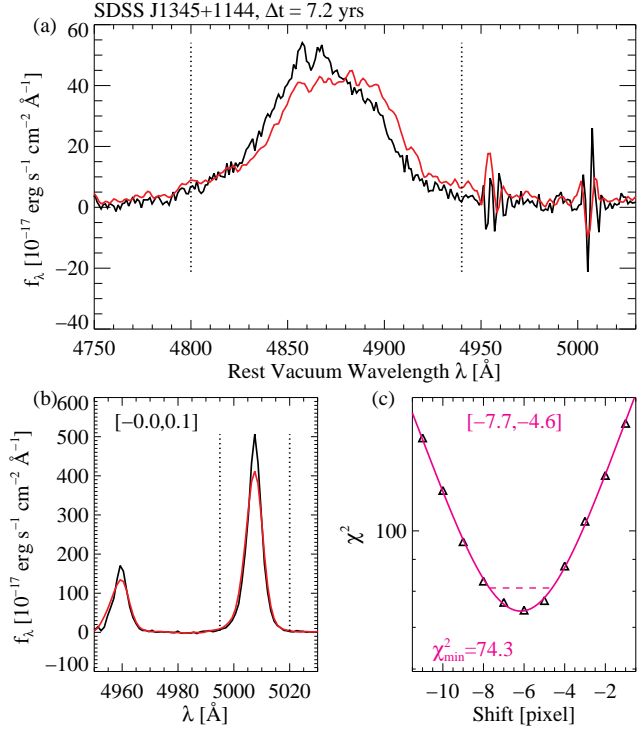}
    \caption{Example of the cross-correlation analysis to
    measure the velocity shift of the broad \hbeta\ between two epochs.
    (a) Broad \hbeta\ spectrum of the original SDSS (black) and followup
    (red) observations. The spectral range of the cross-correlation analysis
    is marked by the dotted vertical lines. The followup spectrum has been
    scaled to match the integrated broad \hbeta\ line flux in the
    cross-correlation range. (b) Same as in panel (a), but for the narrow
    \OIII\ emission lines. Shown in brackets are the 99\% confidence
    ranges (2.5$\sigma$) in units of pixels of the velocity shift of \OIIIb\
    (with 1 pixel corresponding to 69 km s$^{-1}$).  Here and in other figures
    throughout the paper showing the cross-correlation analysis results, negative values
    mean
    that the emission line in the followup spectrum needs to be blueshifted
    to match that in the original SDSS spectrum (i.e., the emission line in the
    followup spectrum is redshifted relative to that in the original spectrum).
    (c) $\chi^2$ for the cross-correlation analysis of the broad \hbeta\ as a
    function of pixels. The solid magenta curve is the sixth-order B-spline fit
    of the 21 grid points centered on the one with the minimal $\chi^2$.
    The dashed horizontal segment indicates the $\Delta \chi^2 = 6.63 (2.5\sigma)$
    range, also indicated in the magenta brackets in units of pixels.}
    \label{fig:cceg1}
\end{figure}

%%%%%%%%%%%%%%%%%%%%%%%%%%%%%%%%%%%%%%%%%%%%%%%%%%%%%%%%%%%%%%
%
%
\section{Measuring Radial Velocity Shift}\label{sec:analysis}

We first describe our approach to quantify the velocity shifts
in emission lines between two epochs (Section \ref{subsec:cc}).
We then discuss measurement uncertainties and caveats (Section
\ref{subsec:simu}).

%%%%%%%%%%%%%%%%%%%%%%%%%%%%%%%%%%
%
\subsection{Cross-correlation Analysis}\label{subsec:cc}

We adopt a cross-correlation analysis (``ccf'' for short)
following the method discussed in Paper I \citep[see
also][]{eracleous11}. Cross-correlation analysis is preferred
over approaches based on line fitting, which are more model
dependent and are less sensitive to velocity shift. As shown
with simulations in Paper I, ccf can in general achieve a
factor of a few better sensitivity in velocity shift than
direct line fitting.

\begin{figure}
  \centering
    \includegraphics[width=90mm]{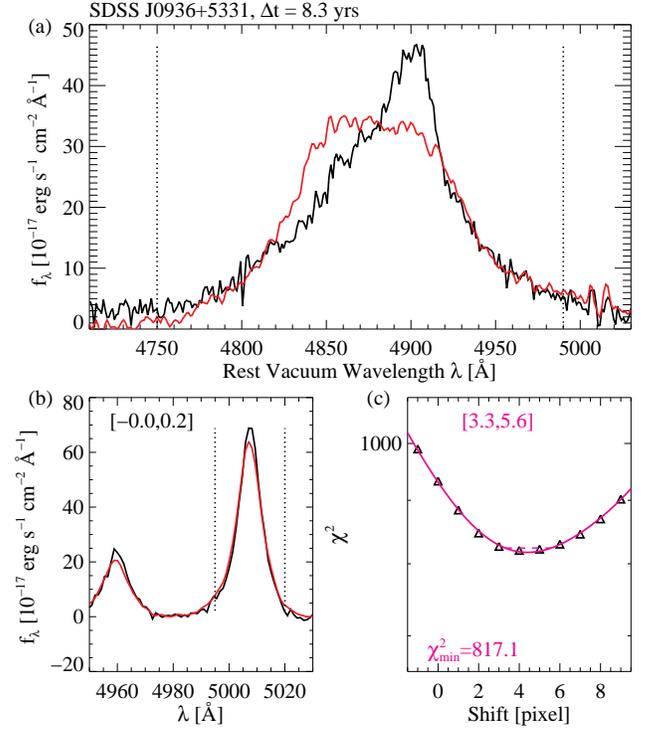}
    \caption{Same as Figure \ref{fig:cceg1}, but for an example where
    the broad \hbeta\ profile changes significantly between two epochs.}
    \label{fig:cceg3}
\end{figure}

The ccf finds the best-fit velocity shift $V_{\rm ccf}$ between
the two epochs based on $\chi^2$ minimization (see Paper I for
details). We have subtracted the pseudo-continua and narrow
emission lines before analyzing the broad-line velocity shifts.
We have fixed the narrow \hbeta\ to \OIII\ ratio to be
consistent between two epochs. This helps minimize the error
caused by incorrect narrow \hbeta\ subtraction due to model
degeneracy. We use \OIII\ narrow emission lines, which are
expected to show zero offset,\footnote{There are three objects
(Table \ref{tab:obs}) whose \OIII\ lines exhibit nonzero
velocity shifts (with absolute values of 50--60 km s$^{-1}$)
between two epochs, which are likely due to either wavelength
calibration errors or slit losses.} to calibrate the absolute
wavelength accuracy; we have subtracted off any nonzero
velocity shifts detected in the \OIII\ lines from the
broad-line velocity shift measurements (assuming that the
nonzero shift in \OIII\ is due to wavelength calibration
errors). In constraining the velocity shift of narrow \OIII ,
we have subtracted the pseudo-continua and broad emission
lines. Figures \ref{fig:cceg1} and \ref{fig:cceg3} show two
examples of the ccf where a velocity shift is detected at the
$>2.5\sigma$ significance level in broad \hbeta ; examples of
no significant velocity shifts can be found in Paper I. Figure
\ref{fig:cceg1} shows a detection in broad \hbeta\ whose line
profiles are consistent within uncertainties between two
epochs, as quantified by various measures of line width and
shape (FWHM and skewness). To double check that the velocity
shift is real and not caused by some subtle changes in line
profiles, we have also repeated the ccf with the
broad-line-only spectrum smoothed with a Gaussian kernel with
standard deviation $\sigma_{{\rm s}}$, and verified that the
velocity shift does not change with varying $\sigma_{{\rm s}}$.
Figure \ref{fig:cceg3} shows a detection in broad \hbeta\ with
significant line profile changes between two epochs as
quantified by line widths. As further explained below in
Section \ref{subsec:detection}, we classify the former case as
a BBH candidate and the latter as due to BLR variability around
a single BH.\footnote{While the dramatic line profile change in
the latter case could also be due to a BBH where both BHs are
active and the systems have two unresolved broad-line peaks
\citep[e.g.,][]{shen10}, this scenario is perhaps unlikely
considering the small parameter space, if any, allowed for such
systems, as discussed in Section \ref{subsec:sum_previous}.}

%%%%%%%%%%%%%%%%%%%%%%%%%%%%%%%%%%
%
\subsection{Uncertainties}\label{subsec:simu}

We now discuss the error budget of the line shift measurements.
Residual errors from wavelength calibration should be minor,
because (1) we have calibrated absolute redshift using
simultaneous constraints on the narrow emission lines based on
\OIII\ ccf; and (2) the relative wavelength accuracy has been
calibrated to within a few percent using multiple lamp
exposures and has been further verified by independent
measurements using a second broad line (\halpha\ or \MgII ).
Errors due to absolute flux calibration should not be a major
issue either, because we have subtracted the pseudo-continua.
The spectra have been normalized to have the same integrated
broad-line flux. Relative spectrophotometric flux calibration
in principle can introduce substantial uncertainty in the
velocity shift measurement, but for most targets we can
constrain such effects by considering independent measurements
from a second broad line.

\begin{figure}
  \centering
    \includegraphics[width=88mm]{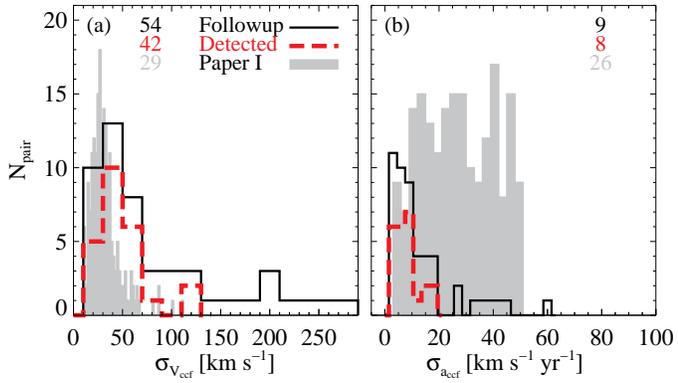}
    \caption{Distribution of the measurement uncertainties of
    (a) broad-line velocity shift and (b) broad-line rest-frame acceleration.
    Numbers indicate median values of measurement uncertainty for different samples.
    }
    \label{fig:sigdist}
\end{figure}

Figure \ref{fig:sigdist} shows distributions of the measurement
uncertainty of broad-line velocity shift $\sigma_{V_{{\rm
ccf}}}$ and acceleration $\sigma_{a_{{\rm ccf}}}$. They were
estimated from the 99\% confidence ranges of the $\chi^2$
curves as listed in Table \ref{tab:obs}. Recall that the
measurement errors in the velocity shift of broad emission
lines increase with increasing line width and decreasing S/N
(Paper I). Compared to the ``superior'' sample\footnote{Defined
in Paper I as pairs of SDSS spectra for which we have derived
meaningful constraints on the velocity shift between two epochs
from the ccf, and whose acceleration uncertainties
$\sigma_{a_{{\rm ccf}}}<50$ km s$^{-1}$ yr$^{-1}$. Here we
compare with the superior sample as an example, because Paper I
derived statistical constraints on the general BBH population
using the superior sample. Similar comparisons can be made with
the ``good'' sample (defined as pairs of SDSS spectra for which
we have derived meaningful constraints on the velocity shift
between two epochs from the ccf, without any cut on
$\sigma_{a_{{\rm ccf}}}$).} of Paper I, the median
$\sigma_{V_{{\rm ccf}}}$ of the followup sample is $\sim85$\%
larger. This is expected for two reasons: (1) the broad \hbeta\
FWHM of the followup sample is on average $\sim25$\% larger
than that of ordinary quasars (Figure \ref{fig:fwhm}); and (2)
the SDSS spectrum S/N of the followup sample is comparable to
that of the superior sample but our new second-epoch spectra
have lower S/N in general. However, the typical
$\sigma_{a_{{\rm ccf}}}$ of the followup sample is $\sim65$\%
smaller than that of the superior sample of Paper I, due to
longer time baselines.

%%%%%%%%%%%%%%%%%%%%%%%%%%%%%%%%%%%%%%%%%%%%%%%%%%%%%%%%%%%%
%
\section{Results}\label{sec:result}

We present the detection frequency of broad-line shifts for
offset quasars in Section \ref{subsec:frequency}. We then
examine the acceleration distribution and compare to the
general quasar population studied in Paper I (Section
\ref{subsec:vs}). In Section \ref{subsec:vacorr} we study the
relation between velocity offset in single-epoch spectra and
acceleration between two epochs.  Finally we discuss the likely
scenario for each individual detection in Section
\ref{subsec:detection}.

%%%%%%%%%%%%%%%%%%%%%%%%%%%%%%%%%%%%%%%%
%
\subsection{Detection Frequency}\label{subsec:frequency}

\begin{figure*}
  \centering
    \includegraphics[width=89mm]{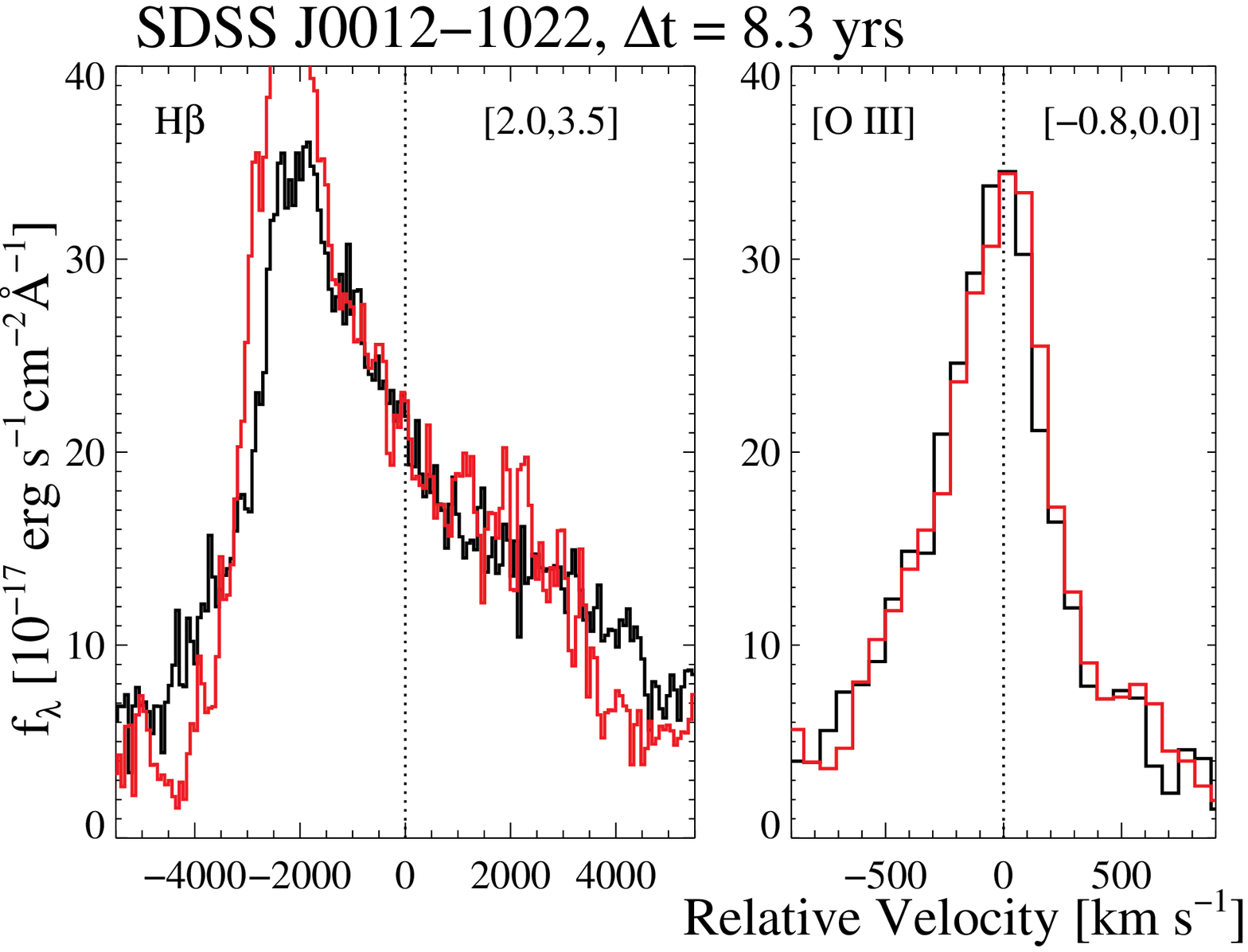}
    \includegraphics[width=89mm]{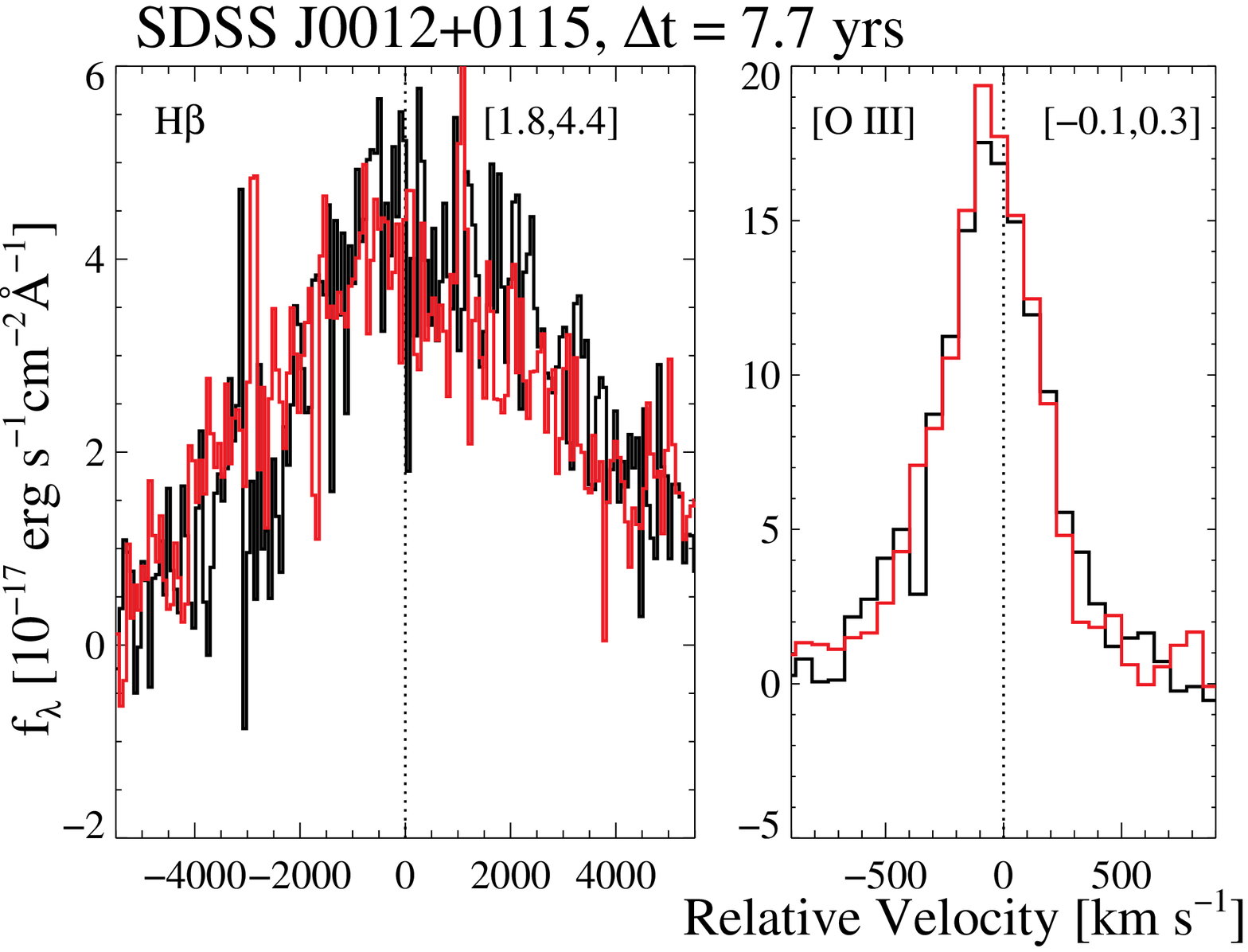}
    \includegraphics[width=89mm]{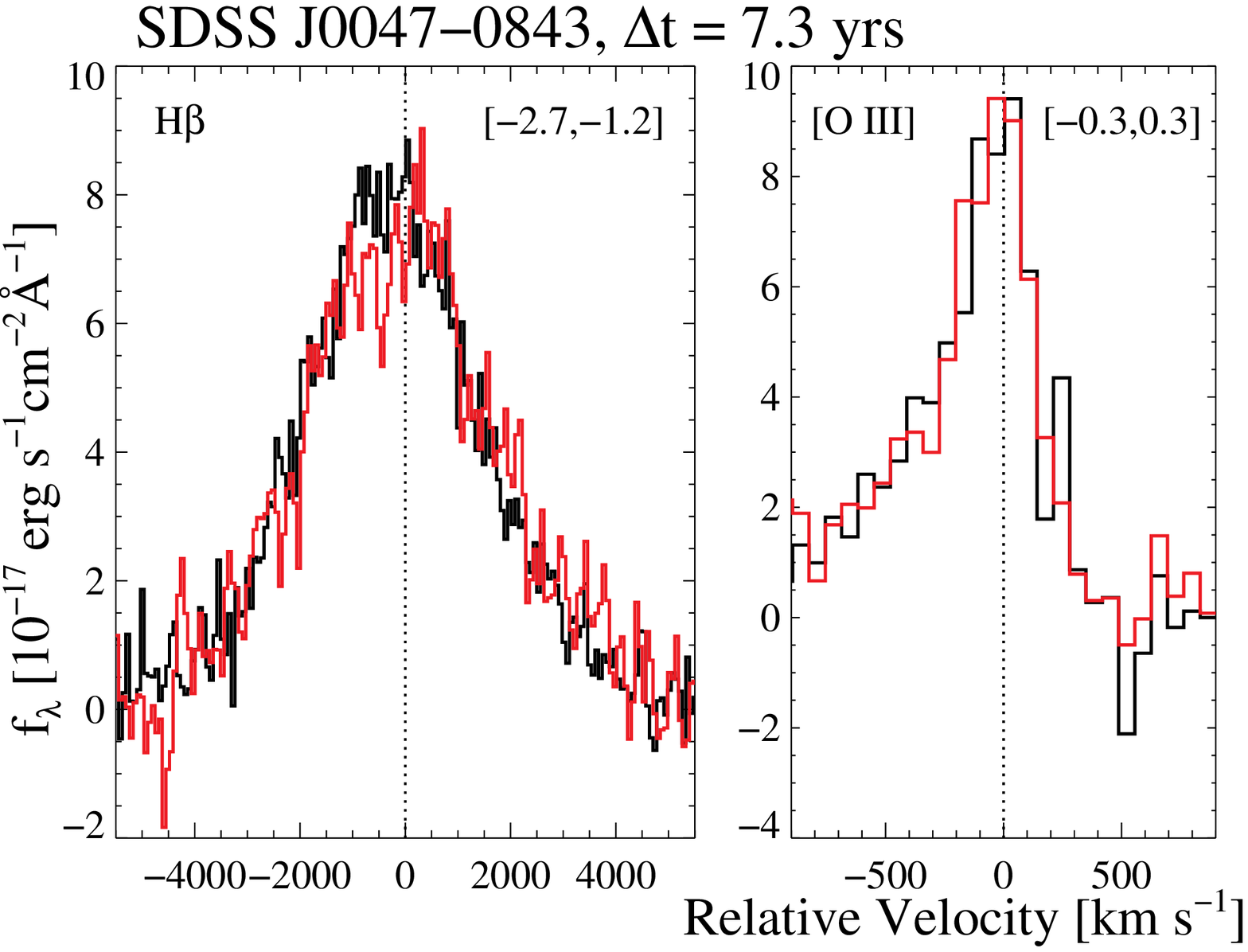}
    \includegraphics[width=89mm]{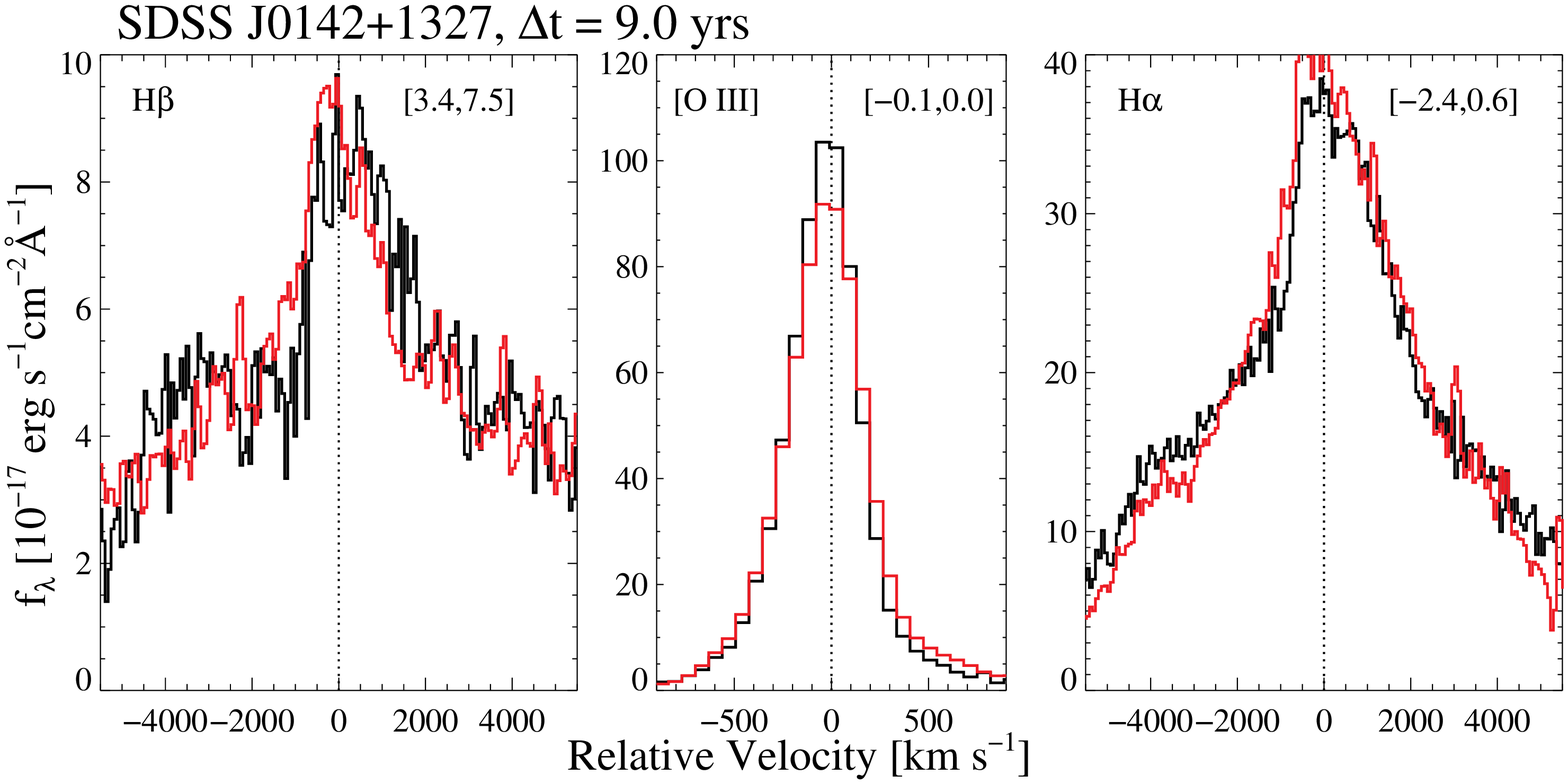}
    \includegraphics[width=89mm]{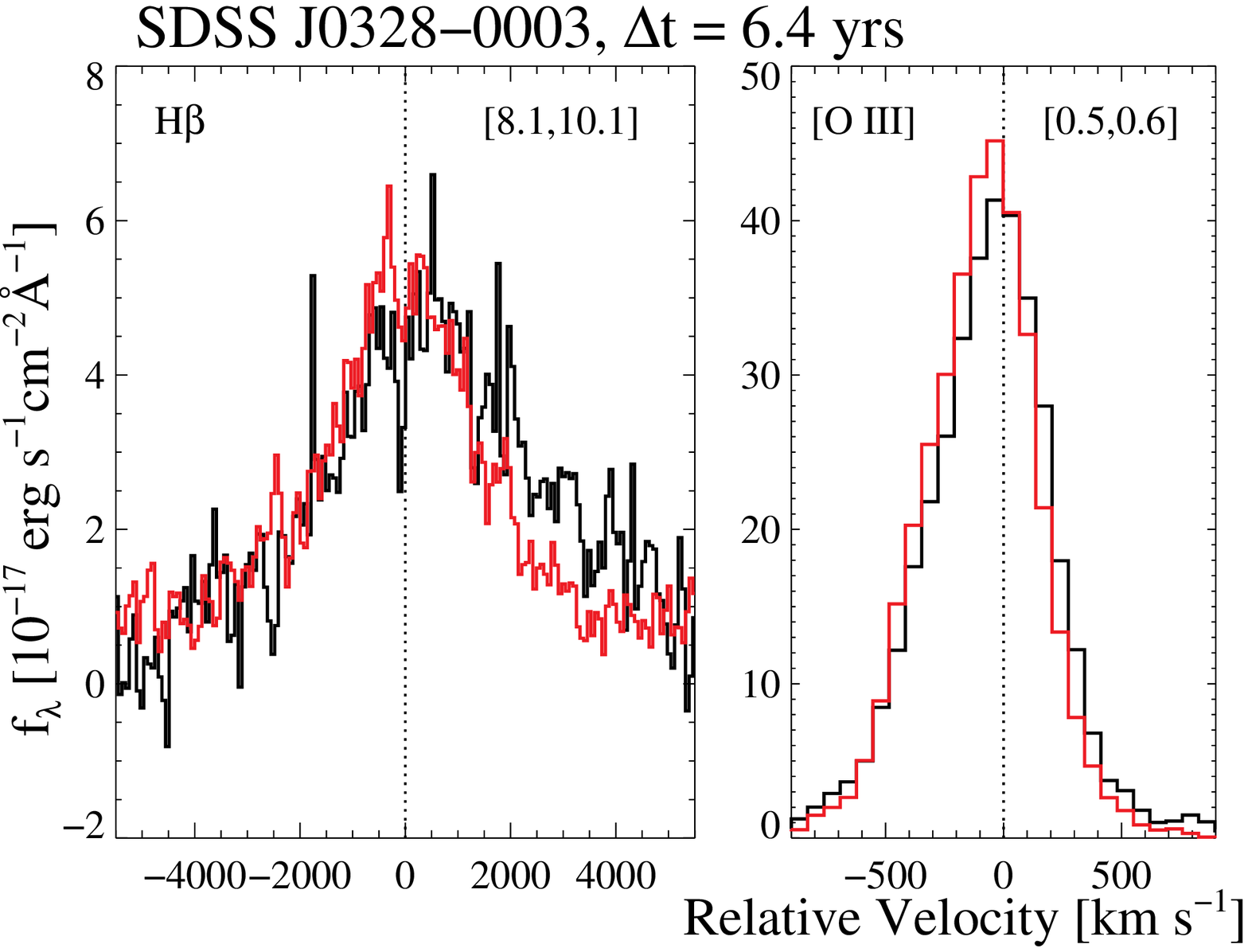}
    \includegraphics[width=89mm]{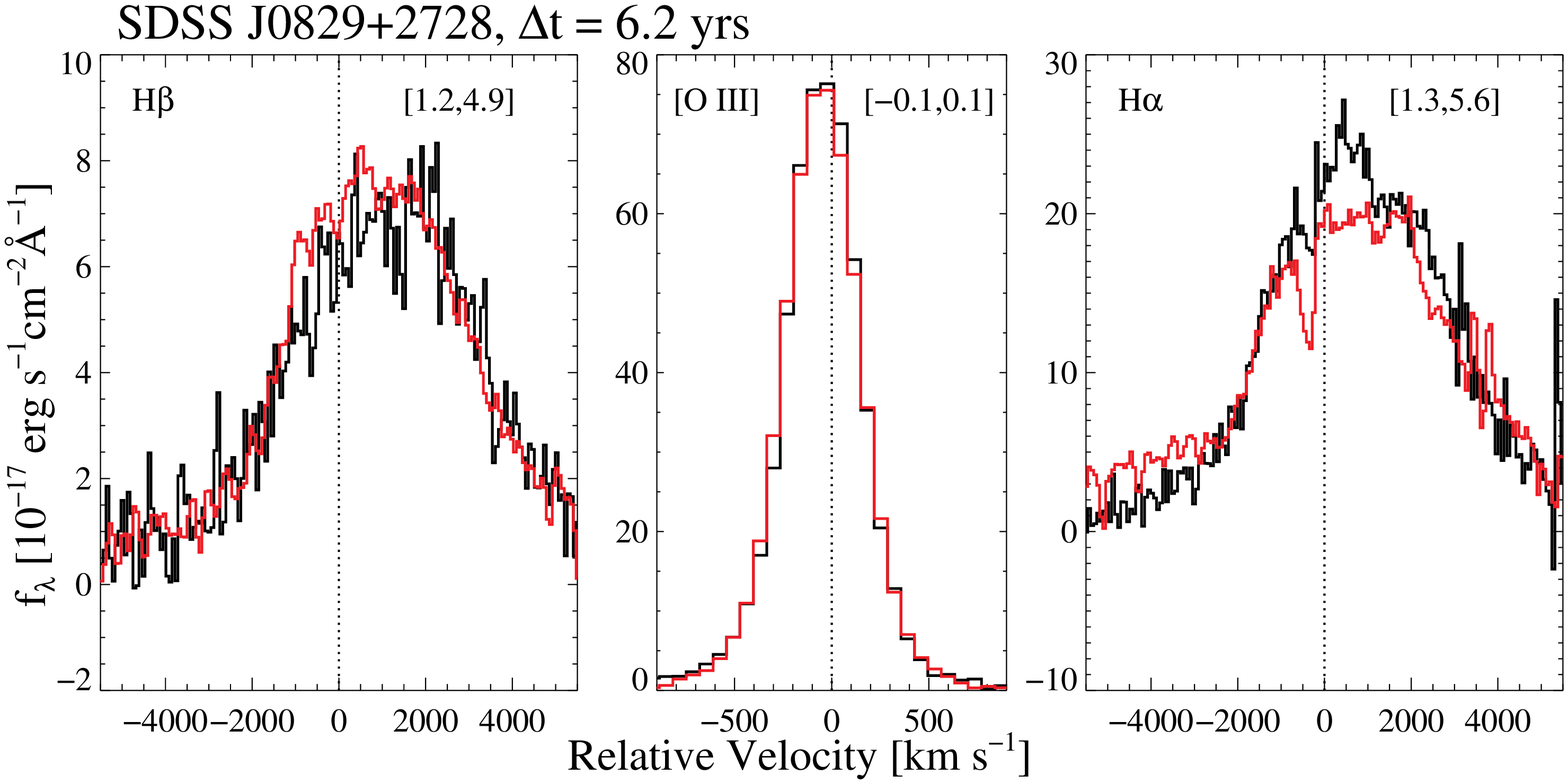}
    \includegraphics[width=89mm]{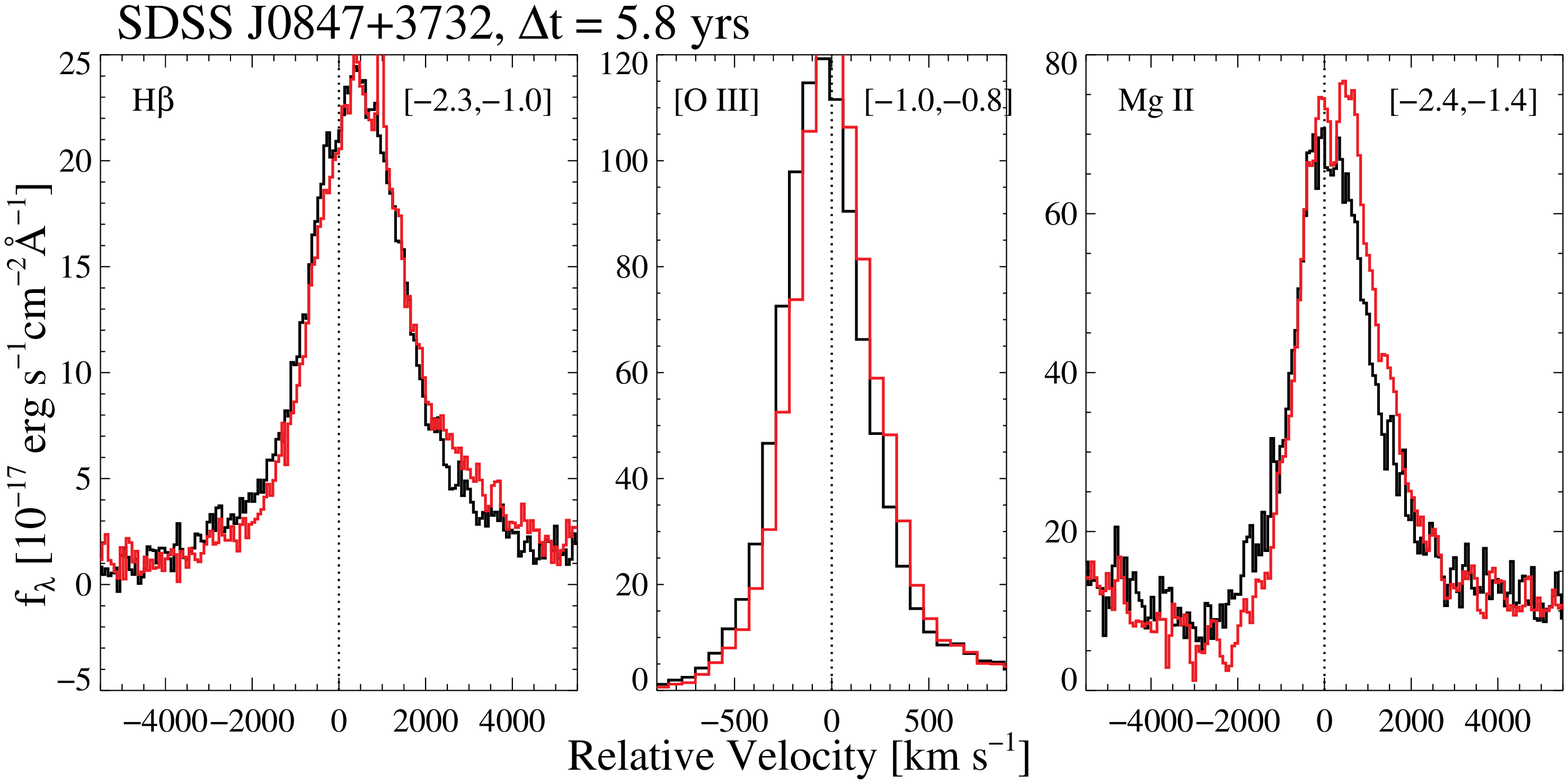}
    \includegraphics[width=89mm]{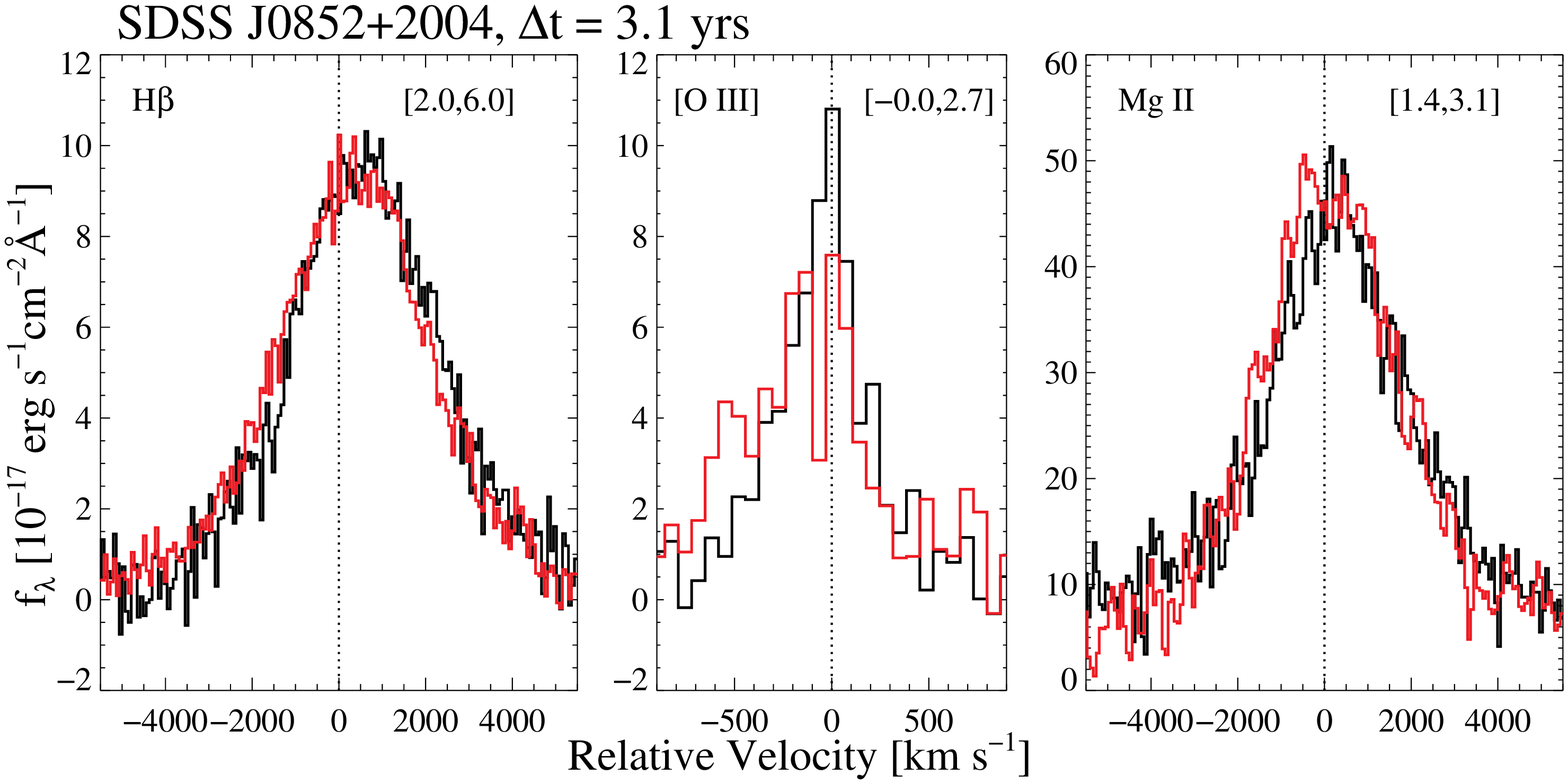}
    \includegraphics[width=89mm]{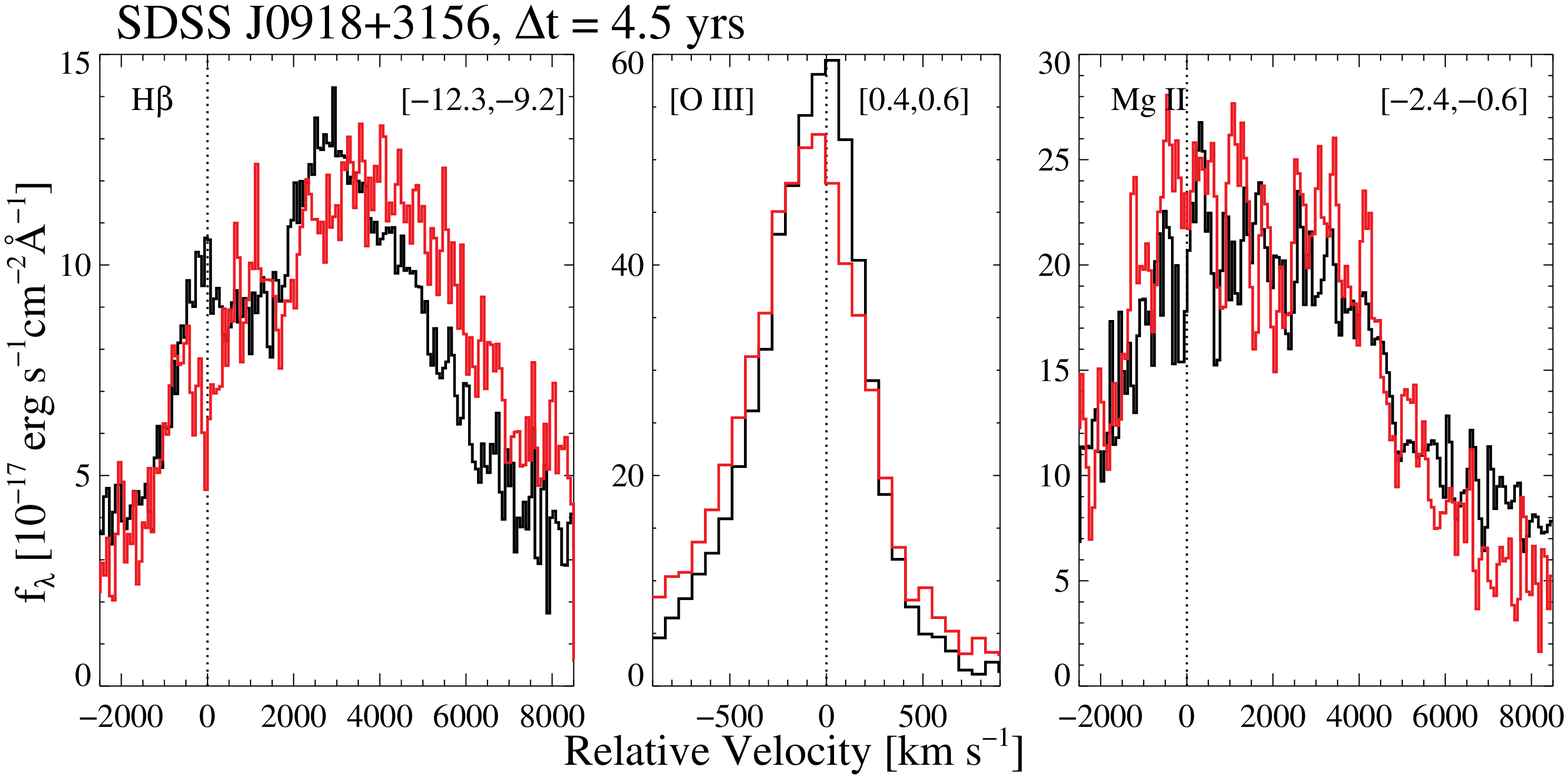}
    \includegraphics[width=89mm]{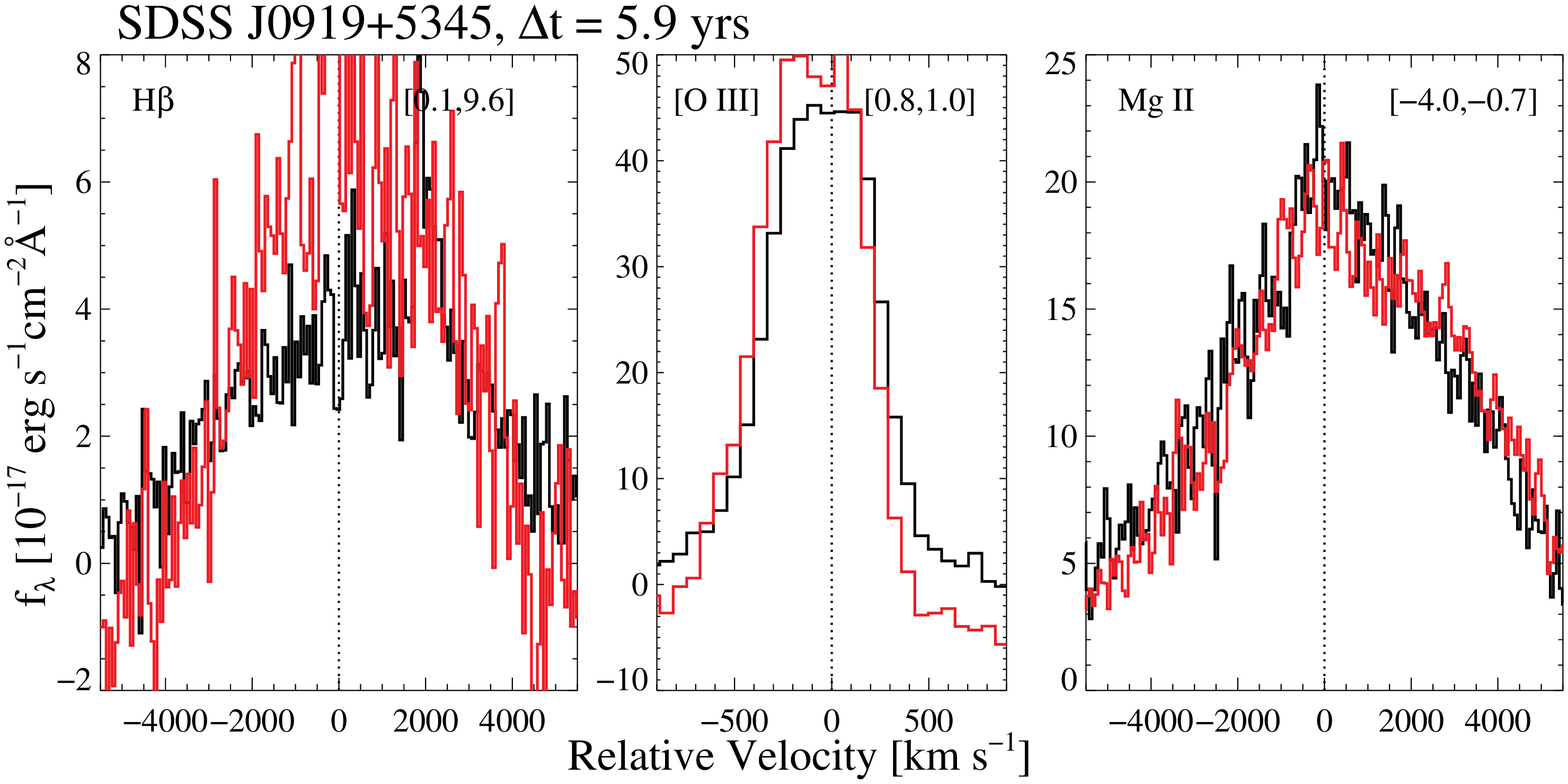}
    \caption{
    Emission line spectra at two epochs for the 24 quasars with detections
    of significant acceleration (together with those shown
    in Figures \ref{fig:vshift2} and \ref{fig:vshift3}).
    Black is for the original SDSS observations and red is for followups.
    The followup spectra are scaled to match the emission line fluxes
    of the SDSS observations.
    For each object, we show the broad \hbeta\ and \OIIIb\ lines in
    velocity space centered at the systemic velocity (noted by dotted lines).
    Also shown are the broad \halpha\ (or \MgII ) results when available.
    In each panel we list the velocity shift between the two epochs
    in brackets enclosing the 2.5$\sigma$ confidence range in units of
    pixels (with 1 pixel being 69 km s$^{-1}$). Here, negative values mean
    that the emission line in the followup spectrum needs to be blueshifted
    to match that in the original SDSS spectrum (i.e., the emission line in the
    followup spectrum is redshifted relative to that in the original spectrum).
    See Section \ref{subsec:detection} for details.}
    \label{fig:vshift}
\end{figure*}

\begin{figure*}
  \centering
    \includegraphics[width=89mm]{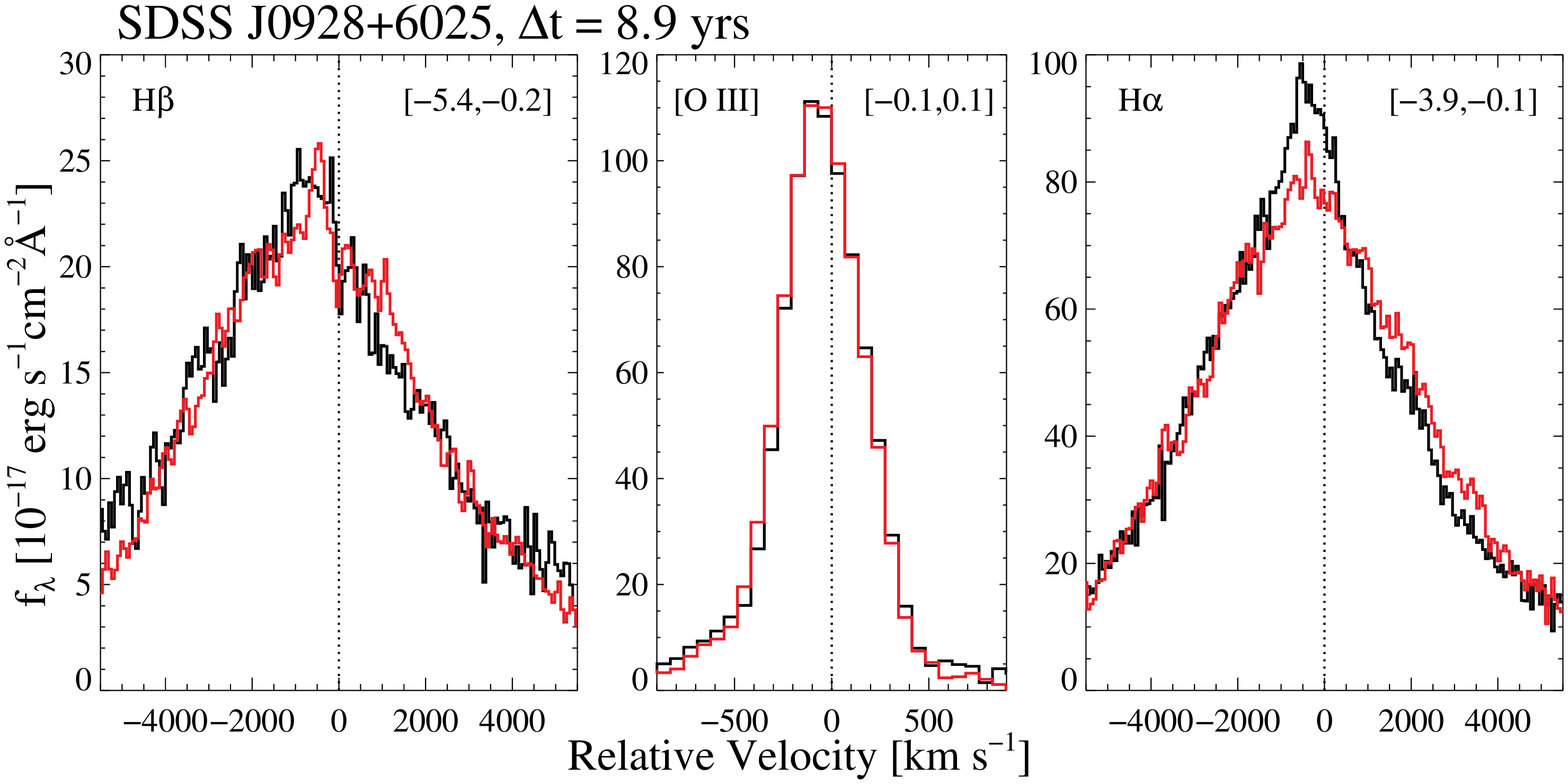}
    \includegraphics[width=89mm]{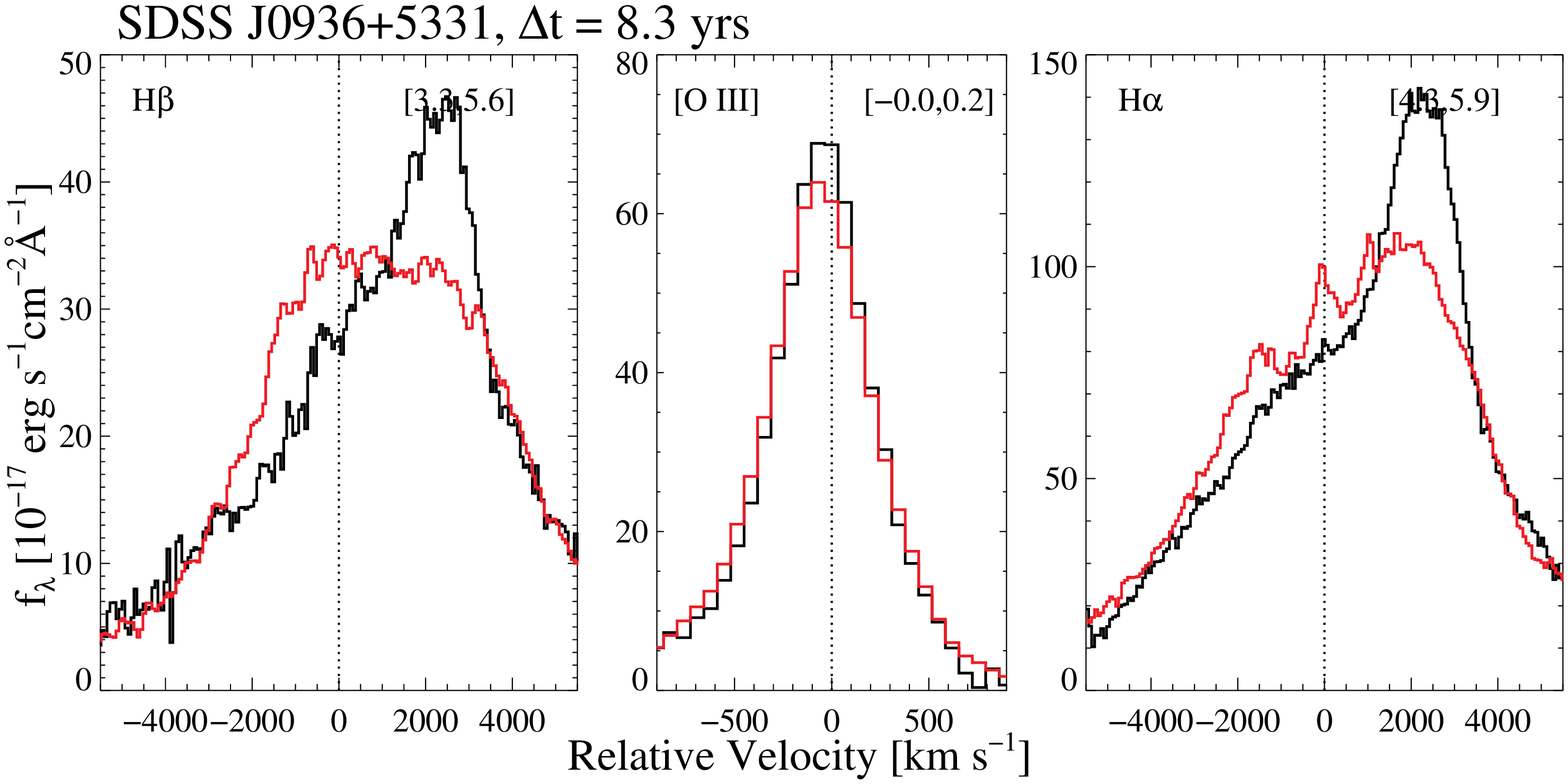}
    \includegraphics[width=89mm]{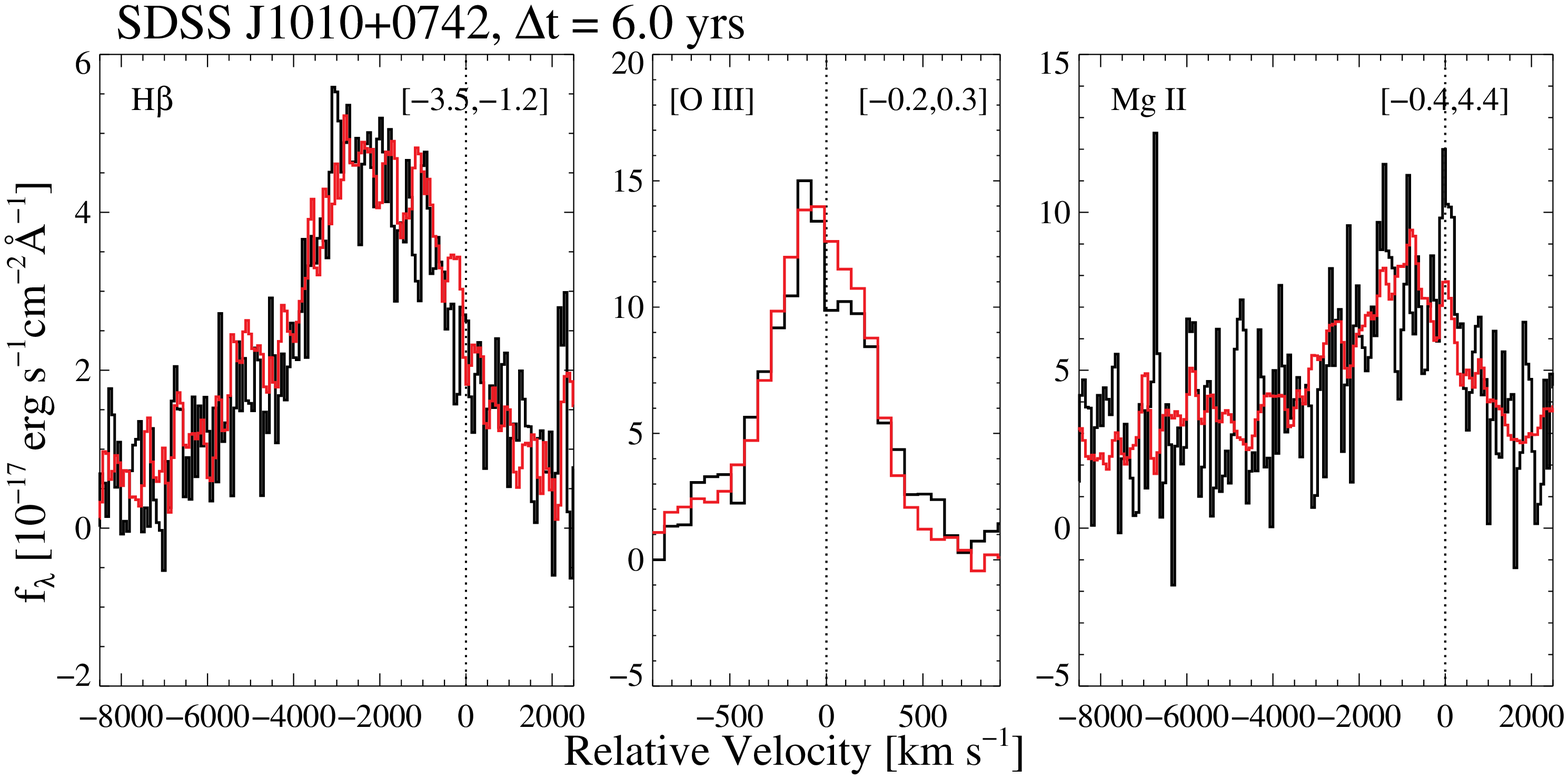}
    \includegraphics[width=89mm]{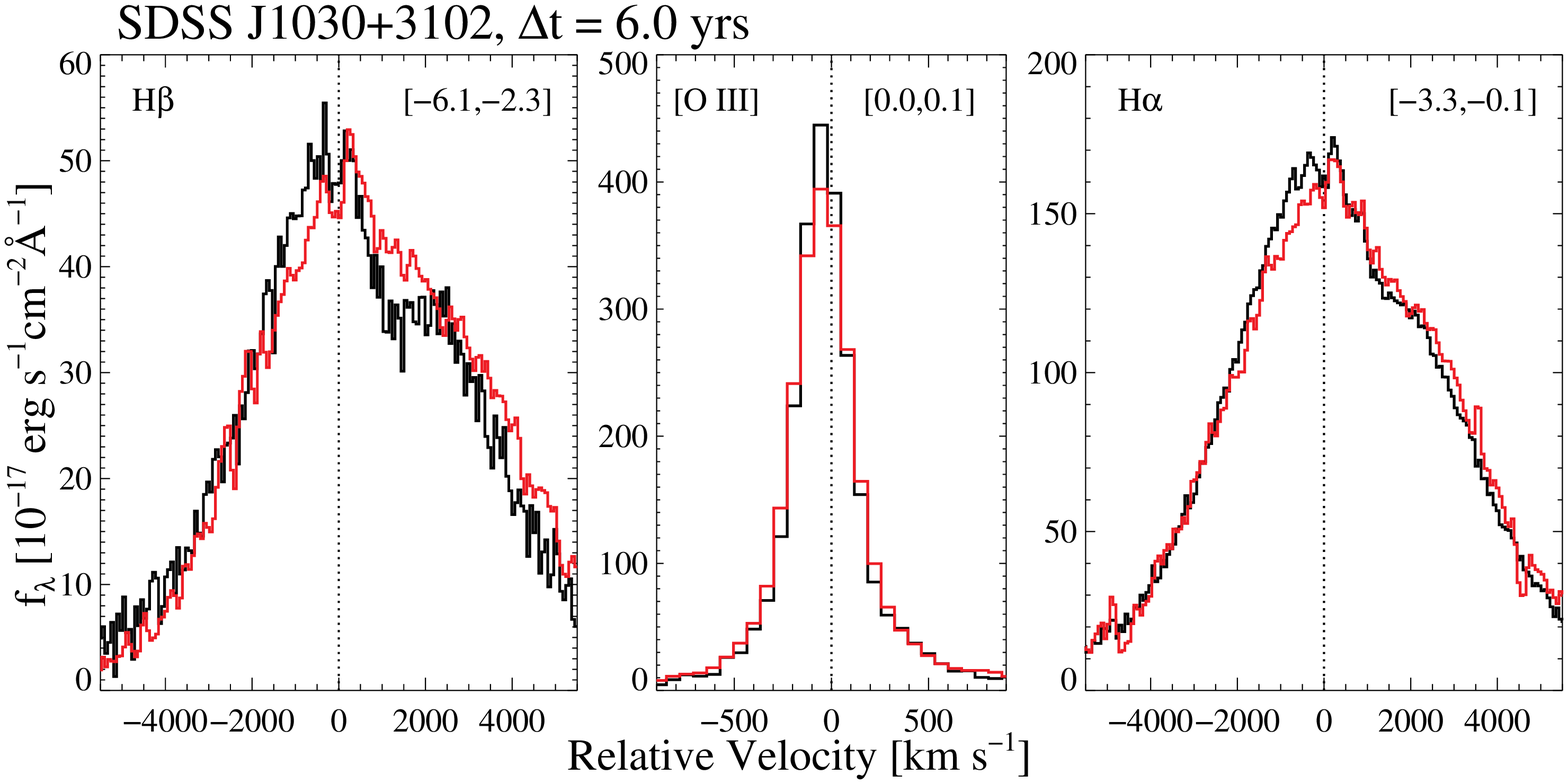}
    \includegraphics[width=89mm]{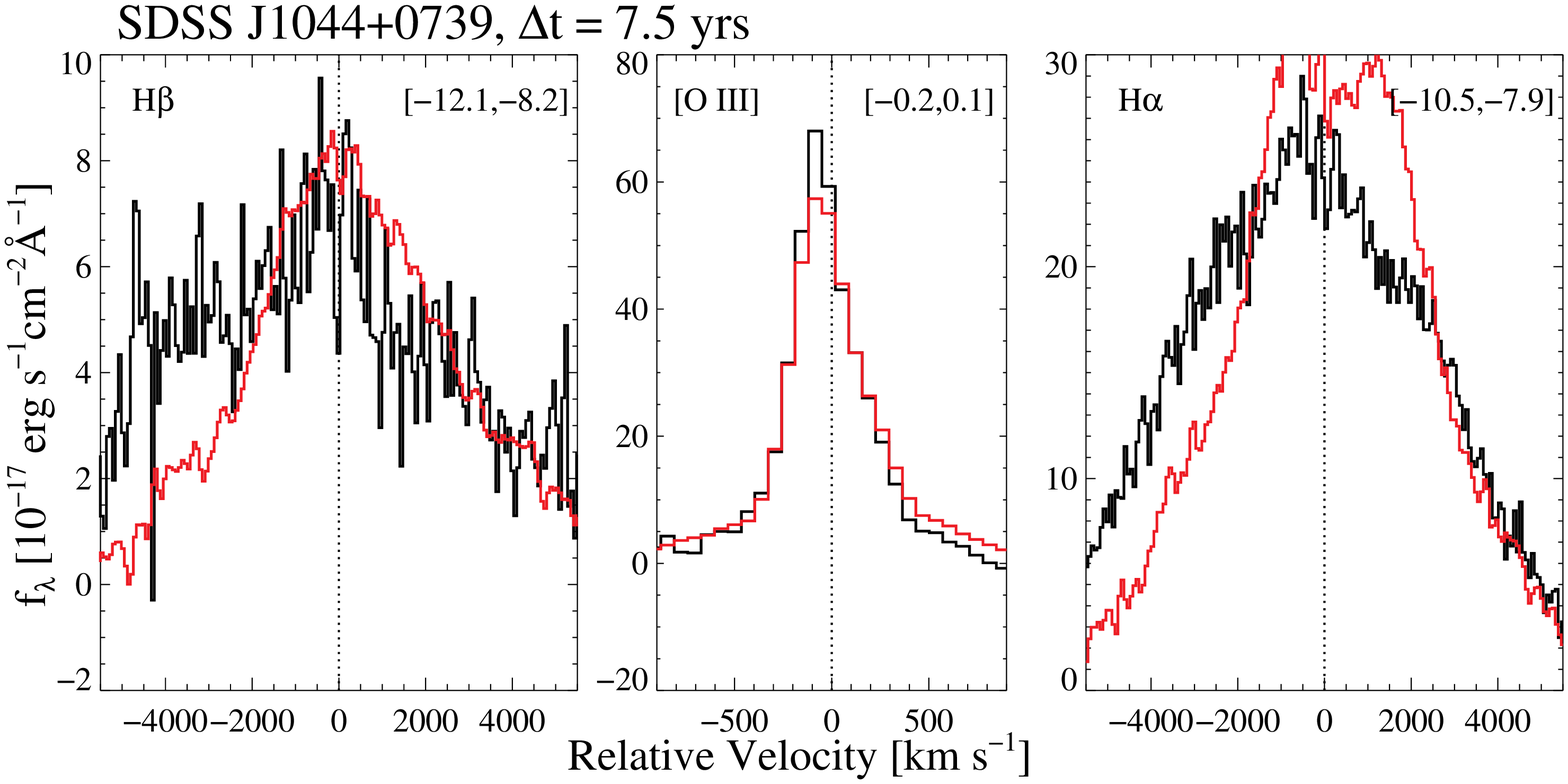}
    \includegraphics[width=89mm]{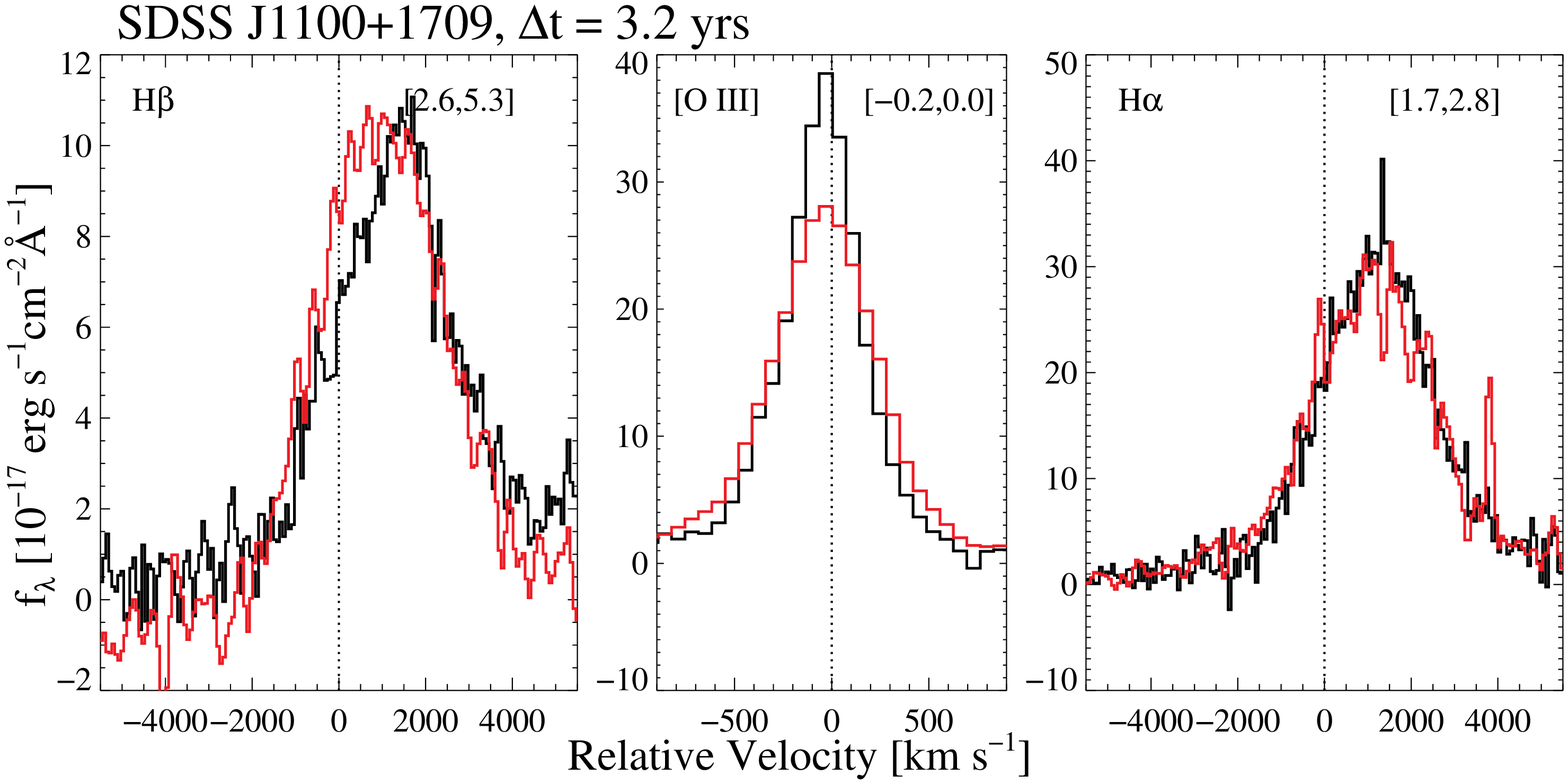}
    \includegraphics[width=89mm]{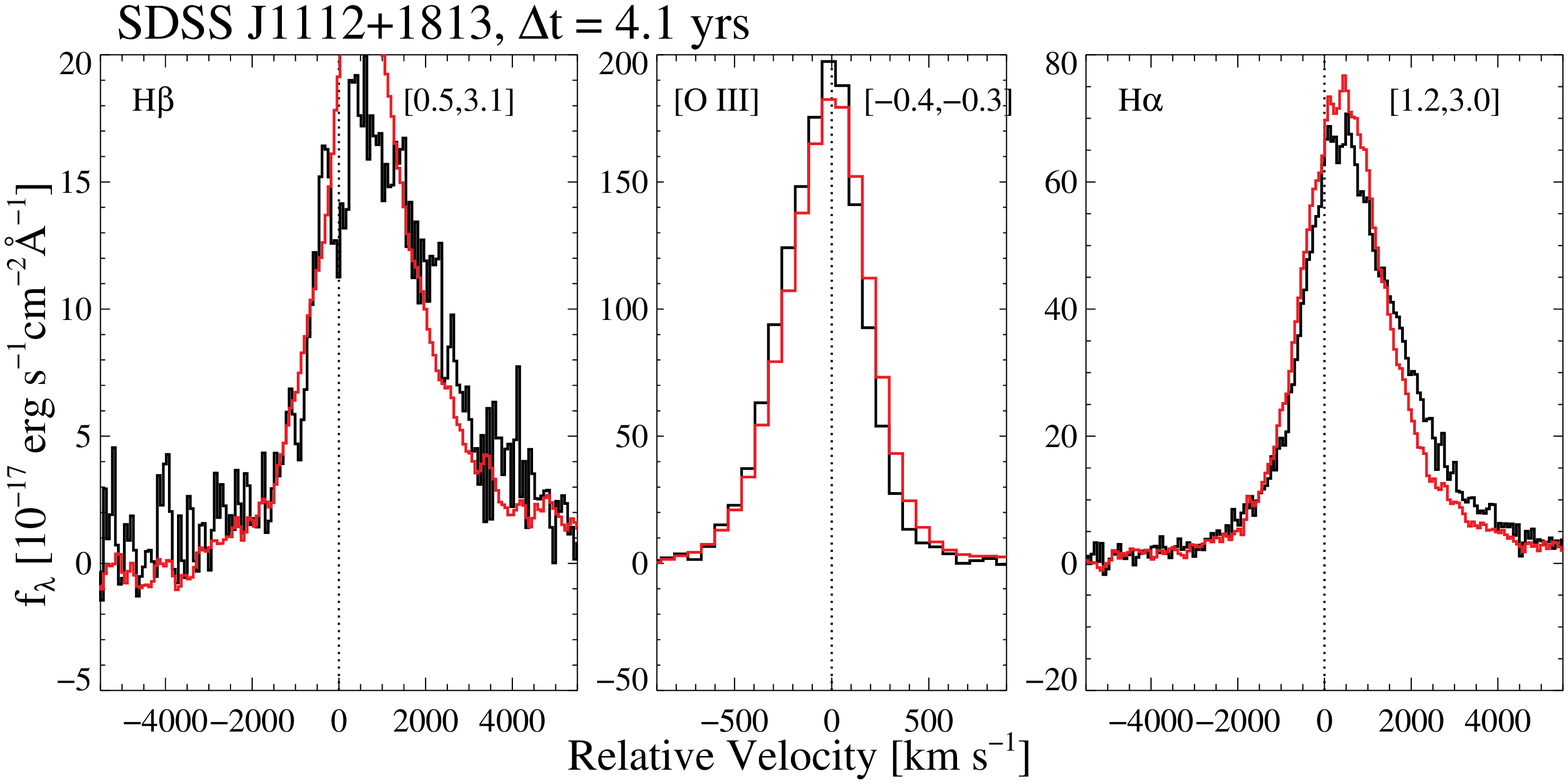}
    \includegraphics[width=89mm]{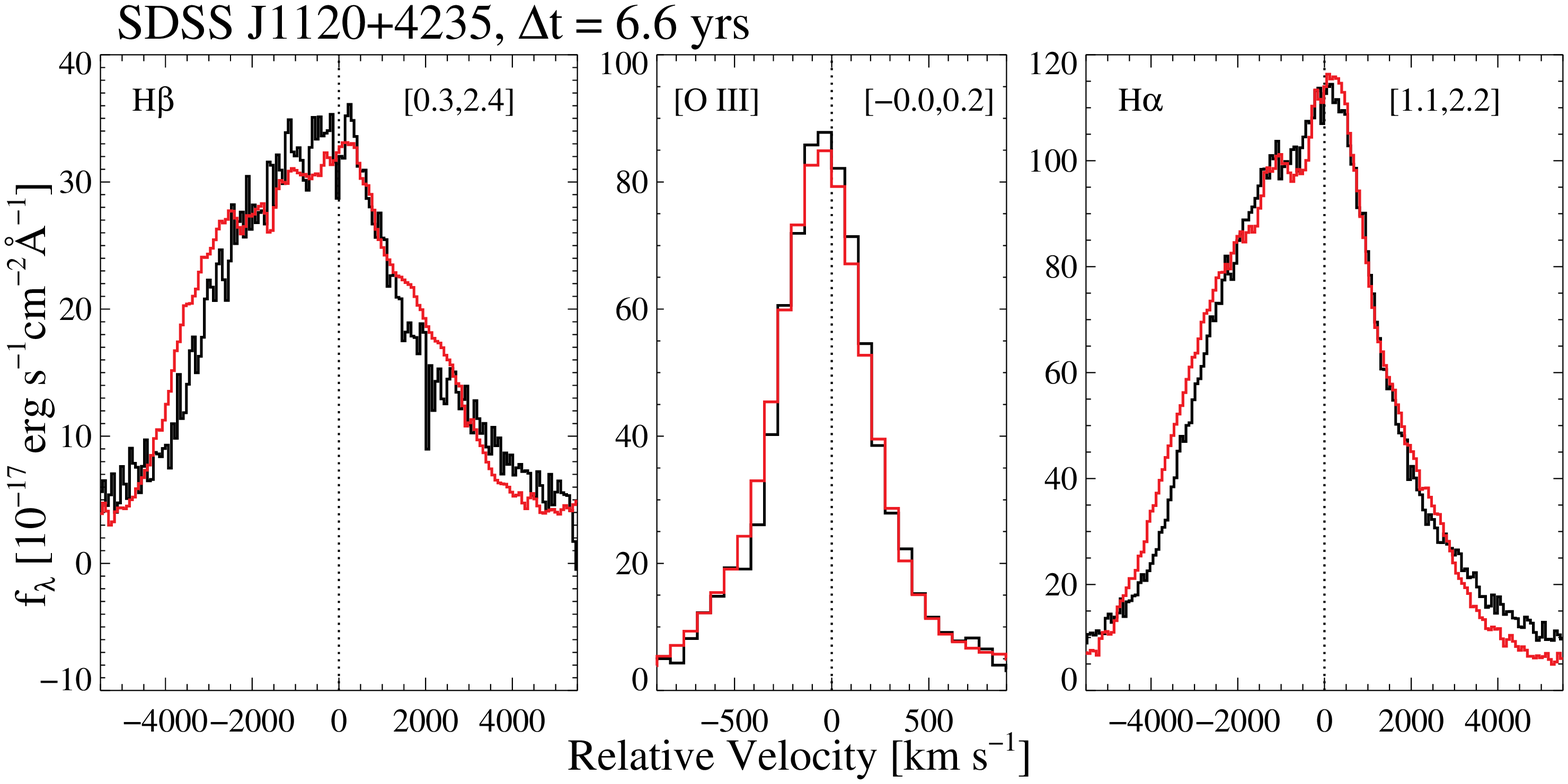}
    \includegraphics[width=89mm]{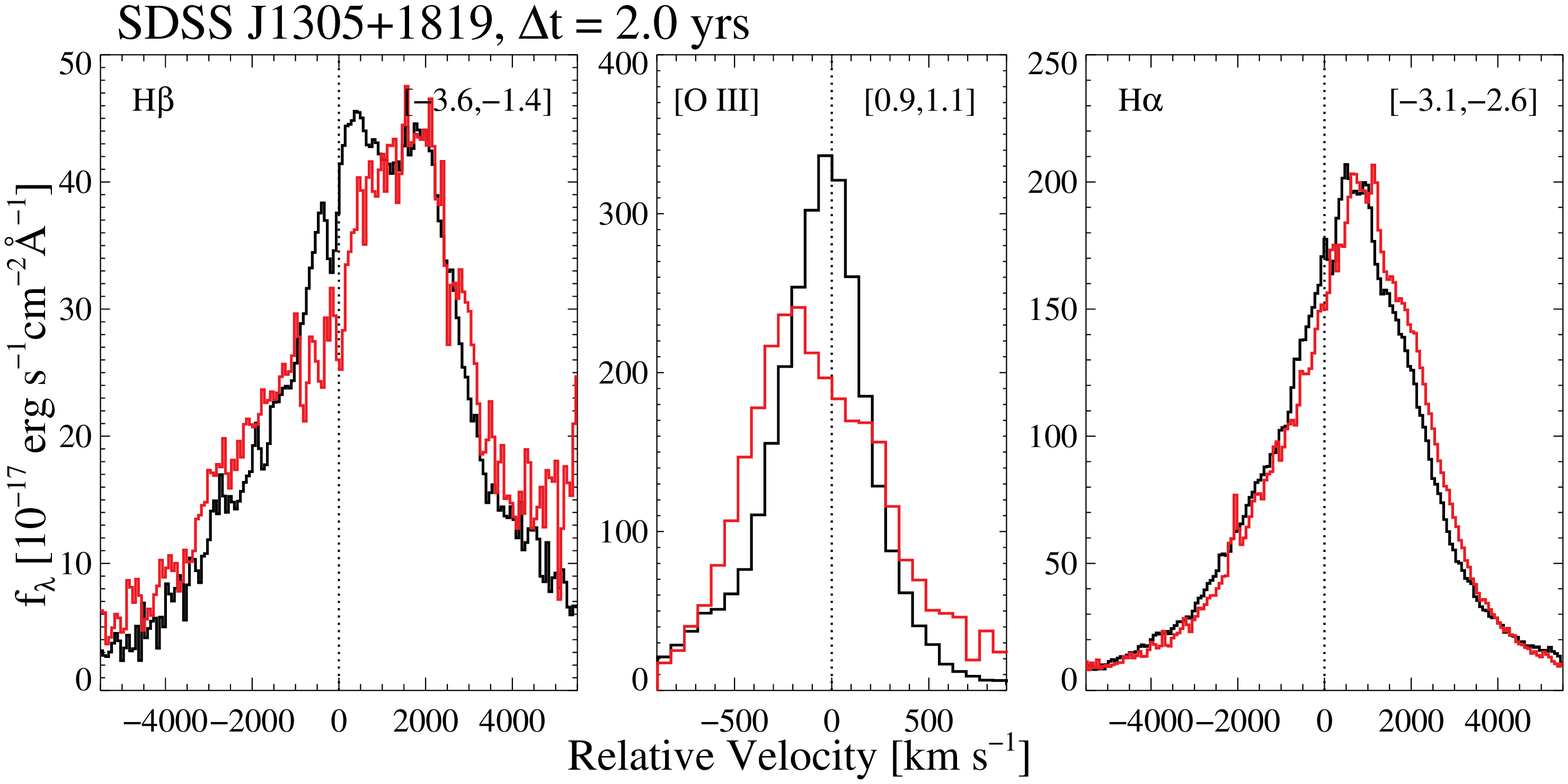}
    \includegraphics[width=89mm]{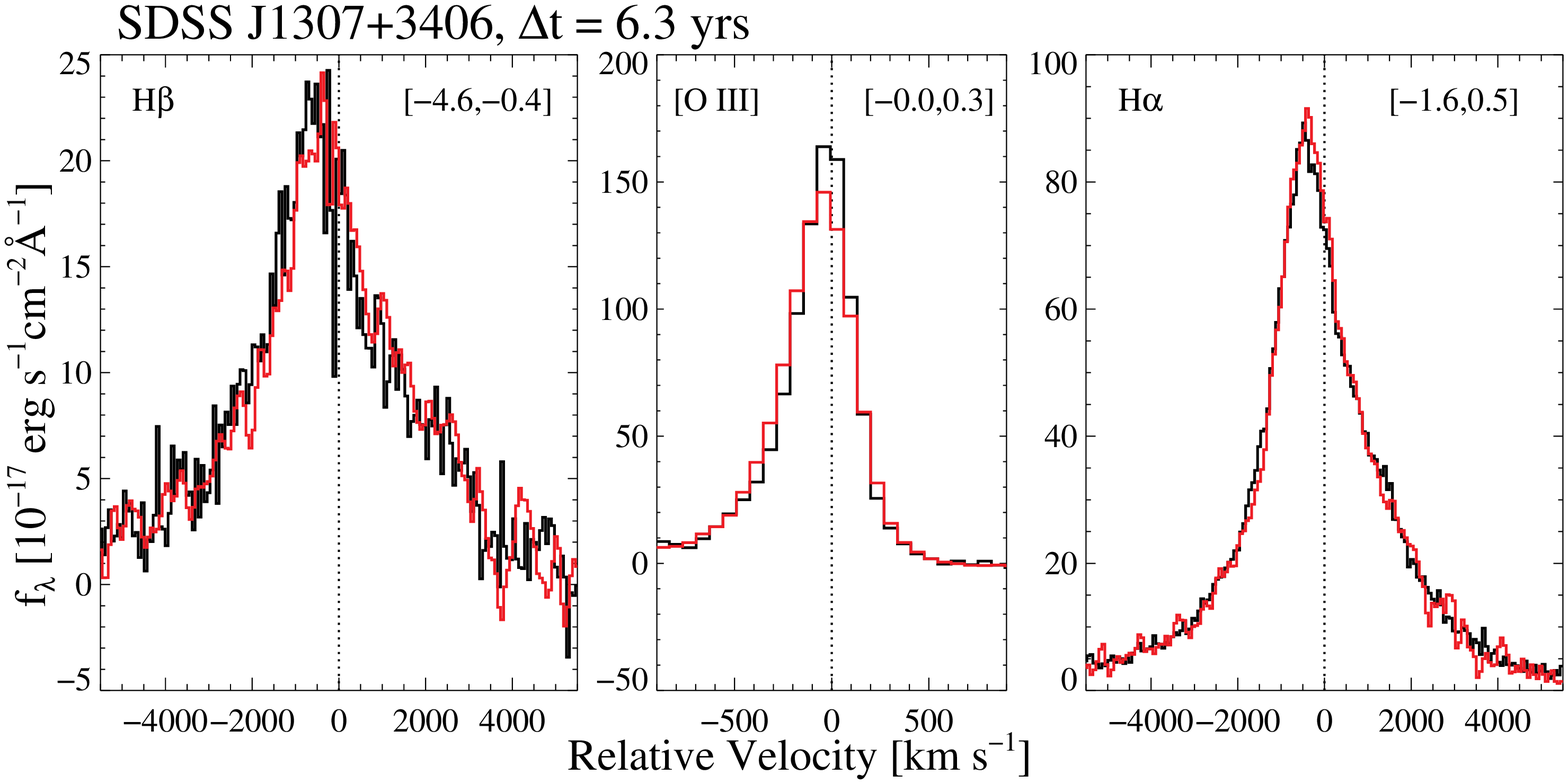}
    \caption{Same as Figure \ref{fig:vshift}, but for another set of detections.}
    \label{fig:vshift2}
\end{figure*}

\begin{figure*}
  \centering
    \includegraphics[width=89mm]{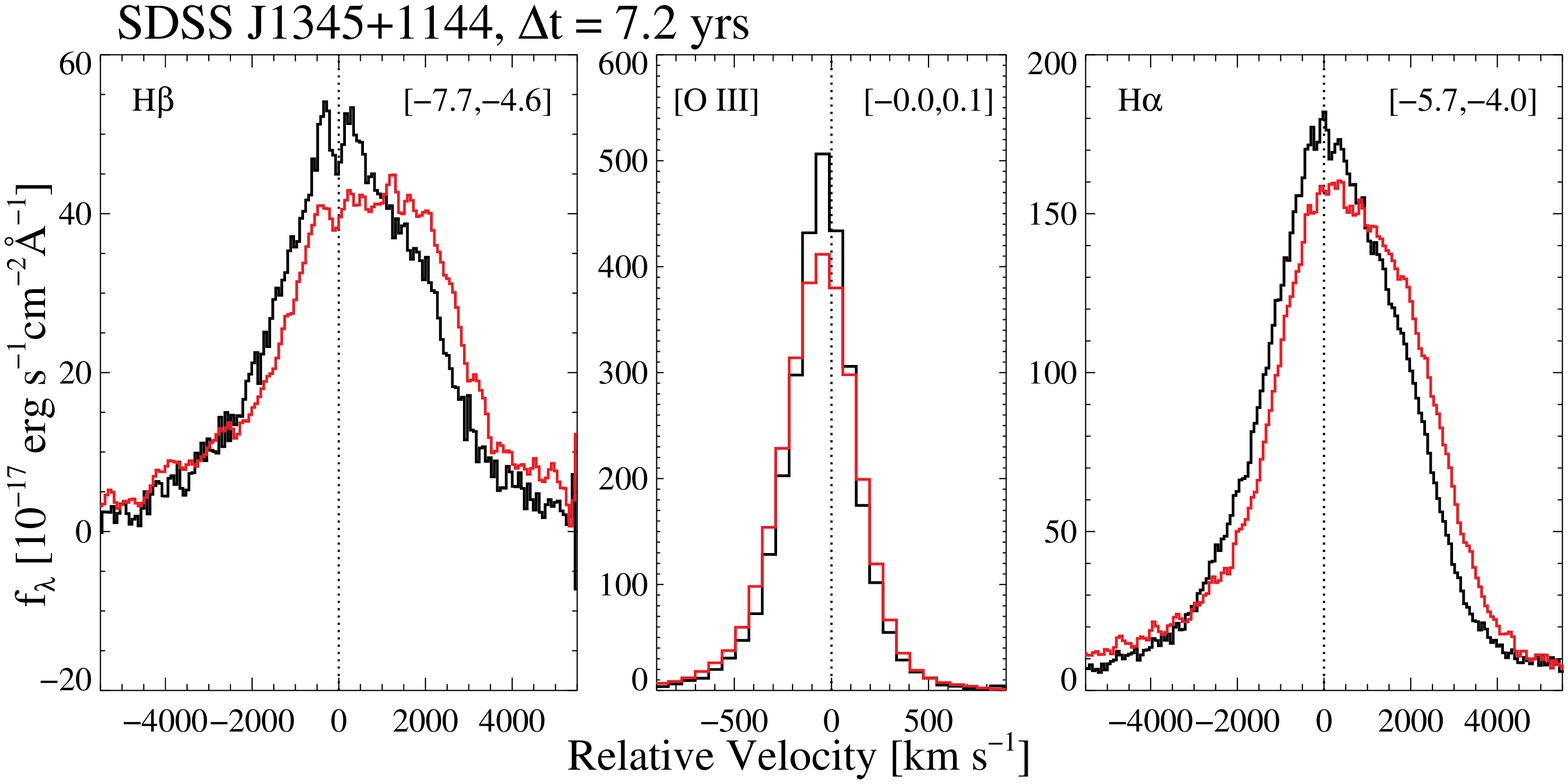}
    \includegraphics[width=89mm]{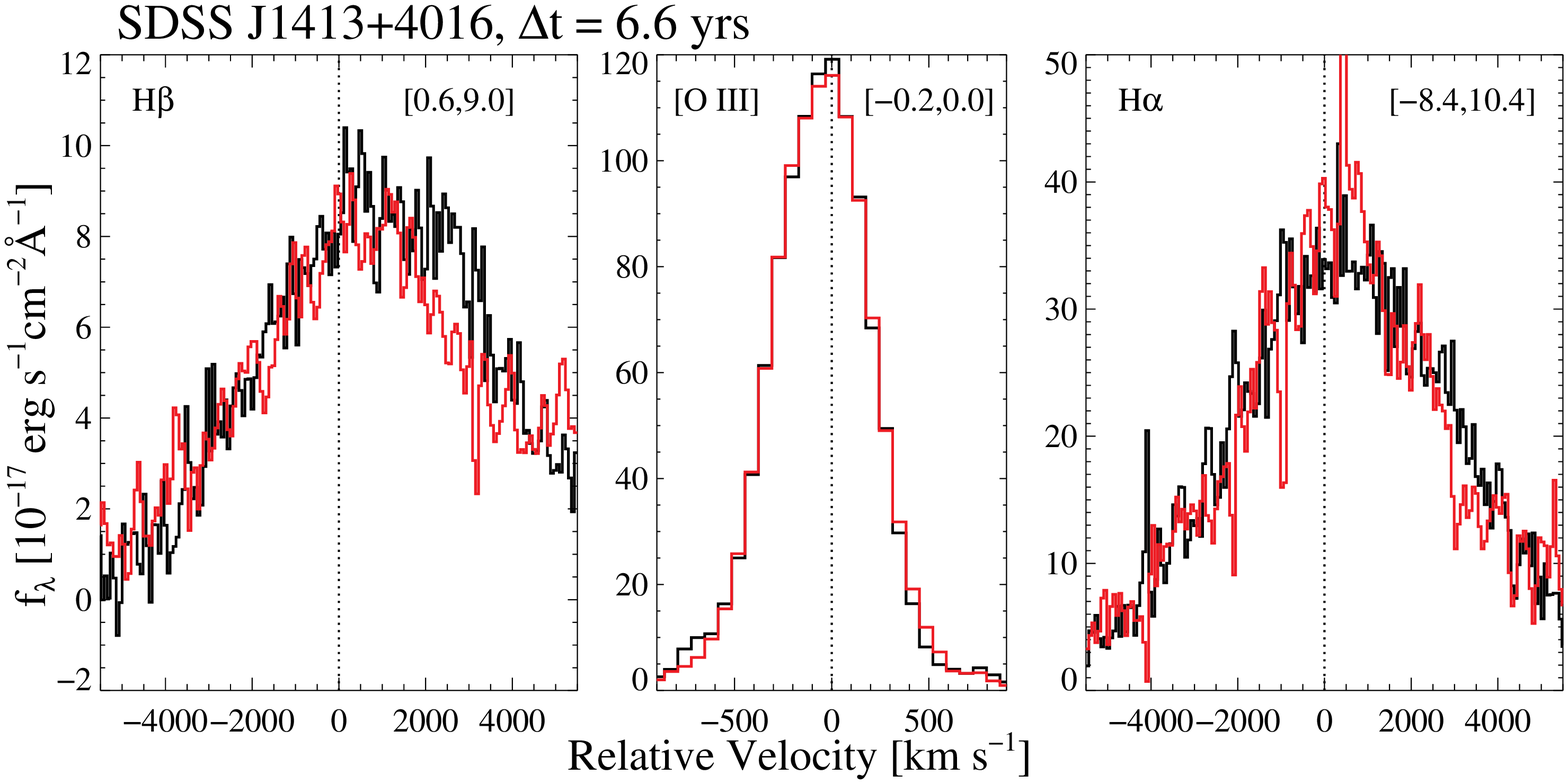}
    \includegraphics[width=89mm]{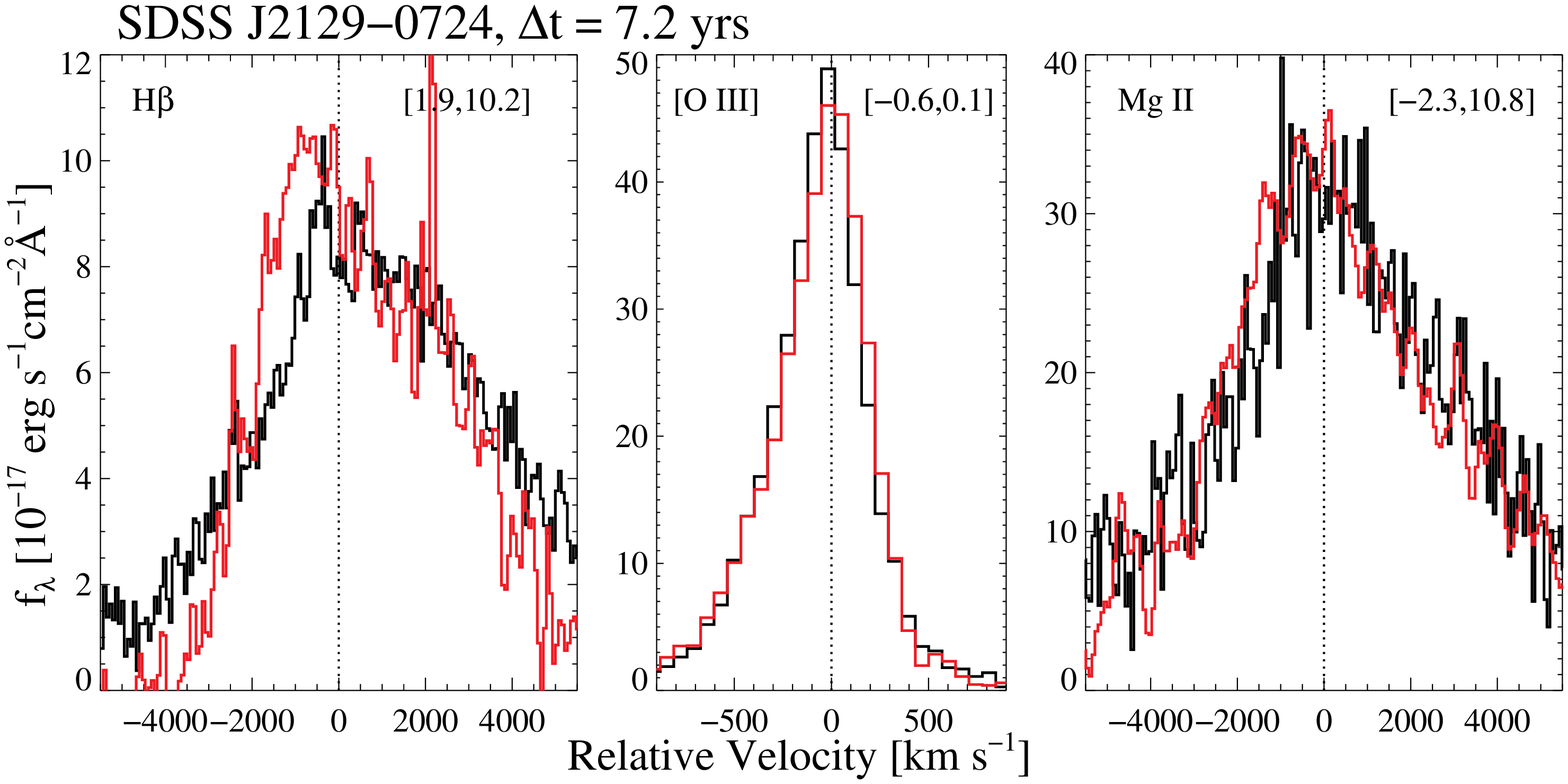}
    \includegraphics[width=89mm]{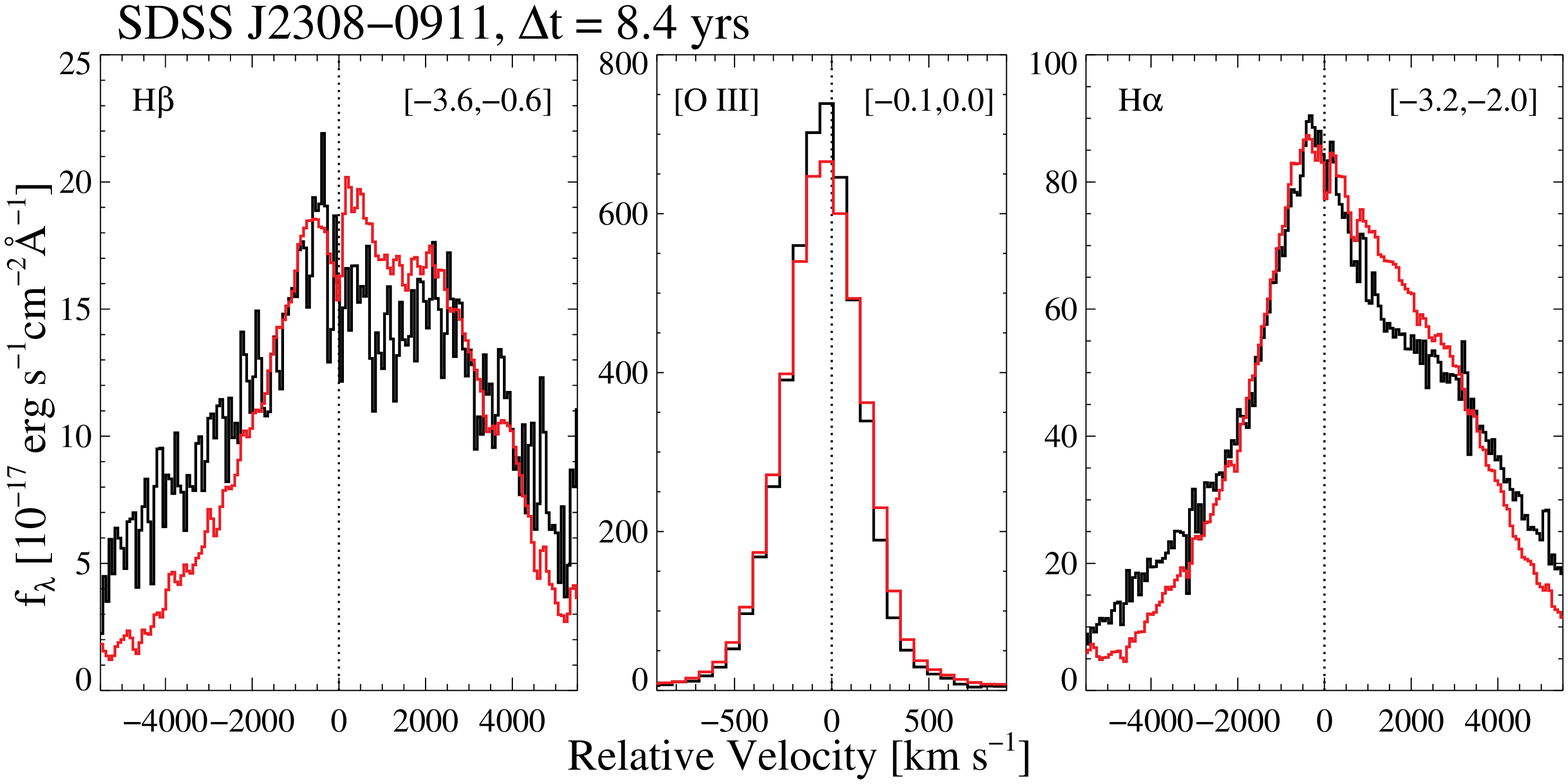}
    \caption{Same as Figure \ref{fig:vshift}, but for another set of detections.}
    \label{fig:vshift3}
\end{figure*}

In Table \ref{tab:obs} we list the velocity shifts measured
from ccf along with their statistical uncertainties (99\%
confidence). Out of the 50 followup targets, 24 show
significant broad-line velocity shifts between the two epochs.
We show the two-epoch broad \hbeta\ lines for all of these 24
quasars in Figures \ref{fig:vshift}--\ref{fig:vshift3}. Also
shown are the \OIII\ lines (for determining the systemic
velocity) and broad \halpha\ (or \MgII ) if available (to check
for consistency).

We now compare the detection fraction with that of the general
quasar population studied in Paper I. We account for
differences in time separation and measurement sensitivity. The
detection fraction in normal quasars (among the ``good''
sample) increases with time separation (Paper I); it is
$\lesssim$ a few percent at $\Delta t<1$ yr, and increases to
$\sim$20\% at $\Delta t>3$ yr.  Naively extrapolating the time
dependence (Figure 11 of Paper I) would imply a detection
fraction of $\sim$30\%\,$\pm$\,10\% at $\Delta t> 5$ yr. This
prediction is marginally lower than the detection fraction we
find in this paper ($\sim$50\%\,$\pm$\,10\%) at $\Delta t> 5$
yr. The difference becomes more prominent considering that the
velocity shift uncertainty of the followup sample is $\sim25$\%
larger than that of the good sample in Paper I (median
$\sigma_{V_{{\rm ccf}}}\sim 50$ compared to $\sim$40 km
s$^{-1}$). Among the detections, the fraction of BBH candidates
(9 out of 24 or $40$\%\,$\pm$\,10\%; see Section
\ref{subsubsec:candidate} for details) is consistent with that
of Paper I (7 out of 30 or 25\%\,$\pm$\,10\%) within
uncertainties, although Paper I had less stringent constraints
on objects that show significant line profile changes, because
contamination by double-peaked broad lines is less common in
the general quasar population.

In summary, these results suggest that the frequency of
broad-line shifts in offset quasars might be higher than that
in the general quasar population on timescales of more than a
few years. Of course, this is expected in the BBH scenario
since both velocity offsets and shifts in these offsets are due
to the orbital motion of the BBH. However, the statistics are
poor and the results are still consistent with having no
difference between the two populations.

\begin{figure}
  \centering
    \includegraphics[width=80mm]{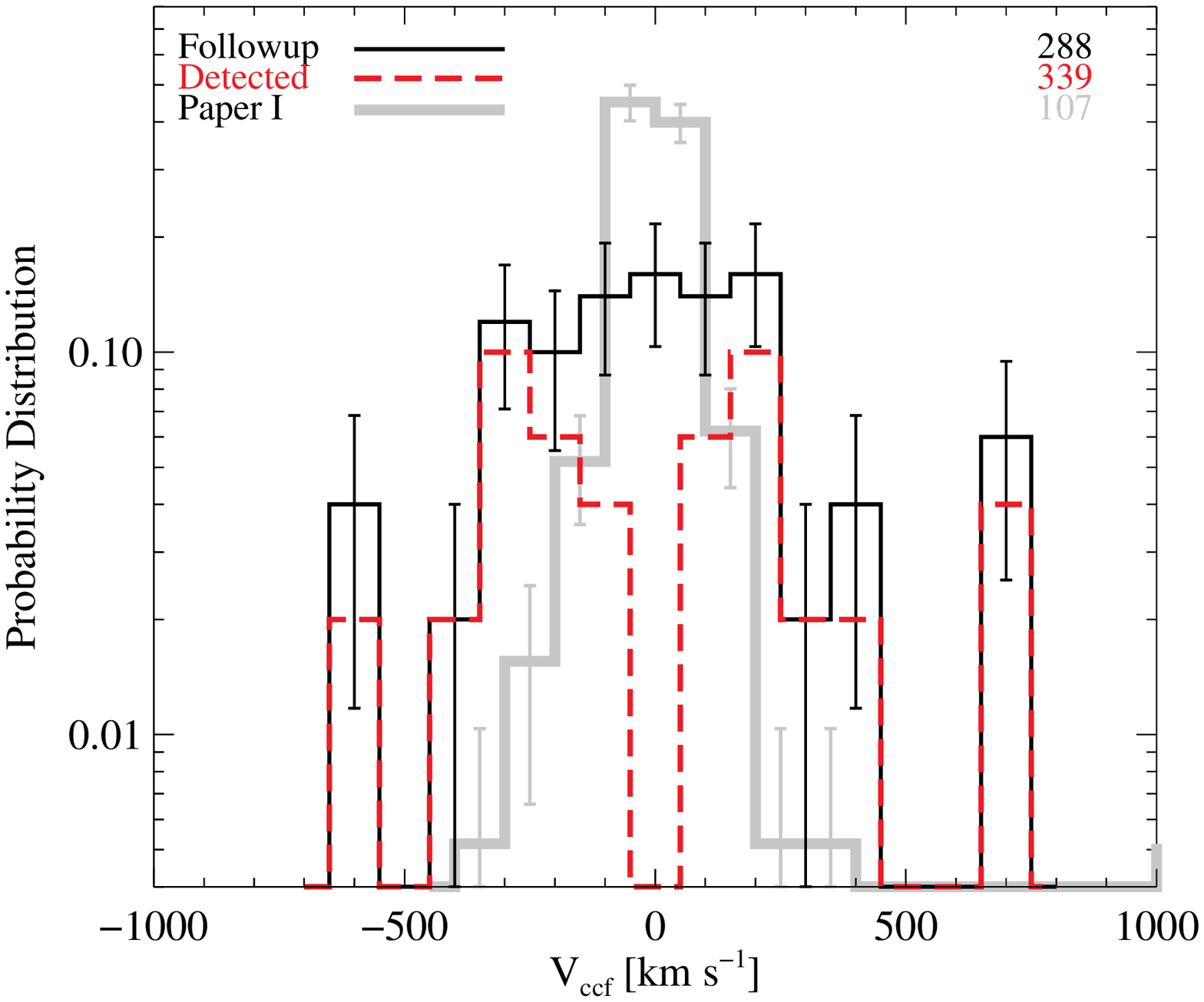}
    \caption{Distribution of the broad-line velocity shift
    between two epochs.
    Error bars indicate Poisson uncertainties.
    Numbers in the upper-right corner indicate standard deviations for different
    samples.
    }
    \label{fig:vshiftdist}
\end{figure}
\begin{figure}
  \centering
    \includegraphics[width=80mm]{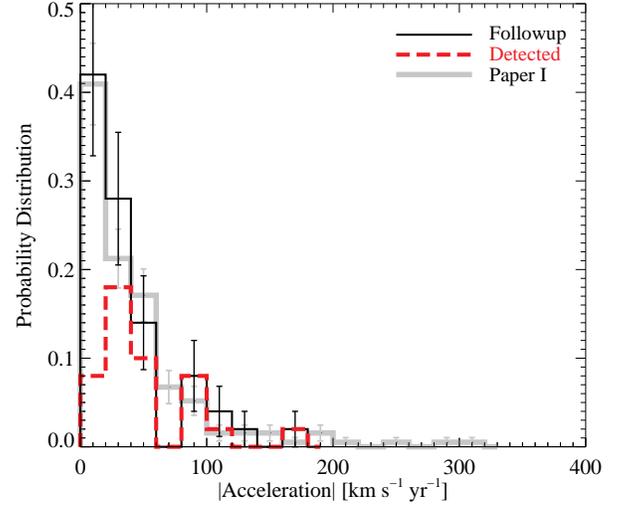}
    \caption{Probability distribution of the rest-frame broad-line
    acceleration between two epochs. Error bars indicate Poisson uncertainties.
    The observed distribution of the followup sample is consistent with
    that found in Paper I within uncertainties.
    Since the followup sample in this paper has smaller measurement errors
    in acceleration than the ``superior'' sample in Paper I (Figure \ref{fig:sigdist}),
    this suggests a broader (by $\sim30$\%) intrinsic acceleration distribution in the
    followup sample of offset quasars than in normal quasars (See Section \ref{subsec:vs}
    for details).
    }
    \label{fig:accdist}
\end{figure}
%%

%%%%%%%%%%%%%%%%%%%%%%%%%%%%%%%%%%%%%%%%
%
\subsection{Acceleration Distribution}\label{subsec:vs}

\begin{figure*}
  \centering
    \includegraphics[width=88mm]{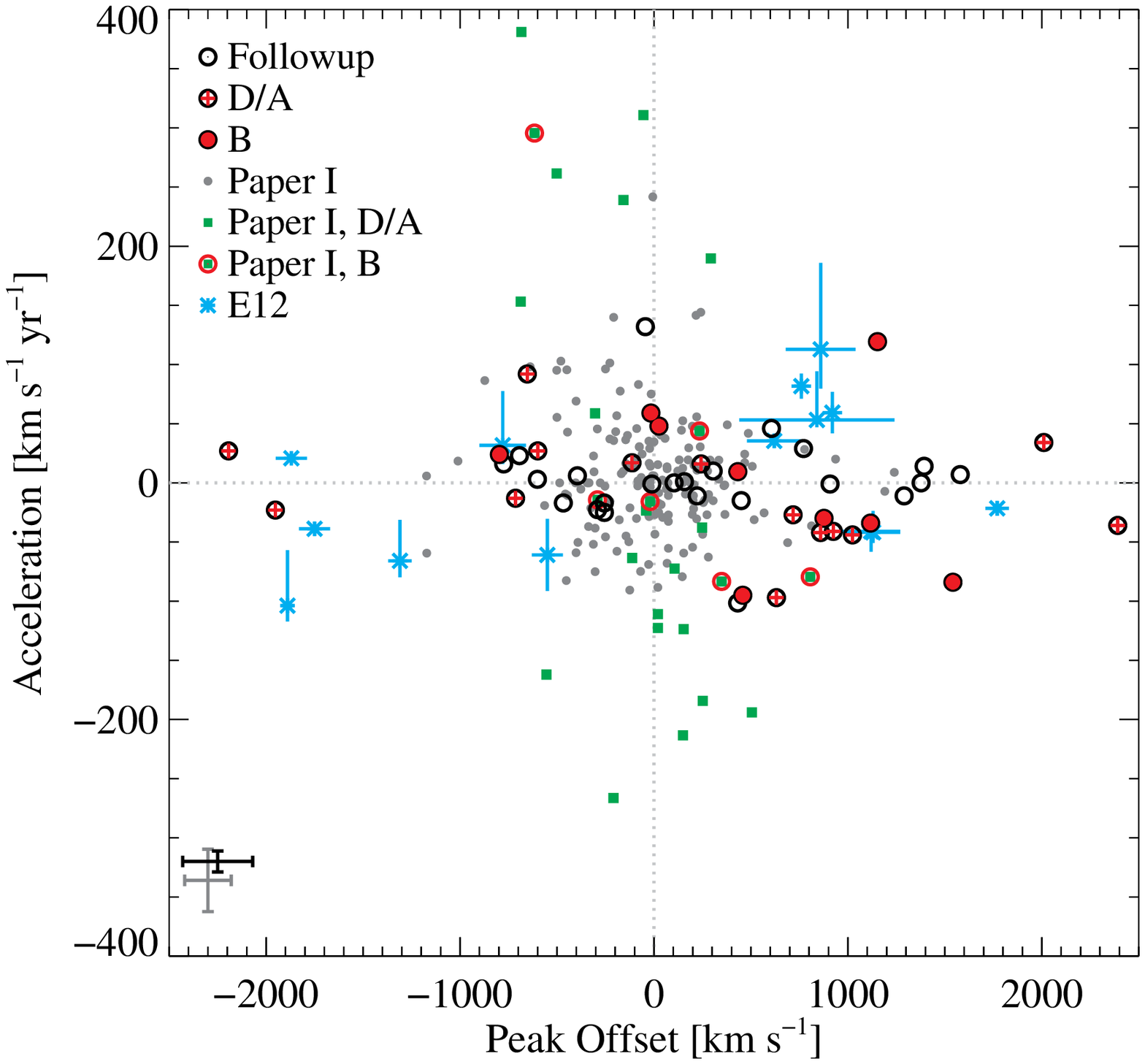}
    \includegraphics[width=88mm]{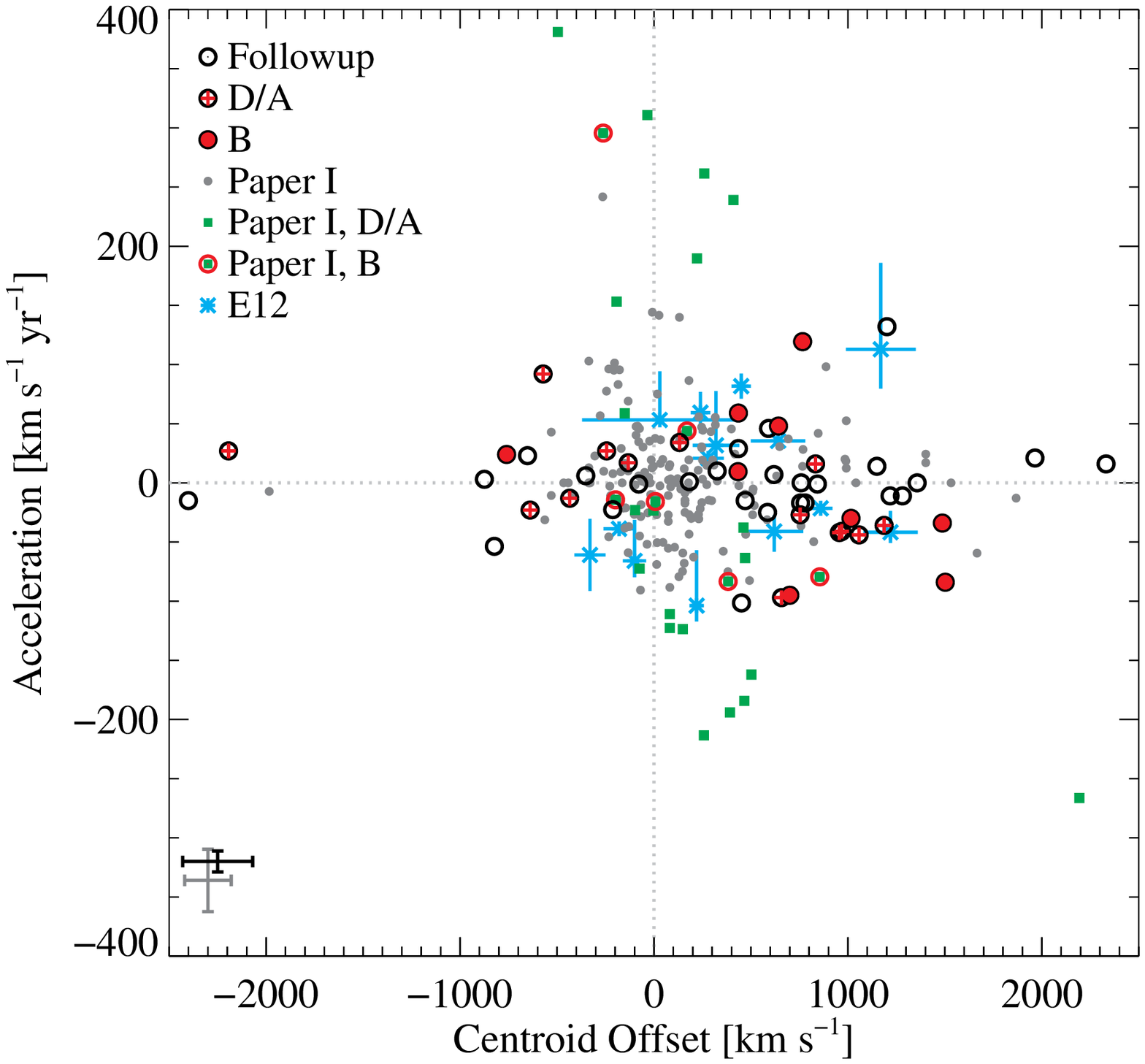}
    \caption{Left: broad \hbeta\ rest-frame acceleration
    vs. peak velocity offset (estimated from the original SDSS spectrum).
    Positive offsets mean redshifted broad emission lines from the systemic velocities;
    positive accelerations indicate more redshifted broad lines
    in the followup spectra than those in the original SDSS spectra.
    We show objects without significant broad-line velocity shifts in
    the followup sample as black open circles (26 objects).
    We also show the 24 cases with significant broad-line velocity shifts, including
    9 binary candidates (labeled with ``B'' and shown as red filled circles), and
    15 BLR variability or ambiguous cases (labeled with ``D/A''
    (which is short for ``disk emitters or ambiguous'') and
    shown as black open circles with red plus signs).
    Shown for comparison are the superior sample (gray dots) and
    the velocity-shift detected sample (green filled squares with red open circles for
    binary candidates, and green filled squares for BLR variability or ambiguous cases) of
    Paper I.
    The black error bars in the lower left corner indicate typical uncertainties of the
    followup sample, whereas the gray error bars denote those of the superior sample in Paper
    I.
    Also shown for comparison is the velocity-shift detected
    sample of \citet[][cyan asterisks with error bars]{eracleous11}.
   Right: same as left panel, but for centroid velocity offset.
   There is a weak correlation (Spearman $P_{{\rm null}}\sim10^{-3}$)
   in both the shift-detected samples of this work and of Paper I,
   except perhaps the few objects with the most extreme offset
   velocities (i.e., $>10^3$ km s$^{-1}$).
   See Section \ref{subsec:vacorr} for details.
    }
    \label{fig:vacccorr}
\end{figure*}

Figure \ref{fig:vshiftdist} shows the observed distribution of
broad-line shifts of the followup sample compared with that of
the superior sample of Paper I. The velocity shift distribution
of offset quasars is 1.7 times wider than that of normal
quasars.  The statement still holds for the intrinsic velocity
shift distribution after accounting for the difference in
measurement errors (Figure \ref{fig:sigdist}).  From $V_{{\rm
ccf}}$ and $\Delta t$, we then estimate $a_{{\rm ccf}}$, the
average acceleration between two epochs, as listed in Table
\ref{tab:obs}. Figure \ref{fig:accdist} shows the distribution
of the absolute value of the measured acceleration of the
followup sample compared with the superior sample of Paper I.

The measured acceleration distribution of the followup sample
is consistent with the superior sample of Paper I. After
accounting for the difference in measurement errors (Figure
\ref{fig:sigdist}), the inferred intrinsic acceleration
distribution of the followup sample is $\sim30$\% wider than
that of the superior sample of Paper I. However, the timescales
being probed are very different ($>$5 yr in this work compared
to an average of $\sim1$ yr in Paper I) so a fair comparison is
difficult.

%%%%%%%%%%%%%%%%%%%%%%%%%%%%%%%%%%%%%%%%
%
\subsection{Velocity Offset Versus Acceleration}\label{subsec:vacorr}

We now examine the relation between broad-line velocity offset
(relative to systemic) and acceleration. Figure
\ref{fig:vacccorr} shows the broad \hbeta\ peak velocity offset
of the original SDSS spectrum versus the radial acceleration
between the two epochs. We show our followup sample, the
superior sample of Paper I, and the sample with detected
velocity shifts from \citet{eracleous11}. We find a tentative
anti-correlation (Spearman $P_{{\rm null}} =10^{-3\pm0.5}$ and
$\rho=-0.13\pm0.01$, where the 1$\sigma$ errors were estimated
from bootstrap tests) between peak $V_{{\rm off}}$ and $a_{{\rm
ccf}}$ in our combined sample (i.e., the ``superior'' sample of
Paper I (193 pairs of observations) plus our followup sample
(50 pairs of observations)). The velocity offset and
accelerations tend to have opposite signs, in the sense that
blueshifted (redshifted) broad emission lines in the
single-epoch spectra on average tend to become less blueshifted
(redshifted) in the followup spectra after a few years. A
tentative anti-correlation is also detected in the subset of
detected objects (i.e., the 54 objects with significant
velocity shifts from Paper I and our followup sample) but is
weaker ($P_{{\rm null}}\sim10^{-2}$ and $\rho \sim -0.3$),
probably due to poorer statistics.

Figure \ref{fig:vacccorr} also shows the same results based on
the centroid offset. Again, there is a tentative
anti-correlation ($P_{{\rm null}}= 10^{-3\pm0.2}$) between
centroid $V_{{\rm off}}$ and $a_{{\rm ccf}}$ both in the
combined sample ($\rho= -0.14\pm0.01$) and in its subset with
detected shifts ($\rho= -0.36\pm0.02$). These weak
anti-correlations are only suggestive and would need to be
confirmed with larger samples. We expect no correlation between
the velocity and the instantaneous acceleration in the BBH
scenario, but see Section \ref{subsec:implication} for further
discussion on the implications of these results in the context
of identifying candidate BBHs.

%%%%%%%%%%%%%%%%%%%%%%%%%%%%%%%%%%%%%%%%
%
\subsection{Individual Detections}\label{subsec:detection}

As listed in Table \ref{tab:obs}, we divide the 24 detections
into three categories following Paper I according to different
possible origins of the observed broad-line velocity shifts:
(1) BBH candidates, (2) broad-line variability, and (3)
ambiguous cases. These categories are only meant to be our best
attempt at an empirical classification and are by no means
rigorous. Below we present these classifications and comment on
individual cases.

%%%%%%%%%%%%%%%%%%%%
%%%%%%%%%%%%%%%%%%%%
\subsubsection{BBH Candidates}\label{subsubsec:candidate}

We categorize nine objects as BBH candidates (Table
\ref{tab:obs}). The criteria are (1) broad-line velocity shifts
are detected between two epochs ($>99$\% confidence); (2) there
is an overall bulk velocity shift (i.e., the measured velocity
shift is not solely caused by a profile change); and (3) the
velocity shifts independently measured from broad \hbeta\ and
broad \halpha\ (or \MgII ) are consistent within uncertainties.
While the presence of broad-line profile changes without bulk
velocity shifts does not necessarily rule out the possibility
of BBHs \citep[e.g.,][]{shen10}, we assign these cases as BLR
variability (e.g., due to disk emitters) to minimize
contamination (Section \ref{subsubsec:other}). This criterion
rejects more candidates when applied to the present sample
compared to the sample in Paper I (20\% compared to 10\%),
because dramatic profile changes are much more commonly seen in
offset quasars than in normal quasars. Below we comment on each
case. In Section \ref{subsec:model} we discuss implications of
our results for the model parameters under the BBH hypothesis.

{\it J082930.60+272822.7}. We detect consistent velocity shifts
over 6.2 yr in broad \hbeta\ and broad \halpha\ with no
significant line profile changes. The radial acceleration
measured from broad \hbeta\ is [$-54$, $-13$] km s$^{-1}$
yr$^{-1}$ (2.5$\sigma$).  This object was also noted by
\citet{eracleous11} and by \citet{tsalmantza11}, but no
second-epoch spectrum was available.

{\it J084716.04+373218.1}. Consistent velocity shifts over 5.8
yr are detected in broad \hbeta\ and \MgII\ with no significant
line profile change. The radial acceleration measured from
broad \hbeta\ is [3, 18] km s$^{-1}$ yr$^{-1}$ (2.5$\sigma$).

{\it J085237.02+200411.0}. Broad \hbeta\ and \MgII\ show
consistent velocity shifts over 3.1 yr with no significant line
profile change. The radial acceleration measured from broad
\hbeta\ is [$-133$, $-43$] km s$^{-1}$ yr$^{-1}$ (2.5$\sigma$).

{\it J092837.98+602521.0}. Broad \hbeta\ and \halpha\ show
consistent velocity shifts over 8.9 yr with no significant line
profile change. The radial acceleration measured from broad
\hbeta\ is [1, 42] km s$^{-1}$ yr$^{-1}$ (2.5$\sigma$).

{\it J103059.09+310255.8}. Broad \hbeta\ and \halpha\ show
consistent velocity shifts over 6.0 yr with no significant line
profile change. The radial acceleration measured from broad
\hbeta\ is [26, 70] km s$^{-1}$ yr$^{-1}$ (2.5$\sigma$).

{\it J110051.02+170934.3}. Broad \hbeta\ and \halpha\ show
consistent velocity shifts over 3.2 yr with no significant line
profile change. The radial acceleration measured from broad
\hbeta\ is [$-113$, $-55$] km s$^{-1}$ yr$^{-1}$ (2.5$\sigma$).
This object was also noted by \citet{eracleous11}, but no
second-epoch spectrum was available.

{\it J111230.90+181311.4}. Consistent velocity shifts over 4.1
yr are detected in broad \hbeta\ and \halpha\ with no
significant line profile change. The radial acceleration
measured from broad \hbeta\ is [$-51$, $-8$] km s$^{-1}$
yr$^{-1}$ (2.5$\sigma$).

{\it J130534.49+181932.9}. Consistent velocity shifts over 2.0
yr are detected in broad \hbeta\ and \halpha\ with no
significant line profile change. The radial acceleration
measured from broad \hbeta\ is [80, 158] km s$^{-1}$ yr$^{-1}$
(2.5$\sigma$).  This object was also noted by
\citet{eracleous11}. The authors found no significant shift in
the \hbeta\ region based on a spectrum taken 1.9 yr after the
original SDSS observation; no second-epoch spectrum was
available for the \halpha\ region.

{\it J134548.50+114443.5}. This object has the largest velocity
shift detected among the BBH candidates in our sample.
Consistent velocity shifts over 7.2 yr are detected in broad
\hbeta\ and \halpha\ with no significant line profile change.
The radial acceleration measured from broad \hbeta\ is [44, 73]
km s$^{-1}$ yr$^{-1}$ (2.5$\sigma$).

%%%%%%%%%%%%%%%%%%%%
%%%%%%%%%%%%%%%%%%%%
\subsubsection{Alternative Scenarios}\label{subsubsec:other}

As listed in Table \ref{tab:obs}, there are five objects for
which we suggest that broad-line variability likely causes the
observed velocity shifts.  Although we tried to eliminate
double-peaked broad emission lines in the original sample
selection based on single-epoch spectra, the rejection was
incomplete. These objects usually have more complicated
broad-line profiles than the BBH candidates.  The line profile
changes between two epochs are more dramatic, similar to some
of the well-monitored double-peaked broad lines which are
generally explained as disk emitters known in the literature.
This is in contrast to the findings in Paper I that significant
profile changes are uncommon in the general quasar population
(over timescales less than a few years). The cases with
dramatic line profile changes in our followup sample exhibit
similar velocity shifts in broad \hbeta\ and broad \halpha\ (or
\MgII ). However, the velocity shifts measured from ccf are
mostly caused by profile changes instead of bulk velocity
shifts. Below we comment on each case.

{\it J001224.02$-$102226.5}. We detect a velocity shift over
8.3 yr in broad \hbeta . The broad \hbeta\ line, and in
particular the broad base of the line profile, got narrower,
similar to the profile change reported in broad \halpha\ by
\citet{Decarli2013}. The radial acceleration measured from
broad \hbeta\ is [$-29$, $-16$] km s$^{-1}$ yr$^{-1}$
(2.5$\sigma$), consistent with the acceleration
($-32^{+17}_{-26}$ km s$^{-1}$ yr$^{-1}$) reported in broad
\halpha\ by \citet{Decarli2013} based on a spectrum taken 8.1
yr after the original SDSS observation. This object was also
noted by \citet{shen10}, \citet{zamfir10}, and
\citet{eracleous11}. \citet{eracleous11} reported a velocity
shift of $+125^{+35}_{-30}$ km s$^{-1}$ in the \hbeta\ region
based on a spectrum taken 6.8 yr after the original SDSS
observation; no second-epoch spectrum was available for the
\halpha\ region.

{\it J093653.84$+$533126.8}. This object has among the most
dramatic profile changes in our followup sample. The redshifted
peaks of broad \hbeta\ and \halpha\ appear to have blueshifted
by $>2000$ km s$^{-1}$ over 8.3 yr, while no obvious shifts
were seen in the bases of the broad lines. \citet{eracleous11}
reported a profile change in the \hbeta\ region similar to our
findings based on a spectrum taken 6.7 yr after the original
SDSS observation; no second-epoch spectrum was available for
the \halpha\ region. \citet{Decarli2013} also reported a
similar profile change in the \hbeta\ and \halpha\ regions
based on a spectrum taken 7.5 yr after the original SDSS
observation.

{\it J104448.81$+$073928.6}. This object is also among those
with the most dramatic profile changes. The line widths of
broad \hbeta\ and broad \halpha\ appear to have narrowed by
$\sim$30\%--40\% over 7.5 yr.

{\it J112007.43$+$423551.4}. Broad \hbeta\ and \halpha\ show
consistent velocity shifts over 6.6 yr, but the shifts are
mostly due to changes in the line profile rather than bulk
velocity shifts. The blueshifted shoulder of broad \hbeta\ also
becomes more prominent in the second-epoch spectrum, which is
reminiscent of double-peaked broad emission lines generally
interpreted as disk emitters.

{\it J230845.60$-$091124.0}. Broad \hbeta\ and \halpha\ show
consistent velocity shifts over 8.4 yr, but the shifts are
mostly due to changes in line profiles rather than bulk
velocity shifts.

We classify the rest of the objects with broad \hbeta\ velocity
shifts as ``ambiguous'' (Table \ref{tab:obs}). These objects
either do not have followup spectra for a second broad line, or
that they do but the measured velocity shifts are inconsistent
with those from broad \hbeta . More observations with wider
spectral coverage and enhanced S/N are needed to better
determine the nature of these objects.

%%%%%%%%%%%%%%%%%%%%%%%%%%%%%%%%%%%%%%%%%%%%%%%%%%%%%%%%%%%%
\section{Discussion}\label{sec:discuss}

%%%%%%%%%%%%%%%%%%%%%%%%%%%%%%%%%%
%
\subsection{Comparison with Previous Work}\label{subsec:compare}

Recently \citet{eracleous11} carried out the first systematic
spectroscopic followup study of quasars with offset broad
\hbeta\ lines. The authors identified 88 quasars from the SDSS
DR7 with offsets of $\gtrsim 1000$ km s$^{-1}$. Compared to our
offset sample, the \citet{eracleous11} sample has larger
broad-line velocity offsets, focusing on the most extreme
objects. Among the 68 objects for which \citet{eracleous11}
conducted second-epoch spectroscopy, 14 were found to show
significant velocity shifts after rejecting objects with
significant line profile changes. This fraction ($\sim20$\% or
14 out of 68) is similar to that of our ``BBH candidates''
category (9 out of 50). \citet{eracleous11} found no relation
between the acceleration and the initial velocity offset,
although their sample size may be too small to see any weak
(anti-)correlation. Another possibility for the null
correlation is that the \citet{eracleous11} sample focuses on
more extreme velocity offsets (i.e., $>10^3$ km s$^{-1}$) than
our samples, where the anti-correlation between $V_{{\rm off}}$
and $a_{{\rm ccf}}$ seems to break down.

The existing data cannot prove that the ``BBH candidates''
reported in this work are indeed BBHs. To further test the BBH
hypothesis would require future observations, most importantly
more followup observations to confirm or reject whether the RV
curve follows the expected binary motion. The time baselines
required to detect a significant fraction of an orbit may be
too long to be practical considering that the expected orbital
periods range from a few decades to a few centuries (Table
\ref{tab:model}), but linear growth of the velocity shift with
time without significant variation in the profile shape would
already be a strong signature. Despite the tentative nature of
the evidence so far, in the following sections we focus the
discussion on implications of our results under the BBH
hypothesis, as it is the main motivation of the current work.

\begin{figure}
  \centering
    \includegraphics[width=88mm]{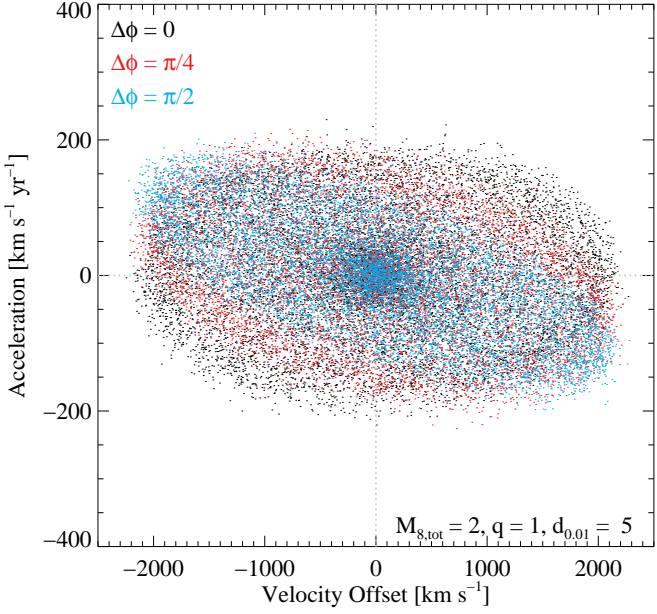}
    \caption{Illustration of the effect of orbital phase in modulating the correlation
    between the observed first-epoch velocity offset and the average acceleration between two
    epochs
    (Figure \ref{fig:vacccorr}). The acceleration is calculated as
    $(V(I, \phi + \Delta \phi) - V(I, \phi))/\Delta t$,
    where $\Delta t = P \Delta \phi/2\pi$ is the time separation between two epochs
    with $P$ being the orbital period.
    Black, red, and cyan color denote the cases when $\Delta t$ is $0$, $1/4$,
    and $1/8$ of $P$, respectively.
    Shown for each $\Delta \phi$ are measurements of $10^4$ mock orbits assuming
    random orbital inclination and phase
    for an equal-mass BBH with $2\times10^8~M_{\odot}$ total BH mass,
    and a circular orbit with a binary separation of 0.05 pc (orbital period $P=75$ yr).
    Gaussian noises with standard deviations $80$ km s$^{-1}$ and $20$ km s$^{-1}$ yr$^{-1}$
    have been added to the velocity offsets and accelerations, respectively.
    A correlation similar to the observed ones (Figure \ref{fig:vacccorr}) arises
    when $\Delta t$ is a non-negligible fraction of $P$ (further shown in Figure
    \ref{fig:testcorrdt}). See Section \ref{subsec:implication} for more discussion,
    and the Appendix for more details on model assumptions and parameter definitions.
    }
    \label{fig:testcorr}
\end{figure}
\begin{figure}
  \centering
    \includegraphics[width=88mm]{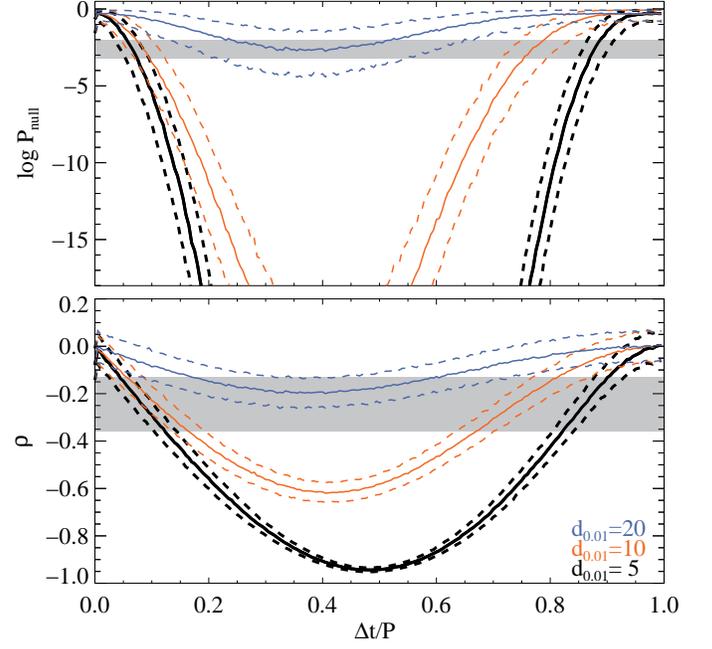}
    \caption{Spearman correlation coefficients $\rho$ and $P_{{\rm null}}$ of $a$ and $V$
    as a function of $\Delta t/P$. At a given $\Delta t/P$, $\rho$ and $P_{{\rm null}}$ are
    calculated using 243 data points (matching the size of our combined sample)
    drawn from $10^3$ realizations of $10^4$ mock orbits generated assuming zero eccentricity
    and random orbital inclination and phase.
    Gaussian noises with standard deviations $80$ km s$^{-1}$ and $20$ km s$^{-1}$ yr$^{-1}$
    have been added to the velocity offsets and accelerations, respectively.
    Solid curves denote the median values whereas the dashed
    curves indicate 1$\sigma$ confidence intervals.
    Gray shaded areas represent ranges of the observed correlation
    coefficients (measured from the combined sample using peak or centroid velocities).
    As in Figure \ref{fig:testcorr}, the baseline model (shown in black) assumes
    an equal-mass BBH with $2\times10^8~M_{\odot}$ total BH mass,
    and a binary separation of 0.05 pc (orbital period $P=75$ yr).
    Shown in color are models with larger separations.
    See Section \ref{subsec:implication} for more discussion.
    }
    \label{fig:testcorrdt}
\end{figure}
%%

%%%%%%%%%%%%%%%%%%%%%%%%%%%%%%%%%%
%
\subsection{Implications for Identifying BBH Candidates}\label{subsec:implication}

There is a tentative anti-correlation between broad-line
velocity offset and acceleration (Section \ref{subsec:vacorr}).
In the BBH scenario, both RV and acceleration are determined by
the binary separation, masses of the two BHs, orbital phase
$\phi$, and orbital inclination $I$ to the line of sight (LOS).
The similar distributions in virial BH mass, quasar luminosity
and broad \hbeta\ FWHM (Figures \ref{fig:zmag} and
\ref{fig:fwhm}) suggest that the intrinsic properties of
offset-line quasars are similar to those of normal quasars. It
is therefore unlikely that the selection of larger offset
velocities has induced any major bias in terms of BH mass and
accretion rate. On the other hand, selecting for larger
velocity offsets in single-epoch spectra could preferentially
yield BBHs with smaller binary separations and/or larger values
of $|\sin I|$ and $|\sin \phi|$.

Any model in which the broad-line offsets are due to BBHs makes
several clean predictions (assuming random orbital phases and a
selection function that depends only on the absolute value of
the velocity offset): (1) the distribution of velocity offsets
should be symmetric around zero (after correcting for
gravitational redshift); (2) the distribution of accelerations
should be symmetric around zero; (3) there should be no
correlation between velocity $V(I, \phi)$ and the instantaneous
acceleration $a(I, \phi)$, i.e., for fixed $V$ the probability
of seeing $a$ and $-a$ should be the same (since $\pi -\phi$ is
just as likely as $\phi$); and (4) there should be a weak
anti-correlation between $|V(I, \phi)|$ and $|a(I, \phi)|$
since they are out of phase by $90^{\circ}$ at least for
circular orbits.

For our offset sample, (1) is violated, but as we discussed
(Section \ref{subsubsec:property}), this is likely a selection
bias; (2) is satisfied; both (3) and (4) are violated. A
possible explanation is that we measure the average
acceleration over the interval $\Delta t$, rather than the
instantaneous acceleration, and these are not the same if
$\Delta t$ is a significant fraction of the orbital period $P$.
Figure \ref{fig:testcorr} illustrates such an effect. We show
the measurements of $10^4$ mock orbits assuming zero
eccentricity and random orbital inclination and phase for an
equal-mass BBH with $2\times10^8~M_{\odot}$ total BH mass, and
a binary separation of 0.05 pc (orbital period $P=75$ yr).
Gaussian noises with standard deviations 80 km s$^{-1}$ and 20
km s$^{-1}$ yr$^{-1}$ (typical for the measurements in our
combined sample) are added to the velocity offsets and
accelerations, respectively. Different colors denote the cases
when $\Delta t$ is 0, 1/4, and 1/8 of $P$. An anti-correlation
arises when $\Delta t$ is a non-negligible fraction of $P$,
similar to that observed in Figure \ref{fig:vacccorr}.

Figure \ref{fig:testcorrdt} shows more quantitative results. It
plots the Spearman correlation coefficients $P_{{\rm null}}$
and $\rho$ of the acceleration $a$ and velocity $V$ as a
function of $\Delta t/P$. At a given $\Delta t/P$, the
coefficients are calculated using 243 data points (matching the
size of our combined sample) drawn from $10^3$ realizations of
$10^4$ mock orbits assuming zero eccentricity and random
orbital inclination and phase. As in Figure \ref{fig:testcorr},
the baseline model assumes an equal-mass BBH with
$2\times10^8~M_{\odot}$ total BH mass, and a binary separation
of 0.05 pc (orbital period $P=75$ yr). Again, Gaussian noises
with standard deviations 80 km s$^{-1}$ and 20 km s$^{-1}$
yr$^{-1}$ are added to the velocity offsets and accelerations,
respectively. Also shown are models with the same masses but at
larger binary separations, which result in larger correlation
coefficients (i.e., weaker anti-correlations) because of the
smaller S/N due to the smaller velocity and acceleration
amplitudes. The coefficients observed in our samples are shown
as gray shaded areas. The comparison between the data and the
baseline model suggests $\Delta t/P\gtrsim 0.05$--0.1 since
there must be some dilution of the correlation due to objects
in the data that are not BBHs. Thus the observed
velocity--acceleration anti-correlation may suggest that the
typical orbital period is $\sim10$--20 times the typical time
baseline (which is $\sim1$ yr for the superior sample in Paper
I, or $\sim3$--10 yr for the followup sample in this work), or
$\sim30$--200 yr for our followup sample, which is consistent
with our estimates for individual candidates (see Section
\ref{subsec:model}). These numbers are only crude estimates.
Nevertheless, the result does suggest that the time baselines
of our observations are a significant fraction of the orbital
period, regardless of whether the velocity shifts are due to
BBHs or some other periodic feature such as an orbiting clump
or cloud in the BLR.

An alternative explanation why quasars with offset broad
emission lines are more likely to exhibit accelerations might
be BLR variability is likely to produce both velocity offsets
and accelerations, and the statistical properties of the
stochastic process producing the velocity offsets and
accelerations leads to an observed anti-correlation between
velocity offset and acceleration measured over a nonzero time
interval.
%

%% estimates for BBH model parameters

%----------------------------------------------------------------------------------------------
%\begin{landscape}
\begin{deluxetable*}{lcccccccccc}
%\tabletypesize{\footnotesize} %
\tabletypesize{\scriptsize} %
%\tablewidth{0pc} %
\tablewidth{\textwidth} %
\tablecaption{Estimated Model Parameters for the Binary Black Hole Candidates
\label{tab:model} %
} %
\tablehead{
 & & &
 \multicolumn{4}{c}{$q=2$} & \multicolumn{4}{c}{$q=0.5$} \\
\cline{4-7}  \cline{8-11} \\
\colhead{} & \colhead{${\rm log} M_{1}$} &
\colhead{$R_{{\rm BLR}}$} & \colhead{$f_r^{-1} R_{{\rm BLR}}$} & \colhead{$d$} &
\colhead{$P$} & \colhead{$t_{{\rm gr}}$} & \colhead{$f_r^{-1} R_{{\rm BLR}}$} & \colhead{$d$} &
\colhead{$P$} & \colhead{$t_{{\rm gr}}$} \\
\colhead{~~~~~~SDSS Designation~~~~~~} & \colhead{($M_{\odot}$)} &
\colhead{(pc)} & \colhead{(pc)} & \colhead{(pc)} &
\colhead{(yr)} & \colhead{(Gyr)} & \colhead{(pc)} &
\colhead{(pc)} & \colhead{(yr)} & \colhead{(Gyr)} \\
\colhead{(1)} & \colhead{(2)} & \colhead{(3)} &
\colhead{(4)} & \colhead{(5)} & \colhead{(6)} &
\colhead{(7)} & \colhead{(8)} & \colhead{(9)} &
\colhead{(10)} & \colhead{(11)} %
} %
\startdata
082930.60$+$272822.7\dotfill & 8.6 & 0.046 &0.15~~ &0.24~~& 320 &~~~~5& 0.11~~& 0.071 &~~72 &~~0.3  \\
084716.04$+$373218.1\dotfill & 8.1 & 0.054 &0.17~~ &0.29~~& 720 & 260 & 0.12~~& 0.13~~& 290 & 70~~~ \\
085237.02$+$200411.0\dotfill & 8.4 & 0.058 &0.18~~ &0.13~~& 160 &~~~~1& 0.13~~& 0.063 &~~74 &~~0.6  \\
092837.98$+$602521.0\dotfill & 8.9 & 0.071 &0.22~~ &0.46~~& 580 &~~~~7& 0.16~~& 0.21~~& 260 &~~3~~~ \\
103059.09$+$310255.8\dotfill & 8.7 & 0.043 &0.13~~ &0.25~~& 300 &~~~~3& 0.098 & 0.11~~& 130 &~~1~~~ \\
110051.02$+$170934.3\dotfill & 8.2 & 0.041 &0.13~~ &~0.092& 120 &~~~~2& 0.093 & 0.026 &~~27 &~~0.1  \\
111230.90$+$181311.4\dotfill & 7.9 & 0.029 &0.091  &0.12~~& 230 &~~30 & 0.066 & 0.032 &~~47 &~~1~~~ \\
130534.49$+$181932.9\dotfill & 8.3 & 0.029 &0.090  &0.10~~& 120 &~~~~1& 0.065 & 0.043 &~~49 &~~0.4  \\
134548.50$+$114443.5\dotfill & 8.1 & 0.031 &0.097  &0.11~~& 180 &~~~~7& 0.071 & 0.051 &~~77 &~~2~~~ \\
\enddata
\tablecomments{Column 2: virial BH mass estimate for the active BH taken from \citet{shen11};
Column 3: BLR size estimated from the 5100 \angstrom\ continuum
luminosity assuming the empirical $R$-$L_{5100}$ relation in
\citet{Bentz2009}; Columns 4--11: estimated BBH model parameters, assuming orbital inclination
$I=45^{\circ}$ (slightly smaller than the median inclination for random
orientations ($I=60^{\circ}$) because edge-on quasars are more likely to be obscured)
and varying mass ratio $q=2$ or $0.5$, where $q\equiv
M_2/M_1$.
$f_r^{-1} R_{{\rm BLR}}$ (Columns 4 and 8) is the estimated lower
limit for the binary separation $d$ (Columns 5 and 9),
if we require that the BLR size is smaller than the Roche radius (see the Appendix for details).
The condition $d>f^{-1}_{r} R_{{\rm BLR}}$ is met for $q=2$ in all cases considering uncertainties.
Columns 6 and 10: binary orbital period $P$; Columns 7 and 11: orbital decay timescales
due to gravitational radiation. See Sections \ref{subsec:model} and \ref{subsec:bbhgw}
and the Appendix for more discussion.}
\end{deluxetable*}
%\clearpage
%\end{landscape}
%-------------

%%%%%%%%%%%%%%%%%%%%%%%%%%%%%%%%%%
%
\subsection{Constraints on BBH Model Parameters}\label{subsec:model}

We now discuss the implications of our results for BBH model
parameters.  As demonstrated in Paper I for the general quasar
population, the observed distribution of broad-line radial
accelerations can be used to place constraints on a
hypothetical BBH population. However, this exercise cannot be
straightforwardly implemented for offset-line quasars, because
(1) as discussed above, there is likely a selection bias in
orbital phase. This needs to be properly accounted for in order
to translate the observed acceleration distribution into
unbiased constraints on BBH parameters. However, the specific
selection function is unclear because of the complication due
to visual selection; and (2) the size of the followup sample is
too small to derive robust statistical constraints as we did in
Paper I.

In view of these difficulties, we adopt here a different
approach: we derive model parameters for the purported BBH
candidates, using plausible assumptions for the orbital
inclination $I$ and mass ratio $q\equiv M_2/M_1$. Following
Paper I, we assume that the BBH is on a circular orbit and that
only BH number 1 is active and powering the observed BLR. We
then set the velocity $V_1$ equal to the average of the
broad-line velocity offsets from the SDSS and followup spectra,
and determine the acceleration $a_1$ from the difference in
these offsets. We adopt the mass $M_1$ from the virial BH mass
estimate and the radius of the BLR from the 5100 \angstrom\
continuum luminosity, both taken from \citet{shen11}. Then we
find the orbital phase $\phi$ and binary separation $d$ as
described in the Appendix. We then make the following
consistency checks. The binary separation $d$ should at least
satisfy $d>R_{{\rm BLR}}$ for the model to be self-consistent.
If we further require that the BLR size is smaller than the
Roche radius, the binary separation must also satisfy
$d>f^{-1}_{r} R_{{\rm BLR}}$, where $f_r$ is the average radius
of the Roche lobe in a circular binary system (Equation
(\ref{eq:fr})).

Table \ref{tab:model} lists model parameters for the nine BBH
candidates.  The baseline model assumes $I=45^{\circ}$,
slightly smaller than the median inclination for random
inclinations ($I=60^{\circ}$) because edge-on quasars are more
likely to be obscured, and $q=2$, because some theoretical work
\citep[e.g.,][]{Dotti2006} suggests that the less massive BH in
a BBH is more likely to be active. For the baseline model, both
conditions on the separation are met for all
candidates\footnote{For two objects the listed $d$ is slightly
smaller than $f^{-1}_{r} R_{{\rm BLR}}$ at face value, but the
result is also consistent with $d>f^{-1}_{r} R_{{\rm BLR}}$
considering uncertainties.}; $d$ is $\sim$2--7 times larger
than the BLR size $R_{{\rm BLR}}$. As discussed in Paper I, one
caveat in these arguments is that our adopted BLR size is
estimated from the $R$--$L$ relation \citep[][]{Bentz2009}
found by reverberation mapping studies, which represents the
emissivity-weighted average radius and could be an
underestimate of the actual size. Nevertheless the bulk of the
line emission should come from within the adopted BLR size.
Models with smaller mass ratios (e.g., $q=0.5$) are less
favored, because the Roche condition is violated in most cases
(six out of nine), unless the inclination angle is higher than
the random case (i.e., $I>60^{\circ}$). Models with much larger
mass ratios (i.e., $q\gg1$) are possible but perhaps also less
likely, because the resulting total BH mass would be too large
for the galaxies to obey the $M_{{\rm BH}}$--$\sigma_{\ast}$
relation observed in local inactive galaxies
\citep[e.g.,][assuming that quasar host galaxies follow the
same relation as local inactive galaxies and that the gas
velocity dispersion can be used as an approximate surrogate for
the stellar velocity dispersion $\sigma_{\ast}$]{Gultekin2009}.

These tests suggest that there is some permitted parameter
space in the BBH model that explains our observations for the
nine suggested BBH candidates. While they neither validate the
BBH hypothesis nor prove our oversimplified model assumptions,
these tests do show that our models are self-consistent, and
suggest that they provide a viable explanation of the data.

%%%%%%%%%%%%%%%%%%%%%%%%%%%%%%%%%%
%
\subsection{Implications for BBH Orbital Evolution}\label{subsec:orbit}

The model parameters that we derived in the previous section
are based on the assumption that the binary separation is
larger than the BLR size so that the broad-line bulk velocity
would trace binary orbital motion. Under this assumption, our
followup sample is most sensitive to $d\lesssim 1$ pc given the
acceleration measurement accuracy (median 1$\sigma$ error of
$\sim10$ km s$^{-1}$ yr$^{-1}$) and the estimated BH masses.
However, this assumption is not necessarily true, even if the
observed broad-line velocity shift is indeed caused by BBH
orbital motion. If the binary separation is smaller than
typical BLR sizes (e.g., $d<0.05$ pc), the relation between
orbital frequency and the temporal broad-line velocity shift
would be more complicated
\citep[e.g.,][]{Bogdanovic2008,shen10}.

Stellar dynamics simulations suggest that the orbital evolution
of BBHs in spherical galaxies may stall at $\sim$pc scales
\citep[e.g.,][the so-called ``final-parsec''
problem]{begelman80,milosavljevic01,yu02,Vasiliev2013}, but the
barrier may be overcome in more realistic models of triaxial or
axisymmetric galaxies \citep[e.g.,][but see
\citealt{Vasiliev2013}]{yu02,Merritt2004,Preto2011,Khan2013} or
in gaseous environments (e.g., \citealt{gould00,Escala2005};
but see \citealt{Lodato2009,Chapon2013}). If some of the
suggested sub-pc BBH candidates (Section
\ref{subsubsec:candidate}) were confirmed with future long-term
spectroscopic monitoring, and their gravitational radiation
decay timescales were less than the Hubble time (discussed
below in Section \ref{subsec:bbhgw}), the presence of such
systems would be evidence that the final-parsec barrier can be
overcome at least in some quasar host galaxies.

%%%%%%%%%%%%%%%%%%%%%%%%%%%%%%%%%%
%
\subsection{Implications for GW Source Density}\label{subsec:bbhgw}

Assuming that the suggested candidates (Section
\ref{subsubsec:candidate}) are indeed BBHs in which one member
is active, our results would indicate a lower limit for the
fraction of sub-pc BBHs in SDSS quasars at $z<0.83$ as
$(N_{{\rm offset}}/N_{{\rm qso}}) (N_{{\rm BBH}}/N_{{\rm
followup}}) = (399/20774)(9/50) \sim 3.5\times 10^{-3}$.
Considering that the space density and average duty cycle of
SDSS quasars with $L_{{\rm Bol}}>10^{45}$ erg s$^{-1}$ at
$z\sim0.5$ (which are typical values of our sample) are
$\sim2\times10^{-6}$ Mpc$^{-3}$ and $\sim4\times10^{-3}$
\citep[e.g.,][]{shen09}, respectively, this would imply that
the space density of sub-pc BBHs at $z\sim0.5$ is $n_{{\rm
sub-pc\, BBH}}>(2\times10^{-6})(3.5\times
10^{-3})/(4\times10^{-3})$ Mpc$^{-3}$ $\sim 2\times10^{-6}$
Mpc$^{-3}$ for systems with masses comparable to those of SDSS
quasars (assuming that the duty cycle for BBHs is the same as
for single BHs, where duty cycle is defined as the probability
of a galaxy harboring a quasar). This is far smaller than the
density of luminous galaxies,
$\sim5\times10^{-3}\mbox{\,Mpc}^{-3}$, but this is not
surprising since our detection method is sensitive to only a
small fraction of the likely BBH parameter space.

\begin{figure}
  \centering
    \includegraphics[width=80mm]{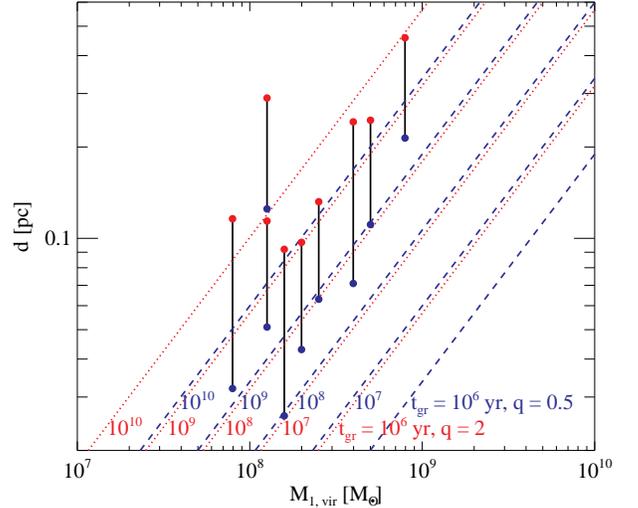}
    \caption{Virial BH mass (for the active member) vs. estimated binary
    separations in the BBH model for the nine BBH candidates in Table \ref{tab:model}.
    The red (blue) points denote mass ratio $q=2$ ($q=0.5$).
    Each solid line connects the two different estimates for the same object.
    The red dotted (blue dashed) lines are constant orbital decay
    timescales due to gravitational radiation, $t_{{\rm gr}}$ (Equation (\ref{eq:tgr})), for
    circular binaries with mass ratio $q=2$ ($q=0.5$).
    $t_{{\rm gr}}$ is of order the age of the universe at the quasar redshifts for $q$ of
    order unity
    (see Section \ref{subsec:bbhgw} for more discussion).
    }
    \label{fig:gwtime}
\end{figure}

Figure \ref{fig:gwtime} shows the BBH candidates in the
$M_1$--$d$ plane, using the virial estimate $M_1$ for the
active BH 1 and the binary separation $d$ as discussed in
Section \ref{subsec:model} assuming mass ratio $q = 0.5, ~2$
and orbital inclination $I=45^{\circ}$ (Table \ref{tab:model}).
The red dotted (blue dashed) line shows the orbital decay time
due to gravitational radiation in a circular binary with a mass
ratio $q = 2$ ($q=0.5$), which is given by \citep{Peters1964}
\begin{equation}\label{eq:tgr}
t_{{\rm gr}} =
\frac{5}{256}\frac{c^5}{G^3}\frac{d^4}{q(1+q)M_1^3}~.
\end{equation}
Most of the purported BBH candidates have GW decay times
$t_{{\rm gr}}\gtrsim 10^9$ yr (Table \ref{tab:model}). $t_{{\rm
gr}}$ could be an overestimate for the actual orbital decay
time, if other more efficient decay processes (either stellar
or gaseous) are at work. For $q\sim1$, $t_{{\rm gr}}$ is of
order the age of the universe at the quasar
redshifts.\footnote{As discussed in Section \ref{subsec:model},
$q \ll 1$ is less favored for the nine detections shown in
Figure \ref{fig:gwtime}, given the Roche lobe requirement; $q
\gg1$ is still viable for the nine detections which would yield
$t_{{\rm gr}}$ much longer than the Hubble time; however, this
possibility is perhaps less likely since the resulting total BH
masses would be too large for the quasars to obey the same BH
mass--bulge scaling relations as local inactive galaxies.} None
of the detected systems has short gravitational decay time
(e.g., $t_{\rm gr}\ll$ the Hubble time, so that the probability
of detecting them is very small). This is reassuring evidence
that the BBH interpretation is at least self-consistent.

%%%%%%%%%%%%%%%%%%%%%%%%%%%%%%%%%%%%%%%%%%%%%%%%%%%%%%%%%%%%
\section{Summary and Future Work}\label{sec:sum}

Long-term spectroscopic monitoring of statistical quasar
samples is used to test the hypothesis that some of them may
contain sub-pc BBHs. The approach is to look for temporal RV
shifts of the broad lines to constrain the purported binary
orbital motion. In Paper I, we have studied the general quasar
population using multiple spectroscopic observations from the
SDSS.  Here in the second paper in the series, we focus on
objects which are pre-selected to have a supposedly higher
likelihood of being BBHs.  We have selected a sample of 399
quasars from the SDSS DR7 whose broad \hbeta\ lines are
significantly (99.7\% confidence) offset from the systemic
redshift determined from narrow emission lines.  The velocity
offset has been suggested as evidence for BBHs, but
single-epoch spectra cannot rule out alternative scenarios such
as accretion disk emitters around single BHs or recoil BHs.

As a pilot program, we have obtained second-epoch optical
spectra from MMT/BCS, ARC 3.5~m/DIS, and FLWO 1.5~m/FAST for 50
of the 399 offset-line quasars, separated by $\sim5$--10 yr
(rest frame) from the original SDSS observations.  We summarize
our main findings in the following.

\begin{enumerate}

\item[1.] We have adopted a $\chi^2$-based
    cross-correlation method to accurately measure the
    velocity shifts between two epochs, with particular
    attention paid to quantifying the uncertainties (Paper
    I).  The velocity shifts have been measured from the
    broad-line components after subtracting the
    pseudo-continua and narrow emission lines with reliable
    spectral decompositions \citep{shen11} and systematics
    control \citep{shen13}. The velocity-shift zero point
    has been calibrated using simultaneous observations of
    the \OIII\ emission lines. We have detected significant
    (99\% confidence) RV shifts in broad \hbeta\ in 24 of
    the 50 followup targets and have placed limits on the
    rest (Section \ref{subsec:detection}). For the detected
    cases, the absolute values of the measured
    accelerations are $\sim10$ to $\sim200$ km s$^{-1}$
    yr$^{-1}$, with a typical measurement uncertainty of
    $\sim10$ km s$^{-1}$ yr$^{-1}$.

\item[2.] Following Paper I, we have divided the 24
    detections into three categories, which include 9 ``BBH
    candidates'', 5 ``BLR variability'', and 10
    ``ambiguous'' cases. For BBH candidates, we require
    that the measured velocity shift is caused by an
    overall shift in the bulk velocity, rather than
    variation in the broad-line profiles. We further
    require that the velocity shifts independently measured
    from a second broad line (either broad \halpha\ or
    \MgII ) are consistent with those measured from broad
    \hbeta .

\item[3.] Compared to the general quasar population (i.e.,
    with no significant broad-line velocity offsets)
    studied in Paper I, our results suggest that the
    frequency of broad-line shifts in offset quasars is
    marginally higher on timescales of more than a few
    years ($\sim$50\%\,$\pm$\,10\% for $\Delta t>5$ yr with
    median velocity shift uncertainty of 50 km s$^{-1}$ for
    offset quasars, compared to $\sim$30\%\,$\pm$\,10\% for
    $\Delta t>5$ yr with median velocity shift uncertainty
    of 40 km s$^{-1}$ for normal quasars), after accounting
    for differences in time separation and measurement
    sensitivity (Section \ref{subsec:frequency}). However,
    the statistics are poor and the results are consistent
    with having no difference between the two populations.
    Offset-broad-line quasars also show larger radial
    accelerations averaged over a few years than normal
    quasars (Section \ref{subsec:vs}), with an intrinsic
    width of the acceleration distribution $\sim$30\%
    broader than that of the superior sample of Paper I,
    although the timescales being probed are very different
    ($>$5 yr in this work compared to an average of $\sim$1
    yr in Paper I) so a fair comparison is difficult.

\item[4.] Combining our followup sample with the
    ``superior'' sample defined in Paper I, we have found a
    tentative (Spearman $P_{{\rm null}}\sim10^{-3}$)
    anti-correlation between the broad-line velocity offset
    (both in peak and centroid) in the first-epoch spectra
    and the acceleration between the first and second epoch
    (Section \ref{subsec:vacorr}). The velocity offset and
    acceleration tend to show opposite signs, in the sense
    that blueshifted (redshifted) broad emission lines in
    the first-epoch spectra are likely to become less
    blueshifted (redshifted) in the second-epoch spectra
    after a few years. However, the correlation is weak and
    would need to be confirmed with larger samples in
    future work.

\item[5.] We have discussed implications of our results
    under the BBH hypothesis (Section
    \ref{subsec:implication}). If the velocity offsets and
    accelerations are due to BBHs, then the observed
    anti-correlation in the velocity offset and the average
    acceleration between two epochs can be explained as
    orbital phase modulation, when the time baseline is a
    non-negligible fraction of (and is smaller than) the
    orbital period. The selection of significant broad-line
    velocity offsets in single-epoch spectra would boost
    the chance of detecting accelerations after a few
    years, given fixed measurement accuracy.

\item[6.] We have estimated orbital parameters for the nine
    BBH candidates assuming that the orbits are circular
    and that only one BH is active (Section
    \ref{subsec:orbit}). For binaries with roughly equal BH
    masses, the estimated binary separations are
    $\lesssim0.1$ pc to $\sim0.5$ pc with orbital periods
    of a few decades to a few centuries (Table
    \ref{tab:model}). The gravitational radiation decay
    timescales are of order the age of the universe at the
    quasar redshifts (Section \ref{subsec:bbhgw}). While
    the models are clearly oversimplified and the BBH
    hypothesis lacks verification, our results suggest that
    the binary models provide a viable explanation of the
    data. If confirmed with future observations, the
    frequency of the BBH candidates among all quasars would
    set a lower limit for the space density of sub-pc BBHs
    at $z\sim0.5$ as $\sim2\times10^{-6}$ Mpc$^{-3}$ (for
    systems with BH masses comparable to those of SDSS
    quasars, assuming that the duty cycle for BBHs is the
    same as for single BHs).

\end{enumerate}

This work and Paper I represent a first step toward identifying
sub-pc BBHs in quasars and in sorting out the origins for the
observed broad emission line velocity offsets. Given that the
orbital periods are expected to be a few decades at least
(Table \ref{tab:model}), detecting a complete binary orbit
would be a challenging exercise, but a continuing constant
acceleration without changes in the profile shape would already
be a strong signature for a BBH. The spectroscopic monitoring
work also needs to be extended to quasars with lower
luminosities, where our method would be sensitive to systems
with a shorter period, at a fixed BH mass (because the expected
BLR size would be smaller, thus allowing systems with smaller
binary separations). In addition, more observations are needed
for a larger sample to confirm the tentative anti-correlation
between broad-line offset and acceleration suggested by the
pilot study here, and to explore the diversity of BBH
candidates. Future followup observations with longer time
baselines would effectively mitigate measurement uncertainty
for the radial acceleration, similar to the situation measuring
stellar proper motions, where the error drops as $T^{-3/2}$
when observing at a steady rate for an interval $T$. For
example, spectra taken in summer 2014 would improve the S/N by
a factor of $\sim30$\%, and spectra taken in 2016 would improve
the S/N by $\sim60$\%. Other future work includes systematic
assessment of the quasar and host galaxy properties
\citep[e.g., spectral energy
distributions;][]{Tanaka2013,Roedig2014,Generozov2014} for
offset-broad-line quasars and comparison studies against normal
quasars or double-peaked broad-line quasars with extreme
velocity offsets, to search for indirect evidence as
complementary tests of the BBH hypothesis.

In addition, our pilot followup program has also detected
objects with kinematically offset broad emission lines whose
offset velocities (and profiles) stay remarkably stable to
within a few km s$^{-1}$ yr$^{-1}$ over many years (rest
frame). This is markedly different from the expectation and
empirical evidence from observations of double-peaked
broad-line objects generally interpreted as disk emitters, and
could signal a recoiling BH. More work \citep[e.g., based on
photoionization arguments to examine whether the ionizing
source is significantly displaced from the galactic
core;][]{bonning07,shields09} is needed to examine the
possibility of recoiling BHs, which will be the subject of a
future paper.

%------------------------------------------------------------------------------
\acknowledgments

We thank Mike Eracleous, Luis Ho, and Alice Shapley for useful
discussion, Perry Berlind, Michael Calkins, and Bill Wyatt for
assistance with FLWO 1.5\,m/FAST queue observations and data
retrieval, and Michael Strauss for his support during the
course of this work. We also thank an anonymous referee for a
prompt and careful report. Support for the work of X.L. and
Y.S. was provided by NASA through Hubble Fellowship grant
numbers HST-HF-51307.01 and HST-HF-51314.01, respectively,
awarded by the Space Telescope Science Institute, which is
operated by the Association of Universities for Research in
Astronomy, Inc., for NASA, under contract NAS 5-26555. This
work was supported in part by NSF grant AST-1312034 (A.L.) and
NASA grant NNX11AF29G (S.T.).

Funding for the SDSS and SDSS-II has been provided by the
Alfred P. Sloan Foundation, the Participating Institutions, the
National Science Foundation, the U.S. Department of Energy, the
National Aeronautics and Space Administration, the Japanese
Monbukagakusho, the Max Planck Society, and the Higher
Education Funding Council for England. The SDSS Web site is
http://www.sdss.org/.

The SDSS is managed by the Astrophysical Research Consortium
for the Participating Institutions. The Participating
Institutions are the American Museum of Natural History,
Astrophysical Institute Potsdam, University of Basel,
University of Cambridge, Case Western Reserve University,
University of Chicago, Drexel University, Fermilab, the
Institute for Advanced Study, the Japan Participation Group,
Johns Hopkins University, the Joint Institute for Nuclear
Astrophysics, the Kavli Institute for Particle Astrophysics and
Cosmology, the Korean Scientist Group, the Chinese Academy of
Sciences (LAMOST), Los Alamos National Laboratory, the
Max-Planck-Institute for Astronomy (MPIA), the
Max-Planck-Institute for Astrophysics (MPA), New Mexico State
University, Ohio State University, University of Pittsburgh,
University of Portsmouth, Princeton University, the United
States Naval Observatory, and the University of Washington.

Facilities: ARC 3.5\,m (DIS), FLWO:1.5\,m (FAST), MMT (BCS),
Sloan

%-------------------------------------------------------------

%\appendix
\begin{appendix}

\section{Binary Black Hole Model Assumptions}

Following Paper I, we consider a BBH on a circular orbit, where
only BH 1 (with mass $M_1$) is active and powering the observed
broad emission lines. The typical size of the BLR around a
single BH with mass $M_1$ is $R_{{\rm BLR}}\sim 2.7\times
10^{-2} (L/10^{45} {\rm erg\, s}^{-1})^{1/2}$ pc
\citep{shen10}.
%
%We assume that the total BH mass $M_{{\rm tot}}\equiv M_1 + M_2$ follows the local
%$M_{\bullet}$-$\sigma_{\ast}$ relations in inactive galaxies.
%Since our targets are luminous quasars which outshine their
%host galaxy starlight, $\sigma_{\ast}$ is not measurable from
%the available spectra. As a surrogate, we assume $\sigma_{{\rm
%NLR}}\approx\sigma_{\ast}$ where we use narrow emission lines
%to measure $\sigma_{{\rm NLR}}$.
The orbital period, LOS velocity and acceleration of
the active BH are (Paper I)
%% \begin{equation}
%% \begin{split}
%% P&=2\pi d^{3/2} (GM_{{\rm tot}})^{-1/2} = 9.4 d^{3/2}_{0.01}M^{-1/2}_{8,{\rm tot}}~~~~{\rm yr}, \\
%% V_1 &= \frac{M_2}{M_{{\rm tot}}}\bigg(\frac{GM_{{\rm tot}}}{D}\bigg)^{1/2} \sin I \sin \phi
%% = 6560 \bigg(\frac{M_2}{M_{{\rm tot}}}\bigg)M^{1/2}_{8,{\rm tot}}d^{-1/2}_{0.01}\sin I\sin \phi~~~~{\rm km~s}^{-1}, \\
%% a_1&=\frac{GM_2}{d^2}\sin I \cos \phi = 4400\bigg(\frac{M_2}{10^8M_{\odot}}\bigg)d^{-2}_{0.01} \sin I\cos \phi~~~~{\rm km~s}^{-1}~{\rm yr}^{-1},
%% \end{split}
%% \end{equation}
%%
\begin{equation}\label{eq:orbit}
\begin{split}
P&= 9.4 d^{3/2}_{0.01}M^{-1/2}_{8,{\rm tot}}~~~~{\rm yr}, \\
V_1 &= 6560 \bigg(\frac{M_2}{M_{{\rm tot}}}\bigg)M^{1/2}_{8,{\rm tot}}d^{-1/2}_{0.01}\sin I\sin
\phi~~~~{\rm km~s}^{-1}, \\
a_1&= 4400\bigg(\frac{M_2}{10^8M_{\odot}}\bigg)d^{-2}_{0.01} \sin I\cos \phi~~~~{\rm
km~s}^{-1}~{\rm
yr}^{-1},
\end{split}
\end{equation}
where subscripts 1 and 2 refer to BH 1 and 2, respectively, $I$
is the orbital inclination, $d$ is the binary separation,
$\phi$ is the orbital phase, $M_{{\rm tot}}\equiv M_1 + M_2$,
$M_{8,{\rm tot}}\equiv M_{{\rm tot}}/10^8 M_{\odot}$, and
$d_{0.01}\equiv d/0.01{\rm pc}$.

We have set $V_1$ equal to the average of the velocity offsets
from the SDSS and followup spectra and determined $a_1$ from
the difference in these offsets. Assuming values for the
orbital inclination $I$ and mass ratio $q\equiv M_2/M_1$, we
can solve the above equations for $d$ and $P$. From the last
two relations in Equation (\ref{eq:orbit}), we get a quartic
polynomial in $d_{0.01}$:
\begin{equation}\label{eq:d}
\begin{split}
C_1 d^4_{0.01} + C_2 d_{0.01} -1 &=0 ,
\end{split}
\end{equation}
where $C_1 = (a_1/4400~{\rm km}~{\rm s}^{-1}~{\rm
yr}^{-1})^2(M_1/10^8 M_{\odot} )^{-2} q^{-2} (\sin I)^{-2}$ and
$C_2 = (V_1/6560~{\rm km}~{\rm s}^{-1})^2q^{-2}(1+q)(M_1/10^8
M_{\odot})^{-1} (\sin I)^{-2}$. The equation always has two
real roots (one positive and one negative) and two complex
conjugate roots, and we take the real positive root as the
solution. The resulting $d$ should at least satisfy $d>R_{{\rm
BLR}}$ for the model to be self-consistent. If we further
require that the BLR size must be smaller than the Roche
radius, the result must satisfy the stronger constraint
$d>f^{-1}_{r} R_{{\rm BLR}}$, where $f_r$ is the average radius
of the Roche lobe in a circular binary system
\citep[e.g.,][]{Paczynski1971}:
\begin{equation}\label{eq:fr}
\begin{split}
f_r &= 0.38 - 0.2 \log q,~~~~ 0.05 < q < 1.88 \\
& = 0.46224(1+q)^{-1/3},~~~~q>1.88.
\end{split}
\end{equation}

%\section{Offset Quasars with Significant Radial Velocity Shifts between Two Epochs}
%
%In Figures \ref{fig:vshift}--\ref{fig:vshift3} we show the 24
%quasars with significant (99\% confidence) detections of acceleration.

\end{appendix}

\bibliography{recoil}

\end{document}